\documentclass[fleqn,usenatbib]{mnras}
\usepackage{amssymb}	% Extra maths symbols

\usepackage{newtxtext,newtxmath}
\usepackage[T1]{fontenc}

\DeclareRobustCommand{\VAN}[3]{#2}
\let\VANthebibliography\thebibliography
\def\thebibliography{\DeclareRobustCommand{\VAN}[3]{##3}\VANthebibliography}

\usepackage{graphicx}	% Including figure files
\usepackage{amsmath}	% Advanced maths commands
\usepackage{siunitx}
\usepackage{float}
\usepackage{makecell}
\usepackage{amsfonts,mathbbol}
\usepackage{physics}
\usepackage{gensymb}
\usepackage{longtable}
\usepackage{caption}
\usepackage{subcaption}
\title[Kicks of NSs in binaries]{An Observationally-Derived Kick Distribution for Neutron Stars in Binary Systems.}
\author[T.~N.~O'Doherty et al.]{Tyrone~N.~O'Doherty $^{1}$\thanks{E-mail: tyrone.odoherty@postgrad.curtin.edu.au},
Arash~Bahramian$^{1}$,
James~C.~A.~Miller-Jones$^{1}$,
Adelle~J.~Goodwin$^{1}$,
\newauthor
Ilya~Mandel$^{2,3}$,
Reinhold~Willcox$^{2,3}$,
Pikky~Atri${^4}$,
and Jay~Strader${^5}$
\\
$^{1}$International Centre for Radio Astronomy Research -- Curtin University, GPO Box U1987, Perth, WA 6845, Australia\\
$^{2}$Monash Centre for Astrophysics, School of Physics and Astronomy, Monash University, Clayton, Victoria 3800, Australia\\
$^{3}$The ARC Center of Excellence for Gravitational Wave Discovery – OzGrav, Australia\\
$^{4}$ASTRON, Netherlands Institute for Radio Astronomy, Oude Hoogeveensedijk 4, 7991 PD Dwingeloo, The Netherlands\\
$^{5}$Center for Data Intensive and Time Domain Astronomy, Department of Physics and Astronomy, Michigan State University, East Lansing, MI 48824, USA
}

\date{Accepted XXX. Received YYY; in original form ZZZ}

\pubyear{2022}

\begin{document}
\label{firstpage}
\pagerange{\pageref{firstpage}--\pageref{lastpage}}
\maketitle

% Abstract of the paper
\begin{abstract}

Understanding the natal kicks received by neutron stars (NSs) during formation is a critical component of modelling the evolution of massive binaries. Natal kicks are an integral input parameter for population synthesis codes, and have implications for the formation of double NS systems and their subsequent merger rates. However, many of the standard observational kick distributions that are used are obtained from samples created only from isolated NSs. Kick distributions derived in this way overestimate the intrinsic NS kick distribution. For NSs in binaries, we can only directly estimate the effect of the natal kick on the binary system, instead of the natal kick received by the NS itself. Here, for the first time, we present a binary kick distribution for NSs with low-mass companions. We compile a catalogue of 145 NSs in low-mass binaries with the best available constraints on proper motion, distance, and systemic radial velocity. For each binary, we use a three-dimensional approach to estimate its binary kick. We discuss the implications of these kicks on system formation, and provide a parametric model for the overall binary kick distribution, for use in future theoretical modelling work. We compare our results with other work on isolated NSs and NSs in binaries, finding that the NS kick distributions fit using only isolated pulsars underestimate the fraction of NSs that receive low kicks. We discuss the implications of our results on modelling double NS systems, and provide suggestions on how to use our results in future theoretical works.

\end{abstract}

% Select between one and six entries from the list of approved keywords.
% do not make up new ones.

%%%%%%%%%%%%%%%%%%%%%%%%%%%%%%%%%%%%%%%%%%%%%%%%%%

%%%%%%%%%%%%%%%%% BODY OF PAPER %%%%%%%%%%%%%%%%%%

\section{Introduction}
\label{sec:intro}
Neutron stars (NSs) are ultra-dense remnants left behind after massive stars end their lives in a supernova explosion, and are most often observed as pulsars and in X-ray binaries (XRBs). Pulsars are highly magnetic, rapidly rotating neutron stars that produce beamed radio emission. Typical pulsars are young ($<10^8$ yr) and have spin periods $\sim$ \SI{1}{\second}. However, millisecond pulsars (MSPs) are much older ($ > 10^9$ yr) with spin periods $< 30$ \SI{}{\milli \second} and weaker magnetic fields. The NS XRBs are binary systems with a NS accreting material from a stellar companion. Systems with low-mass companions ($M_c \leq 1$ $ M_\odot$) are called low-mass XRBs (LMXBs) and are long lived ($ > 10^9$ yr) with stable mass transfer occurring during the long lifetime of the companion.

Pulsars, NS LMXBs, and MSPs are all linked in the leading evolutionary model for MSPs (\citealt{Alpar82}; see \citealt{Lorimer08} for a review). Over time, due to the loss of rotational energy or due to propeller-mode accretion from the companion, a pulsar's period decreases. After it has spun down to a period greater than about several seconds, the pulsar ceases to produce significant amounts of radio emission. For solitary pulsars, this marks the end of their observability. However, for the pulsars in low-mass binary systems, accretion onto the NS can commence once the stellar companion is significantly evolved or once the orbit of an ultracompact system has shrunk sufficiently via gravitational wave radiation, initiating Roche lobe overflow (RLOF). In this process matter and angular momentum are transferred to the NS and it is spun up to rotational periods as short as a few ms. At this point the NS is fully `recycled' and observable as a rapidly spinning MSP.

However, many NS binaries do not fit cleanly into the NS LMXB or MSP classification. 
SAX J1808.4-3658 was the first such intermediate system; it was associated with X-ray pulsations with a period of 2.49 \SI{}{\milli \second} \citep{Wijnands98}. The system showed consistent rapid pulses of emission, like a MSP, however, the emission was in the X-ray and not radio. In this system the accreted material is funnelled onto the surface of the NS at the magnetic poles creating X-ray hotspots that come into our line of sight as the NS rotates. Since this discovery, other such accreting millisecond X-ray pulsars (AMXPs) have been discovered (see \citealt{Patruno21} for a review). The discovery of transitional millisecond pulsars (tMSPs) -- systems that have been observed as both accretion-powered NS LMXBs and rotation-powered MSPs -- provided the strongest evidence for the MSP evolution theory to date \citep{Archibald09}. It is believed that these systems are in intermediate stages of evolution and will become regular radio MSPs in the future.

Within the MSP class there are two interesting subclasses, the redbacks and black widows, known as the `spider' pulsars \citep{Roberts13}. These systems are tight binaries ($P_{\mathrm{orb}} < 1$ day) where the MSP is ablating the stellar companion. One of the main differences between the two classes is companion mass; black widow companions are $\ll 0.1$ M$_\odot$ and redback companions are typically $0.1$ M$_\odot < M_c < 0.5 $ M$_\odot$. The evolutionary connection between redbacks and black widows is not clear. Some studies find that redbacks and black widows share a common start to their evolution before diverging (e.g., \citealt{Chen13}, \citealt{DeVito20}), others find that black widows evolve naturally from redbacks (e.g., \citealt{Ginzburg21}), and others argue for a more complex connection between the two spider binary classes (e.g., \citealt{Benvenuto15}).

The masses of the NSs in redbacks can be constrained through good photometric and spectroscopic measurements of the companion star. \citet{Strader19} find a median mass of $1.78 \pm 0.09$ M$_{\odot}$ for neutron stars in redbacks, much higher than the canonical NS mass of $1.4$ M$_{\odot}$. Studies modelling the mass distribution of MSPs in binaries \citep{Antoniadis16}, and more generally NSs in binaries \citep{Alsing18}, find that the NS mass distribution is bimodal. The first mode is at $\approx1.34$ M$_\odot$, and the second is at $\approx1.8$ M$_\odot$. The NSs in redbacks are particularly interesting as almost all are massive and consistent with the upper component of the bimodal NS mass distribution. It is unclear whether the large redback NS masses arise purely from accretion from their companion or if they are born heavy. For black widows, estimating the NS mass is challenging because the companion star is so irradiated that modelling the system becomes very difficult.

NSs in all these systems were likely formed during a supernova, and therefore potentially received a natal kick. It is worth noting that NSs can be formed through accretion-induced collapse (\citealt{Nomoto79}, \citealt{Miyaji80}), however, the fraction of NSs formed through this channel is thought to be very low \citep{Fryer99}. Young, isolated radio pulsars are the class of NS with the most studied natal kicks, with average kicks in the range of $400-500$ \si{\km \per \second} \citep{Lyne94,Hobbs05}. The leading theory that can explain the large space velocity of these isolated pulsars is the `gravitational tug-boat mechanism', whereby asymmetries in the ejected gas facilitates an anisotropic gravitational pull that can accelerate the NS to velocities greater than 700 \SI{}{\km\per\second} \citep{Foglizzo02,Blondin03,Scheck04,Foglizzo06,Scheck06,Foglizzo07,Wongwathanarat10,Wongwathanarat13}. Many other methods exist that can explain lower velocity kicks ($\lesssim$ 200 \SI{}{\km\per\second}), including asymmetries resulting from hyrdodynamic instabilities and anisotropic neutrino emission (see \citealt{Janka12} for a review). For binaries, even a completely symmetric explosion can produce a kick proportional to the fractional mass lost in the supernova, and the pre-supernova orbital velocity \citep{Blaauw61}. This so-called `Blaauw kick' is purely a recoil kick arising due to the rapid mass loss during the supernova explosion.

If the kick is large enough, or greater than half the system mass is ejected in the supernova, the binary will become unbound \citep{Nelemans99}. \citet{Renzo19} found that $86^{+11}_{-9}$\% of NSs born in binaries may end up as isolated, free-floating NSs, when drawing from the kick distribution of \citet{Hobbs05}. Even if the system is not disrupted, the kick can change the binary parameters. However, the natal kick imparted to the NS on formation is hard to measure directly. Instead, we can measure the binary kick; the excess velocity imparted to the system as a result of the explosion. In contrast, the natal kicks of NSs that form from isolated stars can be directly estimated. This is an important difference that needs to be considered when comparing kicks of isolated and binary systems. However, we have no way of knowing whether an isolated NS came from a single star or from a binary.  Indeed, it seems likely the majority of isolated NS progenitors were originally in binaries given the high multiplicity of massive stars (e.g., \citealt{Sana12}, \citealt{Moe17}). There is an obvious selection effect introduced by the isolated NSs that came from binaries; their natal kick must have been large enough (of order the pre-supernova orbtial velocity or larger) to disrupt the binary. Studying how natal kicks differ between young, isolated NSs (from either isolated stars or disrupted binaries) and NSs in binaries (where the binary survived the supernova) will help to better understand the natal kicks NSs can receive on formation.
 
Natal kicks are identified as having a significant impact on the future binary evolution with, for example, \cite{Belczynski99} finding the number of compact object mergers changes by a factor of 30 when the kick velocity varies within the then-current observational bounds. Natal kicks remain a challenging component of population synthesis to this day. Some population synthesis codes (e.g., \textsc{STARTRACK}, \citealt{Belczynski08}; 
\textsc{SEVN}, \citealt{Spera19}) rely on the Maxwellian kick distribution with 1D dispersion $\sigma = 265$ \SI{}{\km\per\second} (mode at 370 \SI{}{\km\per\second}), fit to pulsar proper motions by \citet{Hobbs05}. However, thanks to improvements in the precision of measurements of pulsar proper motions and distances, \citet{Verbunt17} found that this single Maxwellian prescription may be overestimating the pulsar velocities. Furthermore, \citet{Mandel20} found that using the \citet{Hobbs05} kick distribution results in less than 0.2\% of all NSs receiving kicks under 50 \SI{}{\km\per\second}, which appears at odds with the number of pulsars that are retained in globular clusters. The overestimated high velocity kicks are more likely to result in binary disruption, which in turn means fewer binaries remain bound in simulations, reducing the number of systems that can evolve into X-ray binaries, MSPs, double NSs (DNSs), or other compact object binaries. Furthermore, if the binary remains bound, the kick will change the binary's parameters. This is particularly important for compact object binaries, like DNSs, that can coalesce producing gravitational waves (GW). A large kick could alter the binary parameters such that it will not coalesce within a Hubble time, whereas a small kick may leave the binary in a configuration where coalescence will occur, resulting in observable GWs. Other population synthesis codes (e.g., \textsc{COMPAS}, \citealt{Vigna-Gomez18}; \textsc{MOBSE}, \citealt{Giacobbo18a, Giacobbo19}; \textsc{COMBINE}, \citealt{Kruckow18}; \textsc{COSMIC}, \citealt{Breivik20}; \textsc{POSYDON}, \citealt{Fragos22}) use the \citet{Hobbs05} Maxwellian with the addition of another component at lower velocities. The distribution and magnitude chosen for the lower component is motivated by theoretical modelling of electron-capture supernovae (ECSNe; e.g., \citealt{Nomoto84}, \citealt{Nomoto87}, \citealt{Podsiadlowski04}) that are suggested to produce kicks $\lesssim 50$ \SI{}{\km\per\second}. Reducing NS kicks will also impact predictions on isolated BH speeds and intact BH binaries because population synthesis codes often assume that BH kicks are simply momentum-scaled NS kicks. With the recent discovery of gravitational waves (e.g., see \citealt{Abbott19,Abbott21}, and references therein), understanding the impact kicks have on binary evolution is critical to modeling the underlying population that will evolve into merging compact object binaries. Many studies modelling DNS formation and evolution (e.g., \citealt{Tauris17,Vigna-Gomez18,Andrews19}) find that the natal kicks of NSs are a critical component of recreating the Galactic DNS population.

Historically, many of the observational constraints on natal kick velocity distributions have focused on young, isolated NSs (e.g., \citealt{Lyne94,Arzoumanian02,Brisken03,Hobbs05,Faucher-06,Verbunt17,Igoshev20,Igoshev21,Kapil22}). It is significantly easier to estimate the kicks of young objects than old objects. The young pulsar can be assumed to be in the same region in which it was born. The kicks of these young systems can then be estimated by measuring their current space velocity and subtracting contributions from Galactic rotation and the Sun's relative motion. Note that because they are isolated pulsars, it is only possible to measure their proper motion and not radial velocity, so any kick information that is extracted can only be based on two-dimensional (2D) velocities. \citet{Arzoumanian02}, \citet{Brisken03}, \citet{Verbunt17}, and \citet{Igoshev20} found that the velocity distribution of young, isolated pulsars is bimodal, whereas both \citet{Hobbs05} and \citet{Faucher-06} found the velocity distribution to be unimodal. \citet{Igoshev21} combined Be XRBs, a type of high-mass XRB (HMXB) with a NS accretor, with young, isolated pulsars and also found a bimodal velocity distribution.

\begin{figure*}
	\includegraphics[width=\textwidth]{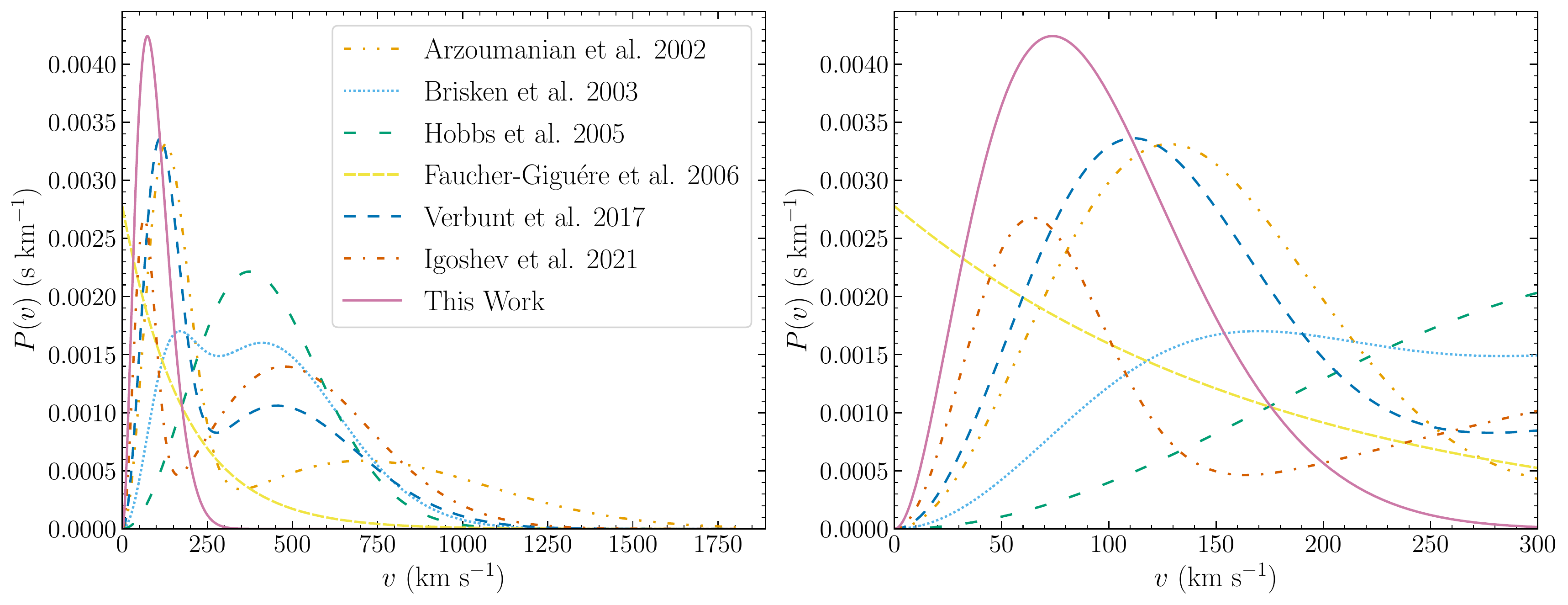}
    \caption{Probability density functions (PDFs) comparing different models of NS kick velocities from previous studies to the distribution derived in this paper for NSs in binaries. Note that all PDFs excluding `This Work' come from observations of young, isolated pulsars. In this work, we present a kick velocity distribution for NSs retained in binary systems. The PDF for `This Work' has been normalised to 0.5 to aide in the comparison to the other PDFs. The right panel shows the same distributions, however, only on the interval from 0 to 300 \si{\km \per \second}.
    }
    \label{fig:previous_work_comparison}
\end{figure*}

Figure \ref{fig:previous_work_comparison} visually compares the different models discussed in the previous paragraph, as well as the model we develop in this work for the binary kicks of NSs in binaries. It is evident that inferring the true underlying kick distribution is a challenging problem. Additionally, the distributions coming from the literature are likely overestimating the true NS natal kick velocities. The population of young, isolated pulsars are more likely to have received larger kicks than NSs that remain in binaries post-supernova, as discussed above. For young, isolated pulsars, if the NS formed from an isolated star then the estimated velocity is reflective of the natal kick. If the NS formed in a binary, the observed speed is not purely attributed to the natal kick since some energy has been spent in the unbinding. However, this difference is small if the kick is much larger than the pre-supernova orbital velocity; \citet{Kapil22} find that predicted velocities of single neutron stars are very similar whether those neutron stars are born in isolation or in binaries. Understanding the contribution of the old NSs that remained in binaries post-supernova is thus important for inferring the true NS kick distribution.

Studying the natal kicks of old NSs is also challenging. Due to their age they have likely moved from the region in which they were born, and experienced acceleration in the Galactic potential. Therefore, applying the same methodology as for young NSs estimates their current peculiar velocity. Peculiar velocity is not a conserved quantity due to acceleration in the Galactic potential as the system orbits the Galaxy, and thus is not necessarily representative of the kick. It is important to note that these velocities should not be used as the kicks NS binaries receive on formation. Substantial work has been done studying the current velocities of MSPs (e.g., \citealt{Toscano99,Hobbs05,Lommen06,Gonzalez11,Desvignes16,Lynch18}). As for isolated pulsars, these studies use 2D velocities from proper motion. The mean 2D velocity for MSPs is typically within $70-120$ \SI{}{\km\per\second}, with \citet{Lynch18} finding somewhat higher average speeds of $152 \pm 48$ \SI{}{\km\per\second} or $234 \pm 143$ \SI{}{\km\per\second}, depending on the method of estimating distance to the MSPs. Understanding the kicks these systems received is critical as the kick distributions derived from young, isolated pulsars do not appropriately consider the lower velocity kicks that do not disrupt these binary systems.

Recently, \citet{Atri19} developed a novel method for estimating the binary kicks of old binary systems. The technique uses the system's distance, proper motion, and systemic radial velocity to robustly estimate the potential binary kick distribution for each system. See Section \ref{sec:pikky_kick} for a more detailed discussion of this method. Whilst \citet{Atri19} focused on black hole low-mass XRBs, this technique is equally applicable to estimating the binary kicks of old NSs like MSPs and NS LMXBs.

In this paper, we compile a sample of 145 old NSs and estimate the potential binary kick distribution they could have received at birth, comparing them first to different classes of old NSs, and then to other NSs and black holes. Section \ref{sec:methods} describes our approach to estimating distances and our method of estimating the kicks. In Section \ref{sec:data}, we discuss the different classes of old NSs we investigate in this paper. Our results are presented in Section \ref{sec:results}. We discuss interpretations of our results as well as their limitations in Section \ref{sec:discussion} and give recommendations for those who wish to incorporate this work in population synthesis, binary evolution, or modelling the population of GW sources.

\section{Methods}
\label{sec:methods}

\subsection{Distances}
\label{sec:distance}

Estimating distances to Galactic sources is, typically, not a straightforward task, and many methods exist. Here we go over the relevant methods for our work, presented in order of accuracy. High significance very long baseline interferometry (VLBI) or optical parallax is the only model-independent method of distance determination, and hence the most accurate. However, at low significance where a distance prior is required it is no longer model-independent. We define a reliable parallax as $\frac{\pi}{\sigma_{\pi}} > 5$, where $\pi$ represents the measured parallax and $\sigma_{\pi}$ represents the uncertainty in the measured parallax. To accurately estimate the uncertainties associated with a parallax-based distance we use a Bayesian inference approach \citep{Astraatmadja16}. The Bayesian approach requires both a likelihood function and an appropriate prior, whereby the distance posterior, $P(D|\pi)$, is calculated as
\begin{equation}
    P(D|\pi) \propto P(D) P(\pi|D),
\end{equation}
where $\pi$ is the measured parallax, $D$ is the distance to the source, $P(D)$ is the distance prior, and $P(\pi|D)$ is the parallax likelihood function. Assuming the measured parallax and its error follow a Gaussian distribution, we can define the likelihood function as
\begin{equation}
    P(\pi|D) = \frac{1}{\sqrt{2\pi}\sigma_{\pi}} \exp{-\frac{1}{2}\frac{(1/D - \pi)^2}{\sigma_{\pi}^2}},
\end{equation}
where $\sigma_{\pi}$ is the parallax measurement uncertainty. As almost all the systems studied in this work are NS LMXBs, or are thought to have evolved from them, we follow the approach of \citet{Gandhi19} and \citet{Atri19} for producing a prior for distances of LMXBs. Using the following equations from \citet{Grimm02}, we create a model of the Galactic space densities of LMXBs that considers the Galactic thin disk, bulge, and halo:

\begin{equation}
\begin{aligned}
\mathrm{\rho_{Bulge}} = &~ \rho_{\mathrm{0,Bulge}}\cdot\left(\frac{\sqrt{r^{2}+\frac{z^{2}}{q^{2}}}}{r_{0}}\right)^{-\gamma}\cdot\exp\left({-\frac{r^{2}+\frac{z^{2}}{q^{2}}}{r_{\mathrm{t}}^{2}}}\right),\\
\mathrm{\rho_{Disk}} = &~ \mathrm{\rho_{0,Disk}}\cdot\exp\left({-\frac{r_{\mathrm{m}}}{r_{\mathrm{d}}}-\frac{r}{r_{\mathrm{d}}}-\frac{|z|}{r_{\mathrm{z}}}}\right),\\
\mathrm{\rho_{Sphere}} = &~ \mathrm{\rho_{0,Sphere}}\cdot\frac{\exp\left({-b\cdot(\frac{R}{R_{\mathrm{e}}})^{\frac{1}{4}}}\right)}{(\frac{R}{R_{\mathrm{e}}})^{\frac{7}{8}}},
\end{aligned}
\end{equation}
with
\begin{equation}
\begin{aligned}
r = &~ \sqrt{R^2_0 + D\cos^2b - 2dR_0\cos{b}\cos{l}},\\
R = &~ \sqrt{R^2_0 + D^2 - 2dR_0\cos{b}\cos{l}},\\
z = &~ D\sin{b}.
\end{aligned}
\end{equation}

\noindent
Here \textit{l} and \textit{b} are the source's Galactic coordinates, and $q$, $r_\mathrm{0}$, $r_\mathrm{t}$, $r_\mathrm{m}$, $r_\mathrm{d}$, $r_\mathrm{z}$, and $R_e$ are scale parameters derived by, and tabulated in, \citet{Grimm02}. Following \citet{Atri19}, we use values of $1.1$, $2.6$, and $13.0$ $\mathrm{M_{\odot}}$ pc$^{-3}$ for $\rho_{\mathrm{0,Bulge}}$, $\mathrm{\rho_{0,Disk}}$, and $\mathrm{\rho_{0,Sphere}}$, respectively, as appropriate for the Milky Way Galaxy. Therefore, the distance prior is this LMXB space density model multiplied by $D^2$:
\begin{equation}
    P(D,l,b) \propto (\mathrm{\rho_{Bulge}} + \mathrm{\rho_{Disk}} + \mathrm{\rho_{Halo}}) \cdot D^2.
\end{equation}
While $P(D,l,b)$ is a function of Galactic coordinates as well as distance, the solid angle encompassing the coordinate uncertainties is orders of magnitude smaller than the scales on which the Galactic model impacts the PDF. Therefore, we consider the coordinates to be fixed for a source and do not consider the uncertainty associated with them. As such, we evaluate the prior $P(D,l,b)$ only for different $D$. This prior is only applied to distances coming from a parallax measurement.

Throughout this work, we quote the median of the distance posterior as the final distance estimate for the system. The uncertainty on distance is calculated using the 15.9 and 84.1 percentiles, such that the upper and lower uncertainties enclose 68.3\% of the posterior, which corresponds to 1$\sigma$ uncertainties for a Gaussian distribution. We use the same method of point and interval estimation for all posteriors in this work, unless otherwise specified.

If a reliable parallax is not available for a source, we turn to optical lightcurve fitting. This technique involves estimating the stellar companion's intrinsic luminosity, correcting for extinction, and then comparing the predicted absolute magnitude to the observed apparent magnitude (e.g., \citealt{Romani11}, \citealt{Kaplan13}). Where this is not feasible, we turn to the \citet{Cordes02} (NE2001) and \citet{YMW16} (YMW16) dispersion measure (DM) models to estimate distances.

The NE2001 and YMW16 DM models split the Galactic interstellar medium (ISM) into thin- and thick-disk components and include more complex features. An important component of DM models is the integrated column density of the entire Galactic disk that depends on the scale height of the thick disk and mean electron density at mid-plane. \citet{Ocker20} estimate a new electron density scale height for the thick disk and mid-plane density whilst reviewing the major differences between previous models. The NE2001 model uses a scale height for the thick disk that empirical studies have since shown to be too low (\citealt{Gaensler08}, \citealt{Savage09}), whilst YMW16 uses a scale height consistent with \citet{Gaensler08} and \citet{Savage09}. However, YMW16 uses a total integrated column density that is significantly lower than suggested by these studies, whilst the total integrated column density used by NE2001 is consistent with these studies. Further complicating the comparison is that the two models approach the problem of inhomogeneities in the ISM differently. Therefore, we chose to report both as neither is well accepted as superior (e.g., \citealt{Deller19}; \citealt{Price21}). Hence, to account for the uncertainties in modelling, we assume an uncertainty of 25\% in the estimated distance as is typically assumed in the literature for this approach (we trialled using an uncertainty of 50\%, however, this did not introduce any statistically significant differences).

Where we do not have a reliable parallax for a source, which would allow direct estimation of a distance posterior PDF, we assume the distance estimate and errors follow a Gaussian distribution (e.g., when distance is based on DM). We then use this to construct a PDF for the distance to the source.

\subsection{Kicks}
\label{sec:pikky_kick}
\citet{Atri19} estimate the potential binary kick of a system using its measured parallax, proper motion, and systemic radial velocity. They attempted to account for the effects of acceleration in the Galactic potential by using a 3D treatment. They determine the 3D Galactocentric orbits of the system of interest, and evolve the orbit back 10 Gyr, recording the peculiar velocity of the system every time it crosses the Galactic plane. Note that the technique is not overly sensitive to the choice of 10 Gyr as evolving over 1 Gyr gives very similar results, see \citet{Atri19}. Using the input distributions of position (RA and Dec), proper motion ($\mu_{\alpha}\cos\delta$ and $\mu_{\delta}$), distance (D), and systemic radial velocity ($\gamma$), they used Monte-Carlo (MC) simulations to estimate the peculiar velocities at plane crossing using $\sim$5,000 random draws. As in \citet{Atri19}, we use the MWPotential2014 model from \citet{galpy} as the Galactic potential in which orbits are evolved. The distance distribution is either the parallax distance posterior, $P(D|\pi)$, or the Gaussian probability function from other distance-determination methods. These estimated peculiar velocities form the `potential kick velocity' (PKV) distribution. This distribution of velocities is used as a proxy for the kick the system (not the formed compact object) could have received at birth as a result of formation of the compact object. It is important to note one of the main underlying assumptions is that the system was formed in the Galactic plane. As all systems in this work were selected such that they are in the Galactic field, we implicitly assume they formed in the plane. However, it is worth noting that some of the systems could have been ejected from globular clusters (e.g., \citealt{King03}). The code detailing the \citet{Atri19} process of estimating the PKV distribution can be found at \url{https://github.com/pikkyatri/BHnatalkicks}.

However, this method introduces a mild bias against large kicks. Each MC realisation will produce an array of PKVs, and these PKV arrays are concatenated for all MC realisations to create the PKV distribution. A lower kick will typically result in more plane crossings than a large kick. And as such, MC realisations that produce low PKVs will contribute more data points to the distribution. To resolve this issue, we sample the same number of PKVs (i.e., plane crossings) from each MC realisation. We perform the analysis similar to \citet{Atri19}, however now keeping track of the number of plane crossings and each PKV for each MC realisation. A low number of plane crossings corresponds to a large kick, but also a large distance from the Galactic plane ($\gg 10$ kpc), making the probability of us seeing the system in its current location in the Galaxy very unlikely. We take the lowest number of plane-crossings and sample the same number of PKVs from each MC realisation. If the lowest number of plane-crossings is $<20$, we discard all MC realisations with $< 20$ crossings and sample 20 PKVs from each MC realisation.

\subsubsection{Validation}
\label{sec:validation}
Validation tests were carried out as follows; we generated 25 synthetic systems in the Galactic plane (not including the Galactic bulge) in bound orbits, and gave them a kick that was randomised in both magnitude (ranging from 15 \si{\km \per \second} to 424 \si{\km \per \second}) and direction. These systems were then evolved through the Galactic potential using \textsc{Galpy} \citep{galpy} (using the same potential as discussed in Section \ref{sec:pikky_kick}) for 10 Gyr. For each of the 25 systems, we took four snapshots of the position and velocity and converted these to observable quantities. The snapshots were selected such that they were at increasing distance (snapshot one was the closest, and snapshot four the farthest, with a maximum allowed distance of 10 kpc from the observer, similar to the maximum distance of the sources in our samples) and were separated in time by at least 200 Myr to sample dfferent Galactocentric orbits. This resulted in a sample of 100 sets of synthetic observable parameters, with a wide range of coordinates, distances, and velocities. Realistic errors of 0.2 \si{mas \per yr} on both components of proper motion, 8 \si{\km \per \second} on radial velocity, and 20\% on distance were applied to the observable parameters. The \citet{Atri19} methodology was then used to estimate the PKV distribution.

For each of the 25 systems, the PKV estimated for each of the four snapshots is consistent. This indicates that the kick estimation method is not sensitive to the time or distance at which the system is observed. There also appears to be no obvious bias in the relation of the estimated PKV with the true kick based on any observable parameter or location in the Galaxy. Furthermore, we find the fraction of systems for which the true kick magnitude falls within a 68\% credible interval to be, on average, 80\%. The true kick magnitude lies within a 95\% credible interval for all 100 estimates. Therefore, we find the \citet{Atri19} kick estimation technique performs as expected, and reliably estimates the true binary kick.

\subsection{Radial Velocity Prior}
\label{sec:RV_prior}
The 3D approach of \citet{Atri19} requires a source's 3D velocity, and hence a measurement of systemic radial velocity. Radial velocity measurements do not exist for a large number of the systems considered in this work, due to a number of reasons such as extinction and faintness of the companion, so we develop a prior for radial velocity to provide an estimate. The massive stars that collapse to form NSs are likely born in the Galactic plane \citep{Urquhart14}. Therefore, a system that received no kick will, in general, move with Galactic rotation. As such, the apparent radial velocity of such a source will be the component of Galactic rotation in the radial direction, as seen from Earth. A system that has received a kick will have had its motion perturbed from this Galactic rotation. Therefore, we devise a prior which uses the radial component of Galactic rotation (see Appendix \ref{appendix:RV_prior}) as the estimate of systemic radial velocity, and some perturbation velocity as its uncertainty (we assume the uncertainties are Gaussian).

We use our sample of systems to estimate the perturbation velocity required for the radial velocity prior. We estimate this velocity using the proper motion, which was been measured for every system. After subtracting out the Galactic rotation there is no reason for the velocity of a system to preferentially be oriented in any specific direction. Hence, the velocity corresponding to each component of proper motion in the direction $l$ and $b$ is estimated, and the contribution due to Galactic rotation is subtracted. Treating the components of proper motion as independent velocities, we then average over all systems and all velocities to calculate the perturbation velocity. In this work our full sample can be split into four subsamples (see Section \ref{sec:data}). We estimate this velocity for each subsample of systems separately.

Therefore, the prior has the following form:
\begin{equation}
    P(\gamma|D) = \mathcal{N}(\gamma_G(D),v_p^2),
\end{equation}
where $\gamma$ is the systemic radial velocity, $D$ is the distance to the source, $\gamma_G(D)$ is the radial component of the velocity of the source's local standard of rest (LSR) relative to the sun (while this is also a function of Galactic coordinates, we consider them fixed for a source for the same reason as in Section \ref{sec:distance}), and $v_p$ is the perturbation velocity of the subsample to which the source belongs.

We verified that this prior does not alter the results significantly by trialling it with systems that have measured systemic radial velocities. We also tested a uniform prior on systemic radial velocity from $-500$ \si{\km\per\second} to $500$ \si{\km\per\second}. This wide, nonphysical prior served to bloat the uncertainties on the PKV estimates but otherwise did not impact the main conclusions of this work.

\subsection{Model Fitting}
\label{sec:model_fit}
Our method for estimating PKVs yields MC samples from the expected PDF for PKV of each system. To infer properties of the underlying distribution of PKV based on our sample, we can use the PKV distributions of each system generated using the \citet{Atri19} methodology to infer properties of the underlying distribution of kicks. We use the bayesian framework outlined by \citet{Mandel10} and \citet{Hogg10} to perform parametric modelling with the following distributions: unimodal truncated Gaussians and bimodal truncated Gaussians following \citet{Atri19}, unimodal Maxwellians (dispersion $\sigma_m$) following \citet{Hobbs05}, and Beta distributions. Negative kick velocities are meaningless in this work as we are estimating the magnitude of the kick. To this end, we use truncated Gaussians for our modelling. These have the form
\begin{equation}
\begin{aligned}
    f(x) = &~ 
    \begin{cases}
    \frac{1}{\sigma} \frac{\phi\left(\frac{x-\mu}{\sigma}\right)}{\Phi\left(\frac{b-\mu}{\sigma}\right) -  \Phi\left(\frac{a-\mu}{\sigma}\right)}, & \text{if } a \leq x \leq b\\
    0, & \mathrm{otherwise}
    \end{cases},\\
    \phi(x) = &~ \frac{1}{\sqrt{2\pi}}\mathrm{exp}\left(-\frac{1}{2}x\right),\\
    \Phi(x) = &~ \frac{1}{2}\left(1+\mathrm{erf}\left(\frac{x}{\sqrt{2}}\right)\right),
\end{aligned}
\end{equation}
where $\mathrm{erf}(x)$ is the error function, $\mu$ and $\sigma^2$ are the mean and variance of the standard Gaussian function, $a$ is the lower bound ($0$ \si{\km \per \second}), and $b$ is the upper bound (we use $2000$ \si{\km \per \second}). Note that $\phi(x)$ and $\Phi(x)$ are the PDF and CDF of the standard Gaussian function, respectively. Therefore in this work we report a mean of $\mu$ and standard deviation of $\sigma_g$ for the unimodal truncated Gaussians, and means of $\mu_1,\mu_2$, standard deviations of $\sigma_{g1},\sigma_{g2}$, and weights $w_1$, and $ w_2 = 1 - w_1$ for bimodal truncated Gaussians. The Beta distribution is defined between 0 and 1, so we use a scaled version to accommodate kicks outside this range. The PDF has the form
\begin{equation}
\label{eq:beta}
    f(x) = \frac{\Gamma(\alpha + \beta)(\frac{x}{s})^{\alpha - 1}(1 - \frac{x}{s})^{\beta - 1}}{\Gamma(\alpha)\Gamma(\beta)s},
\end{equation}
where $\alpha$ and $\beta$ are shape parameters that control the skewness of the distribution, $s$ is the scale parameter with units of \si{\km \per \second}, and $\Gamma$ is the Gamma function. It should be noted that the Beta distribution can be very sensitive to the shape parameters (i.e., skewness and kurtosis of the distribution can change significantly with small changes in the shape parameters). However, in the regime where $\alpha,\beta>1$ and $\alpha<\beta$, the distribution will have positive skew. We also fit bimodal Maxwellians, following \citet{Verbunt17}, however, this model, when fitted to the full sample and individual subsamples, did not yield distinct modes and was indistinguishable from a unimodal Maxwellian model. Therefore, we do not discuss bimodal Maxwellian models further in this work. Furthermore, it is worth noting that Beta distributions with $\alpha>1$ and Maxwellian distributions are continuous, with $P(v < 0) = 0$. This implies $P(v=0)=0$, and therefore, both distributions generally disfavour very low kicks. We do not claim that any of the four distributions are the best possible representation of the underlying distribution, however we seek to compare them and examine what this tells us about the underlying distribution.

Uniform priors were used for all parameters unless otherwise specified. Bounds on the truncated Gaussian means always had a lower limit of 0, and an upper limit that was adjusted to appropriately sample the parameter space. For the bimodal truncated Gaussian model, we enforced $\mu_1 < \mu_2$ to break the symmetry. Bounds on the standard deviation(s) always had a lower limit of 0, and typically had an upper limit of 100 unless the parameter space was insufficiently sampled. The prior for the weight parameter ($w_1$) was between 0 and 1. The Maxwellian parameter ($\sigma_m$) and Beta parameters ($s$, $\alpha$, and $\beta$) are all defined to be greater than 0. In conjunction with this hard boundary on the beta parameters, we also used broad Gaussians as weakly informative priors with the constraint that $\alpha < \beta$ (distribution should have positive skew). To fit these models, we use the Markov Chain Monte Carlo method No U-Turn Sampling (NUTS; \citealt{hoffman14}) as implemented in \textsc{PyMC3} \citep{pymc3}. Convergence was verified using the Gelmin-Rubin diagnostic test \citep{Gelman92}, ensuring $\hat{R}$ close to 1 for all best-fit parameters.

\subsubsection{Model Comparison}
\label{sec:model_comp}
We use Leave-one-out cross-validation (LOO; see \citealt{Vehtari17} for a review) and the corrected Akaike information criterion (AICc; \citealt{Akaike74}, \citealt{Cavanaugh97}, \citealt{Burnham02}, \citealt{Antoniadis16}) statistic to perform model comparison. We prefer LOO over the Widely Accepted Information Criterion (WAIC; \citealt{Watanabe2010}) for direct Bayesian comparison as WAIC has a larger bias than LOO \citep{Vehtari17}. However, \citet{Vehtari17} also note that for small sample sizes the bias in LOO can increase. Therefore, we also estimate AICc which has a penalising term that accounts for low sample sizes. A smaller AICc value suggests a more appropriate fit. A larger LOO weight suggests a preference for that model. 

\section{Sample Selection}
\label{sec:data}
In this work we want to estimate the kicks of old NSs in binaries using the method discussed in Section \ref{sec:pikky_kick}. To this end, we assembled a sample of such systems to which this technique could be applied. This sample totals 145 systems; 14 redback pulsars, 19 NS LMXBs, 17 black widows, and 95 MSPs. The order the systems are introduced is indicative of the quality of estimates of distance, proper motion, and radial velocity across the subsample.

The kick estimation technique we employ in this work assumes the system was born in the Galactic plane and orbits within the Galactic potential. Therefore, we do not consider systems with globular cluster associations as they will move with the cluster's proper motion if retained. Furthermore, the motions of such systems could have been affected through dynamical interactions. Therefore, all systems discussed are assumed to have formed in the Galactic field where we assume the primary cause of peculiar motion is due to the natal kick the system received at the formation of the neutron star.

Below we describe the data collection process for each of the NS binary sub-populations.

\subsection{Redback Pulsars}
\label{sec:redback}
\citet{Strader19} compiled a list of 14 confirmed and 10 candidate redback pulsars in the Galactic field. We use the 14 confirmed redback pulsars as the sample of redback pulsars in this work.

All fourteen redbacks have optical counterparts in \textit{Gaia} early Data Release 3 (eDR3 hereafter; \citealt{GaiaeDR3}) with all sources having estimates of position, proper motion, and parallax \citep{Lindegren21}. Since this work was undertaken \textit{Gaia} Data Release 3 (DR3; \citealt{DR3}) has been released, however, DR3 has the same astrometry as eDR3 and thus does not provide additional useful data for this work. All systems have significant proper motion measurements $>5\sigma$ in eDR3, however, not all systems have high-significance parallaxes. Our approach for estimating the distances to these redback systems is detailed in Section \ref{sec:distance}. The presence of a `literature distance' in Table \ref{tab:rb_table} indicates that the distance used in our analysis came from the literature and was not parallax-based. There are two redbacks for which we deviate from this approach. We elected to go with a parallax-based distance for PSR 1431--0315, as while it is not 5$\sigma$ ($\pi/\sigma_{\pi} = 4.4$), there is no better constrained distance, and the inferred parallax distance agrees with the DM distance well. For PSR J1048+2339 there was no reliable distance estimate. It has a poor parallax measurement in eDR3, and the DM-distance estimates from NE2001 (0.7 kpc) and YMW16 (2.0 kpc) are very different. Therefore, for this system, we perform all calculations using both DM-distance estimates. There are only measured systemic radial velocities for 13 of the 14 redbacks. For the remaining system, PSR J1957+2516, we use the radial velocity prior discussed in Section \ref{sec:RV_prior}.

\begin{center}
    \begin{table*}
        \begin{tabular}{lccccccccc}
            \hline
            Source & $\mu_\alpha\cos{\delta}$ & $\mu_\delta$ & $P_b$ & $\gamma$ & $\pi$ & $d_\pi$ & $d_{\mathrm{lit}}$ & PKV & References\\
                   & (mas yr$^{-1}$)          & (mas yr$^{-1}$) & (days) & (km s$^{-1}$) & (mas)& (kpc) & (kpc) & (km s$^{-1}$) & \\
            \hline
PSR J1023+0038 & $4.760(30)$ & $-17.34(4)$ & $0.19809635690(30)$ & $0.0(2.0)$ & $0.731(22)$ & $1.37^{+0.04}_{-0.04}$ & -- & $153^{+15}_{-28}$ & [1,2,3]\\
PSR J1048+2339 & $-15.45(35)$ & $-11.61(34)$ & $0.250519160(30)$ & $-24(8)$ & $0.5(4)$ & -- & $0.7(2)$ / $2.0(5)^a$ & $69^{+22}_{-20}$ & [4,5,6]\\
XSS J12270--4859 & $-18.77(11)$ & $7.30(9)$ & $0.2878875190(10)$ & $67.0(2.0)$ & $0.49(13)$ & -- & $1.90(10)^b$ & $179^{+16}_{-35}$ & [4,7,8]\\
PSR J1306--40 & $-6.19(14)$ & $4.16(11)$ & $1.09720(16)$ & $32.0(2.0)$ & $0.34(15)$ & -- & $4.7(5)^b$ & $148^{+10}_{-9}$ & [4,9]\\
PSR J1417--4402 & $-4.76(4)$ & $-5.10(5)$ & $5.373720(30)$ & $-15.0(1.0)$ & $0.24(5)$ & $4.51^{+1.20}_{-0.79}$ & -- & $148^{+32}_{-26}$ & [4,10,11]\\
PSR J1431--4715 & $-11.82(13)$ & $-14.52(15)$ & $0.4497391377(7)$ & $-91.0(2.0)$ & $0.56(13)$ & $2.29^{+1.07}_{-0.55}$ & -- & $166^{+67}_{-37}$ & [4,12,6]\\
PSR J1622--0315 & $-13.18(32)$ & $2.30(23)$ & $0.1617006798(7)$ & $-135(6)$ & $0.64(30)$ & -- & $1.10(30)^a$ & $139^{+24}_{-24}$ & [4,13,6]\\
PSR J1628--3205 & $-6.2(5)$ & $-21.43(34)$ & $0.21$ & $-4(7)$ & $0.7(4)$ & -- & $1.20(30)^a$ & $147^{+49}_{-45}$ & [4,14,6]\\
PSR J1723--2837 & $-11.73(4)$ & $-24.050(34)$ & $0.615436473(8)$ & $33.0(2.0)$ & $1.11(4)$ & $0.91^{+0.04}_{-0.04}$ & -- & $147^{+24}_{-29}$ & [4,15,16]\\
PSR J1816+4510 & $-0.06(12)$ & $-4.40(12)$ & $0.36089348170(20)$ & $-99(8)$ & $0.22(10)$ & -- & $4.5(1.7)^b$ & $137^{+15}_{-15}$ & [4,17,18]\\
PSR J1957+2516 & $-4.5(5)$ & $-12.3(1.0)$ & $0.2381447210(7)$ & -- & $2.2(9)$ & -- & $2.7(7)^a$ & $114^{+45}_{-33}$ & [4,19]\\
PSR J2129--0429 & $12.10(7)$ & $10.19(6)$ & $0.63522741310(30)$ & $-64.0(2.0)$ & $0.51(7)$ & $2.01^{+0.33}_{-0.25}$ & -- & $184^{+29}_{-30}$ & [4,20]\\
PSR J2215+5135 & $0.01(24)$ & $2.24(24)$ & $0.172502105(8)$ & $49(8)$ & $0.32(23)$ & -- & $2.90(10)^b$ & $110^{+14}_{-14}$ & [4,21,22]\\
PSR J2339--0533 & $3.92(20)$ & $-10.28(19)$ & $0.19309840181(4)$ & $-49(8)$ & $0.55(18)$ & -- & $1.10(30)^b$ & $100^{+24}_{-21}$ & [4,23,24]\\
            \hline
        \end{tabular}
        \caption{Parameters used in the creation of the PKV distributions for each redback system. No entry in the $\gamma$ column indicates the system did not have a measured systemic radial velocity in the literature, and it was therefore estimated using the method outlined in Section \ref{sec:RV_prior}. If there is an entry in the $d_{\mathrm{lit}}$ column the parallax was not more constraining than any distance in the literature, or of sufficient significance to be reliable, and therefore the parallax distance was not used. PKVs are the potential kick velocity, see Section \ref{sec:results}. The PKV uncertainties correspond to the 15.9 and 84.1 percentiles of the PKV distribution.
        \newline$^a$ indicates a DM distance was used. 
        \newline$^b$ indicates the use of an optical lightcurve distance.\\
        \textbf{References:} [1] \citet{Deller12}; [2]  \citet{Archibald09}; [3] \citet{McConnell15}; [4] \citet{GaiaeDR3}; [5] \citet{Deneva21}; [6] \citet{Strader19}; [7] \citet{Roy15}; [8] \citet{deMartino14}; [9] \citet{Swihart19}; [10] \citet{Camilo16}; [11] \citet{Strader15}; [12] \citet{Bates15}; [13] \citet{Sanpa-Arsa16}; [14] \citet{Ray12}; [15] \citet{Crawford13}; [16] Antoniadis et al., in prep.; [17] \citet{Stovall14}; [18] \citet{Kaplan13}; [19] \citet{Stovall16}; [20] \citet{Bellm16}; [21] \citet{Abdo13}; [22] \citet{Linares18}; [23] \citet{Pletsch15}; [24] \citet{Romani11} }
        \label{tab:rb_table}
    \end{table*}
\end{center}
\subsection{Neutron Star X-ray Binaries}
\label{sec:NSXB}
We considered the sample of NS low-mass X-ray binaries (LMXBs) in the field from \citet{Arnason21}, who cross-matched the catalogues of high-mass and low-mass X-ray binaries by \citet{Liu06,Liu07} with eDR3. This left 18 NS LMXBs that had eDR3 counterparts with measured proper motions. All have total proper motions $>5\sigma$ (proper motion added in quadrature over proper motion uncertainty added in quadrature) excluding MXB 1659--298 and EXO 1747--214, whose proper motion is $>3\sigma$. We include these two systems due to the small size of the sample. To get distances to these systems, six had parallax measurements of at least 5$\sigma$, and seven exhibited photospheric radius expansion (PRE) X-ray bursts. The matter accreted onto the surface of a NS can become sufficiently hot and compressed for unstable thermonuclear ignition, typically resulting in a sudden, rapid X-ray burst (e.g., \citealt{Galloway21}). A subset of X-ray bursts, the PRE X-ray bursts, reach a maximum luminosity (the Eddington luminosity) accompanied by a characteristic signature in the lightcurve and spectral properties. \citet{Kuulkers03} found that these PRE X-ray bursts can be used as empirical standard candles, which are accurate, in principle, to within 15\%. Of the remaining five systems we used a parallax distance for three (2S 0921--630, GX 349+02 (Sco X-2), and EXO 1747--214; note that the parallaxes are lower than 5-sigma significance), an optical lightcurve distance for one (GX 1+4), and the fifth system (1E 1603.6+2600) had no sensible distance and was removed from the sample. The presence of a `literature distance' in Table \ref{tab:nsxb_table} indicates that the distance used in our analysis came from the literature and was not parallax-based. We also include the two AMXPs with measured proper motion (Aql X-1 and IGR J17062--6143), with both of these measurements coming from eDR3. The distances to these two systems were estimated from PRE X-ray bursts. Hence, 19 NS LMXBs were used in the final sample.

\begin{center}
    \begin{table*}
        \begin{tabular}{lccccccccc}
            \hline
            Source & $\mu_\alpha\cos{\delta}$ & $\mu_\delta$ & $P_b$ & $\gamma$ & $\pi$ & $d_\pi$ & $d_{\mathrm{lit}}$ & PKV & References\\
                   & (mas yr$^{-1}$)          & (mas yr$^{-1}$) & (days) & (km s$^{-1}$) & (mas)& (kpc) & (kpc) & (km s$^{-1}$) & \\
            \hline
4U 0614+091 & $1.32(20)$ & $-2.15(17)$ & -- & -- & $0.32(18)$ & -- & $3.2(5)^a$ & $59^{+35}_{-18}$ & [1,2]\\
2S 0921--630 & $-3.163(29)$ & $4.247(27)$ & $9.00260(10)$ & $44.4(2.4)$ & $0.096(23)$ & $9.71^{+2.41}_{-1.63}$ & -- & $74^{+34}_{-18}$ & [1,3,4]\\
4U 1246--58 & $-6.9(6)$ & $-1.8(8)$ & -- & -- & $-0.4(7)$ & -- & $4.3(7)^a$ & $73^{+50}_{-28}$ & [1,5]\\
Cen X-4 & $0.84(15)$ & $-55.68(13)$ & $0.6290522(4)$ & $189.60(20)$ & $0.55(13)$ & -- & $1.30(30)^{a,b}$ & $450^{+94}_{-145}$ & [1,6,7]\\
Sco X-1 & $-7.185(27)$ & $-12.332(19)$ & $0.7873132(5)$ & $-113.60(20)$ & $0.468(22)$ & $2.15^{+0.11}_{-0.10}$ & -- & $213^{+26}_{-40}$ & [1,8]\\
4U 1636--536 & $-5.90(16)$ & $-8.33(12)$ & $0.15804693(16)$ & $-34(5)$ & $0.29(12)$ & -- & $6.0(5)^a$ & $180^{+28}_{-32}$ & [1,9,10]\\
Her X-1 & $-1.212(14)$ & $-7.856(16)$ & $1.7001675900(20)$ & $-65.0(2.0)$ & $0.141(14)$ & $7.25^{+0.80}_{-0.66}$ & -- & $202^{+10}_{-38}$ & [1,11,12]\\
MXB 1659--298 & $-1.6(5)$ & $-1.51(33)$ & $0.296504579(12)$ & $-49(16)^c$ & $0.22(23)$ & -- & $10.5(3.0)^a$ & $364^{+126}_{-147}$ & [1,13,14,15]\\
GX 349+02 (Sco X-2) & $-0.82(14)$ & $-5.19(10)$ & -- & $-250(30)^c$ & $0.10(11)$ & $8.69^{+3.40}_{-1.56}$ & -- & $197^{+84}_{-40}$ & [1,16]\\
IGR J17062--6143 & $-7.4(4)$ & $-1.6(4)$ & $0.02636815(14)$ & -- & $0.2(5)$ & -- & $7.3(5)^a$ & $240^{+32}_{-36}$ & [1,17,18]\\
4U 1700+24 & $-8.73(4)$ & $-5.57(5)$ & $4391(33)$ & $-47.36(6)$ & $1.87(5)$ & $0.53^{+0.02}_{-0.01}$ & -- & $58^{+6}_{-9}$ & [1,19]\\
GX 1+4 & $-3.52(8)$ & $-1.99(6)$ & $1161(12)$ & $-176.73(22)$ & $-0.02(7)$ & -- & $4.3(2.1)^b$ & $178^{+50}_{-37}$ & [1,20]\\
4U 1735--444 & $-3.47(12)$ & $-7.54(7)$ & $0.19383351(32)$ & $-140.0(3.0)$ & $0.12(10)$ & -- & $9.1(1.8)^a$ & $165^{+10}_{-10}$ & [1,21,22]\\
SLX 1737--282 & $0.40(10)$ & $-1.43(6)$ & -- & -- & $0.15(11)$ & -- & $7.3(1.1)^a$ & $140^{+188}_{-72}$ & [1,23]\\
EXO 1747--214 & $-4.9(1.5)$ & $-7.1(1.2)$ & -- & -- & $3.6(1.5)$ & $8.26^{+2.10}_{-1.25}$ & -- & $180^{+78}_{-44}$ & [1]\\
4U 1822--371 & $-9.15(4)$ & $-2.532(32)$ & -- & -- & $0.177(34)$ & $6.97^{+1.16}_{-1.16}$ & -- & $348^{+77}_{-53}$ & [1]\\
Aql X-1 & $-1.8(6)$ & $-5.1(6)$ & $0.7895126(10)$ & $104.0(3.0)$ & $0.21(28)$ & -- & $5.2(8)^a$ & $62^{+7}_{-8}$ & [1,24,25]\\
4U 1954+319 & $-2.158(21)$ & $-6.071(26)$ & -- & -- & $0.303(24)$ & $3.39^{+0.31}_{-0.26}$ & -- & $59^{+58}_{-32}$ & [1]\\
4U 2129+47 & $-2.34(8)$ & $-4.23(8)$ & -- & -- & $0.53(8)$ & $2.06^{+0.41}_{-0.29}$ & -- & $56^{+59}_{-37}$ & [1]\\
            \hline
        \end{tabular}
        \caption{Parameters used in the creation of the PKV distributions for each NS LMXB system. No entry in the $\gamma$ column indicates the system did not have a measured systemic radial velocity in the literature, and it was therefore estimated using the method outlined in Section \ref{sec:RV_prior}. If there is an entry in the $d_{\mathrm{lit}}$ column the parallax was not more constraining than any distance in the literature, or of sufficient significance to be reliable, and therefore the parallax distance was not used. PKVs are the potential kick velocity, see Section \ref{sec:results}. The PKV uncertainties correspond to the 15.9 and 84.1 percentiles of the PKV distribution. Note that Her X-1 is not strictly a LMXB as the companion is estimated to have a mass of $\approx2$ M$_\odot$ \citep{Rawls11}. The three systems 4U 1700+24, GX 1+4, and 4U 1954+319 are symbiotic LMXBs and thus accrete from the companions wind without RLOF. However, recent work by \citet{Hinkle20} suggests that 4U 1954+319 may be a HMXB with a companion mass of $9^{+6}_{-2}$ M$_\odot$ (as opposed to a symbiotic LMXB).
        \newline $^a$ indicates a distance estimated using a PRE X-ray burst.
        \newline $^b$ indicates the use of an optical lightcurve distance.\\
        $^c$ indicates that the systemic radial velocity may be affected by systematics; refer to the original paper.\\
        \textbf{References:} [1] \citet{GaiaeDR3}; [2] \citet{Kuulkers10}; [3] \citet{Ashcraft12}; [4] \citet{Jonker05}; [5] \citet{Zand08}; [6] \citet{gonzalezhernandez2005}; [7] \citet{Casares07}; [8] \citet{Wang18}; [9] \citet{Casares06}; [10] \citet{Galloway06}; [11] \citet{Staubert09}; [12] \citet{Reynolds97}; [13] \citet{Iaria18}; [14] \citet{Ponti18}; [15] \citet{Galloway08}; [16] \citet{Wachter98}; [17] \citet{Strohmayer18}; [18] \citet{Keek17}; [19] \citet{Hinkle19}; [20] \citet{Hinkle06}; [21] \citet{Casares06}; [22] \citet{Augusteijn98}; [23] \citet{Falanga08}; [24] \citet{matasanchez17}; [25] \citet{Jonker04}.  }
        \label{tab:nsxb_table}
    \end{table*}
\end{center}

\subsection{Black Widow Pulsars}
\label{sec:blackwidow}
The sample of black widows was extracted from \citet{Hui19} using their table of Galactic field black widows. From this sample, only the systems with proper motion measurements were retained, of which there are 17. Of the 17, the proper motions for 15 came from pulsar timing, and the final two (PSR J1311--3430 and PSR J1810+1744) came from eDR3. We used DM distances for all black widow systems. We therefore present two individual kick estimates for each system, corresponding to the NE2001 and YMW16 DM models. Systems and their associated proper motions and distances can be found in Table \ref{tab:bw_table}.

\begin{center}
    \begin{table*}
        \begin{tabular}{lcccccccc}
            \hline
            Source & $\mu_\alpha\cos{\delta}$ & $\mu_\delta$ & $P_b$ & $d_{\mathrm{NE2001}}$ & $d_{\mathrm{YMW16}}$ & PKV$_\mathrm{NE2001}$ & PKV$_\mathrm{YMW16}$ & References\\
                   & (mas yr$^{-1}$)          & (mas yr$^{-1}$) & (days) & (kpc) & (kpc) &  (km s$^{-1}$) &  (km s$^{-1}$) & \\
            \hline
J0023+0923 & $-12.63(17)$ & $-5.8(4)$ & $0.13879914382(4)$ & 0.69 & 1.25 & $83^{+50}_{-25}$ & $118^{+53}_{-31}$ & [1,2] \\
J0610--2100 & $9.21(6)$ & $16.73(8)$ & $0.2860160068(6)$ & 3.54 & 3.26 & $197^{+38}_{-43}$ & $200^{+39}_{-44}$ & [3,4] \\
J1311--3430 & $-6.1(1.6)$ & $-5.1(7)$ & $0.0651157335(7)$ & 1.41 & 2.43 & $85^{+56}_{-28}$ & $131^{+68}_{-38}$ & [1,5] \\
J1446--4701 & $-4.00(20)$ & $-2.00(30)$ & $0.27766607699(15)$ & 1.46 & 1.57 & $65^{+61}_{-33}$ & $83^{+79}_{-47}$ & [6] \\
J1641+8049 & $-11.0(1.0)$ & $37.0(3.0)$ & $0.09087396340(10)$ & 1.65 & 3.04 & $302^{+118}_{-92}$ & $383^{+154}_{-124}$ & [7] \\
J1731--1847 & $-1.70(30)$ & $-6.0(3.0)$ & $0.3111341185(10)$ & 2.55 & 4.78 & $109^{+52}_{-40}$ & $186^{+80}_{-60}$ & [6] \\
J1745+1017 & $6.0(1.0)$ & $-5.0(1.0)$ & $0.7302414440(10)$ & 1.26 & 1.21 & $79^{+52}_{-27}$ & $91^{+71}_{-38}$ & [8] \\
J1805+0615 & $8.7(1.3)$ & $12.8(2.9)$ & $0.336872031(5)$ & 2.48 & 3.88 & $208^{+51}_{-52}$ & $243^{+47}_{-60}$ & [9] \\
J1810+1744 & $7.5(5)$ & $-4.2(5)$ & $0.15(0)$ & 2.00 & 2.36 & $112^{+47}_{-30}$ & $135^{+58}_{-37}$ & [1,10] \\
J1959+2048 & $-16.0(5)$ & $-25.8(6)$ & $0.3819666069(8)$ & 2.49 & 1.73 & $265^{+76}_{-72}$ & $216^{+74}_{-58}$ & [11] \\
J2051--0827 & $5.63(4)$ & $2.34(28)$ & $0.09911025490(4)$ & 1.04 & 1.47 & $79^{+52}_{-26}$ & $106^{+62}_{-34}$ & [12] \\
J2052+1218/J2052+1219 & $-4.30(32)$ & $-14.0(6)$ & $0.11461362510(20)$ & 2.44 & 3.91 & $148^{+50}_{-38}$ & $207^{+57}_{-51}$ & [9] \\
J2055+3829 & $5.920(30)$ & $0.79(7)$ & $0.129590372940(10)$ & 4.36 & 4.59 & $219^{+79}_{-62}$ & $237^{+102}_{-68}$ & [13] \\
J2214+3000 & $20.77(8)$ & $-1.46(12)$ & $0.41663294591(20)$ & 1.54 & 1.67 & $182^{+62}_{-50}$ & $204^{+78}_{-56}$ & [1,2] \\
J2234+0944 & $6.96(6)$ & $-32.22(10)$ & $0.41966003706(17)$ & 1.00 & 1.59 & $179^{+65}_{-46}$ & $265^{+107}_{-65}$ & [1,2] \\
J2241--5236 & $17.10(10)$ & $-3.32(5)$ & $0.145672240250(20)$ & 0.51 & 0.96 & $66^{+52}_{-28}$ & $107^{+57}_{-35}$ & [14,15] \\
J2256--1024 & $3.2(1.1)$ & $-8.5(2.7)$ & $0.21288263050(7)$ & 0.65 & 1.33 & $72^{+56}_{-30}$ & $117^{+65}_{-39}$ & [16] \\
            \hline
        \end{tabular}
        \caption{Parameters used in the creation of the PKV distributions for each black widow system. No black widows have measured systemic radial velocities so there is no corresponding column. PKVs are the potential kick velocity, see Section \ref{sec:results}. The PKV uncertainties correspond to the 15.9 and 84.1 percentiles of the PKV distribution.\\
        \textbf{References:} [1] \citet{Manchester05}; [2] \citet{Arzoumanian18}; [3] \citet{Guillemot16}; [4] \citet{Desvignes16}; [5] \citet{Pletsch12}; [6] \citet{Ng14}; [7] \citet{Lynch18}; [8] \citet{Barr13}; [9] \citet{Deneva21}; [10] \citet{Hessels11}; [11] \citet{Arzoumanian94}; [12] \citet{Shaifullah16}; [13] \citet{Guillemot19}; [14] \citet{Jankowski19}; [15] \citet{Keith11}; [16] \citet{Crowter20}. }
        \label{tab:bw_table}
    \end{table*}
\end{center}

\subsection{Millisecond Pulsars}
\label{sec:MSP}
The sample of millisecond pulsars (MSPs) was created using the ATNF Pulasr catalogue\footnote{\url{https://www.atnf.csiro.au/research/pulsar/psrcat/}} \citep[version 1.65, released on 2021/09/09;][]{Manchester05}. We retrieved all pulsars with periods less than 30 ms, measured proper motion, and without globular cluster associations. From these, we removed all redback and black widow pulsars from the other samples, and required the total proper motion significance to be greater than $5\sigma$. This resulted in a sample of 95 MSPs. For all systems, both NE2001 and YMW16 DM distance estimates were used and thus there are also two individual kick estimates for each MSP. Systems and their associated proper motions and distances can be found in Table \ref{tab:msp_table}.

\clearpage
\onecolumn
\begin{center}
\begin{longtable}{lcccccccc}
            \hline
            Source & $\mu_\alpha\cos{\delta}$ & $\mu_\delta$ & $P_b$  & $d_{\mathrm{NE2001}}$ & $d_{\mathrm{YMW16}}$ & PKV$_\mathrm{NE2001}$ & PKV$_\mathrm{YMW16}$ & References\\
                   & (mas yr$^{-1}$)          & (mas yr$^{-1}$) & (days) & (kpc) & (kpc) &  (km s$^{-1}$) &  (km s$^{-1}$)\\
            \hline
J0030+0451 & $-6.21(16)$ & $0.5(4)$ & -- & 0.32 & 0.34 & $45^{+26}_{-12}$ & $48^{+30}_{-13}$ & [1] \\
J0034--0534 & $7.90(30)$ & $-9.2(6)$ & $1.58928182532(14)$ & 0.54 & 1.35 & $51^{+24}_{-16}$ & $112^{+35}_{-29}$ & [2] \\
J0101--6422 & $10.0(1.0)$ & $-12.0(2.0)$ & $1.7875967060(20)$ & 0.55 & 1.00 & $63^{+24}_{-17}$ & $103^{+32}_{-25}$ & [3] \\
J0154+1833 & $10.3(9)$ & $-8.9(1.9)$ & -- & 0.86 & 1.62 & $71^{+27}_{-21}$ & $137^{+51}_{-40}$ & [4,1] \\
J0218+4232 & $5.35(5)$ & $-3.74(12)$ & $2.02884611561(9)$ & 2.67 & 2.93 & $113^{+40}_{-33}$ & $126^{+46}_{-37}$ & [5,2] \\
J0337+1715 & $4.8(5)$ & $-4.4(4)$ & $1.629401788(5)$ & 0.76 & 0.82 & $38^{+23}_{-13}$ & $42^{+26}_{-15}$ & [6,7] \\
J0340+4130 & $-0.53(11)$ & $-3.30(29)$ & -- & 1.73 & 1.60 & $37^{+25}_{-13}$ & $38^{+29}_{-15}$ & [1] \\
J0437--4715 & $121.4385(20)$ & $-71.4754(20)$ & $5.7410459(4)$ & 0.14 & 0.16 & $115^{+36}_{-29}$ & $129^{+41}_{-33}$ & [8] \\
J0509+0856 & $5.40(20)$ & $-4.3(5)$ & $4.9079768930(10)$ & 1.45 & 0.82 & $62^{+21}_{-17}$ & $41^{+26}_{-15}$ & [4] \\
J0613--0200 & $1.860(20)$ & $-10.34(4)$ & $1.198512556715(14)$ & 1.71 & 1.02 & $113^{+41}_{-32}$ & $72^{+27}_{-21}$ & [1,9] \\
J0614--3329 & $0.58(9)$ & $-1.92(12)$ & $53.5846127(8)$ & 1.90 & 2.69 & $74^{+25}_{-17}$ & $94^{+30}_{-21}$ & [10,11] \\
J0621+1002 & $3.23(12)$ & $-0.5(5)$ & $8.31868120(30)$ & 1.36 & 0.42 & $37^{+24}_{-11}$ & $33^{+31}_{-18}$ & [2] \\
J0636+5129 & $3.50(20)$ & $-2.30(20)$ & $0.066551340060(30)$ & 0.49 & 0.21 & $33^{+26}_{-11}$ & $35^{+30}_{-14}$ & [1,9] \\
J0645+5158 & $1.52(4)$ & $-7.42(6)$ & -- & 0.70 & 0.67 & $36^{+24}_{-13}$ & $37^{+29}_{-15}$ & [1] \\
J0711--6830 & $-15.570(30)$ & $14.240(30)$ & -- & 0.86 & 0.11 & $74^{+28}_{-21}$ & $41^{+44}_{-28}$ & [8,1] \\
J0737--3039A & $-3.8(6)$ & $2.13(23)$ & $0.10225156248(5)$ & 0.52 & 1.10 & $36^{+38}_{-24}$ & $41^{+42}_{-24}$ & [12,13] \\
J0740+6620 & $-10.320(31)$ & $-30.87(4)$ & $4.76694461910(10)$ & 0.68 & 0.93 & $113^{+42}_{-33}$ & $162^{+59}_{-49}$ & [1,14] \\
J0751+1807 & $-2.73(5)$ & $-13.40(30)$ & $0.263144270792(7)$ & 1.15 & 0.43 & $91^{+31}_{-26}$ & $41^{+29}_{-16}$ & [2] \\
J0824+0028 & $-4.3(4)$ & $-9.2(1.3)$ & $23.206955708(5)$ & 1.54 & 1.69 & $95^{+36}_{-26}$ & $106^{+40}_{-30}$ & [4] \\
J0900--3144 & $-1.01(5)$ & $2.02(7)$ & $18.7376360594(9)$ & 0.54 & 0.38 & $42^{+37}_{-20}$ & $45^{+42}_{-24}$ & [2] \\
J0931--1902 & $-2.40(22)$ & $-4.30(29)$ & -- & 1.88 & 3.72 & $85^{+29}_{-20}$ & $147^{+49}_{-37}$ & [1] \\
J1012+5307 & $3.0(5)$ & $-26.9(6)$ & $0.604672722901(13)$ & 0.41 & 0.81 & $71^{+22}_{-19}$ & $135^{+43}_{-36}$ & [6,2] \\
J1017--7156 & $-7.31(6)$ & $6.76(5)$ & $6.5119050(20)$ & 2.98 & 1.81 & $77^{+26}_{-19}$ & $59^{+36}_{-19}$ & [15] \\
J1022+1001 & $-14.91(4)$ & $5.610(30)$ & $7.8051348(11)$ & 0.45 & 0.83 & $46^{+25}_{-13}$ & $68^{+23}_{-17}$ & [16,2] \\
J1024--0719 & $-35.30(5)$ & $-48.23(9)$ & -- & 0.39 & 0.38 & $122^{+42}_{-34}$ & $122^{+43}_{-34}$ & [1] \\
J1045--4509 & $-6.07(9)$ & $5.20(10)$ & $4.08352925480(30)$ & 1.96 & 0.34 & $54^{+33}_{-15}$ & $44^{+44}_{-24}$ & [8] \\
J1125--5825 & $-10.00(30)$ & $2.40(30)$ & $76.40321683(5)$ & 2.62 & 1.74 & $56^{+32}_{-19}$ & $48^{+41}_{-22}$ & [15] \\
J1125+7819 & $28.5(7)$ & $-1.2(9)$ & $15.355445959(13)$ & 0.65 & 0.88 & $105^{+26}_{-24}$ & $132^{+33}_{-30}$ & [1,9] \\
J1207--5050 & $6.9(4)$ & $1.4(5)$ & -- & 1.53 & 1.27 & $107^{+29}_{-24}$ & $97^{+29}_{-23}$ & [17,1] \\
J1231--1411 & $-62.03(26)$ & $6.2(5)$ & $1.860143882(9)$ & 0.44 & 0.42 & $112^{+37}_{-31}$ & $110^{+37}_{-31}$ & [10,11] \\
J1300+1240 & $45.50(5)$ & $-84.70(7)$ & $25.2620(30)$ & 0.45 & 0.88 & $201^{+68}_{-54}$ & $355^{+103}_{-118}$ & [18] \\
J1312+0051 & $-22.4(7)$ & $-11.2(1.5)$ & $38.503832800(20)$ & 0.83 & 1.47 & $114^{+38}_{-31}$ & $204^{+74}_{-57}$ & [19] \\
J1400--1431 & $17.0(2.1)$ & $-55(6)$ & $9.5474676743(19)$ & 0.48 & 0.35 & $136^{+41}_{-35}$ & $103^{+31}_{-26}$ & [1,20] \\
J1421--4409 & $-11.6(4)$ & $-7.9(8)$ & $30.746453420(30)$ & 1.57 & 2.08 & $92^{+33}_{-26}$ & $120^{+39}_{-32}$ & [21] \\
J1453+1902 & $0.5(8)$ & $-10.8(1.9)$ & -- & 1.15 & 1.27 & $88^{+24}_{-21}$ & $96^{+27}_{-22}$ & [1] \\
J1455--3330 & $7.98(8)$ & $-1.97(19)$ & $76.174567473(11)$ & 0.53 & 0.68 & $52^{+23}_{-12}$ & $61^{+25}_{-14}$ & [1,9] \\
J1536--4948 & $-7.30(20)$ & $-2.7(5)$ & $62.051498210(20)$ & 1.83 & 0.98 & $57^{+25}_{-17}$ & $41^{+33}_{-17}$ & [17] \\
J1600--3053 & $-0.986(16)$ & $-7.11(6)$ & $14.3484660(30)$ & 1.63 & 2.54 & $72^{+23}_{-20}$ & $112^{+32}_{-30}$ & [1,9] \\
J1603--7202 & $-2.46(4)$ & $-7.33(5)$ & $6.3086296691(5)$ & 1.17 & 1.13 & $42^{+30}_{-17}$ & $44^{+36}_{-19}$ & [8] \\
J1614--2230 & $3.81(12)$ & $-32.5(7)$ & $8.68661942215(7)$ & 1.29 & 1.43 & $225^{+64}_{-61}$ & $251^{+63}_{-67}$ & [1,9] \\
J1640+2224 & $2.078(11)$ & $-11.336(20)$ & $175.460661897(7)$ & 1.16 & 1.51 & $76^{+22}_{-18}$ & $93^{+25}_{-21}$ & [1,9] \\
J1643--1224 & $5.93(11)$ & $3.8(5)$ & $147.01728(7)$ & 2.40 & 0.79 & $125^{+33}_{-29}$ & $59^{+21}_{-13}$ & [1,8] \\
J1658--5324 & $0.2(8)$ & $4.90(23)$ & -- & 0.93 & 0.88 & $57^{+22}_{-12}$ & $57^{+26}_{-13}$ & [22,1] \\
J1709+2313 & $-3.2(7)$ & $-9.7(9)$ & $22.711892380(20)$ & 1.41 & 2.18 & $82^{+24}_{-19}$ & $114^{+30}_{-26}$ & [23] \\
J1710+4923 & $-50.40(20)$ & $-44.70(20)$ & -- & 0.66 & 0.51 & $222^{+71}_{-57}$ & $176^{+55}_{-43}$ & [24,1] \\
J1713+0747 & $4.9150(30)$ & $-3.914(5)$ & $67.8251383185(17)$ & 0.89 & 0.92 & $49^{+23}_{-13}$ & $51^{+27}_{-14}$ & [25] \\
J1719--1438 & $1.9(4)$ & $-11.0(2.0)$ & $0.0907062900(12)$ & 1.21 & 0.34 & $74^{+27}_{-23}$ & $32^{+29}_{-17}$ & [15] \\
J1732--5049 & $-0.41(9)$ & $-9.87(19)$ & $5.2629972182(5)$ & 1.41 & 1.88 & $65^{+24}_{-21}$ & $87^{+30}_{-26}$ & [8] \\
J1738+0333 & $7.07(5)$ & $5.11(10)$ & $0.3547907398724(13)$ & 1.43 & 1.51 & $94^{+27}_{-23}$ & $99^{+29}_{-24}$ & [1,26] \\
J1741+1351 & $-8.980(20)$ & $-7.410(20)$ & $16.3353478283(5)$ & 0.90 & 1.36 & $66^{+23}_{-17}$ & $93^{+28}_{-22}$ & [1,9] \\
J1744--1134 & $18.790(6)$ & $-9.400(30)$ & -- & 0.41 & 0.15 & $50^{+19}_{-13}$ & $34^{+30}_{-15}$ & [8,1] \\
J1751--2857 & $-7.40(10)$ & $-4.3(1.2)$ & $110.74646080(4)$ & 1.11 & 1.09 & $58^{+21}_{-16}$ & $59^{+23}_{-17}$ & [2] \\
J1801--1417 & $-10.89(12)$ & $-3.0(1.0)$ & -- & 1.52 & 1.10 & $97^{+26}_{-24}$ & $75^{+23}_{-19}$ & [2,1] \\
J1802--2124 & $-0.85(10)$ & $4.8(0)$ & $0.698889243381(5)$ & 2.94 & 3.03 & $92^{+24}_{-20}$ & $95^{+26}_{-21}$ & [27] \\
J1804--2717 & $2.56(15)$ & $-17.0(3.0)$ & $11.1287119670(30)$ & 0.78 & 0.81 & $70^{+28}_{-24}$ & $74^{+30}_{-25}$ & [2] \\
J1811--2405 & $0.53(6)$ & $0.0(0)$ & $6.27230196915(11)$ & 1.77 & 1.83 & $29^{+26}_{-13}$ & $32^{+29}_{-15}$ & [28] \\
J1832--0836 & $-7.97(5)$ & $-21.20(20)$ & -- & 1.11 & 0.81 & $134^{+47}_{-37}$ & $101^{+35}_{-28}$ & [1] \\
J1843--1113 & $-1.91(7)$ & $-3.20(30)$ & -- & 1.70 & 1.71 & $37^{+26}_{-14}$ & $39^{+31}_{-15}$ & [2,1] \\
J1853+1303 & $-1.65(4)$ & $-2.89(6)$ & $115.653786446(14)$ & 2.09 & 1.32 & $41^{+33}_{-17}$ & $40^{+38}_{-21}$ & [1,9] \\
J1857+0943 & $-2.655(10)$ & $-5.408(20)$ & $12.32717119157(18)$ & 1.17 & 0.77 & $38^{+32}_{-17}$ & $38^{+38}_{-20}$ & [1,9] \\
J1903+0327 & $-2.06(7)$ & $-5.21(12)$ & $95.174118753(14)$ & 6.37 & 6.12 & $47^{+31}_{-21}$ & $48^{+37}_{-21}$ & [29] \\
J1903--7051 & $-8.8(1.6)$ & $-16.0(2.0)$ & $11.050798330(20)$ & 0.76 & 0.93 & $77^{+27}_{-22}$ & $94^{+33}_{-26}$ & [22] \\
J1905+0400 & $-3.80(18)$ & $-7.3(4)$ & -- & 1.70 & 1.06 & $56^{+28}_{-16}$ & $47^{+33}_{-17}$ & [30,1] \\
J1909--3744 & $-9.5120(10)$ & $-35.782(5)$ & $1.533449474305(5)$ & 0.46 & 0.56 & $98^{+33}_{-29}$ & $123^{+42}_{-35}$ & [31] \\
J1910+1256 & $0.28(5)$ & $-7.29(7)$ & $58.466742057(8)$ & 2.33 & 1.50 & $55^{+28}_{-16}$ & $47^{+36}_{-18}$ & [1,9] \\
J1911--1114 & $-13.75(16)$ & $-9.1(1.0)$ & $2.7165576619(7)$ & 1.22 & 1.07 & $107^{+32}_{-27}$ & $97^{+30}_{-25}$ & [2] \\
J1911+1347 & $-2.85(7)$ & $-3.54(8)$ & -- & 2.07 & 1.36 & $39^{+35}_{-18}$ & $40^{+40}_{-21}$ & [1] \\
J1918--0642 & $-7.149(12)$ & $-5.97(5)$ & $10.91317757989(16)$ & 1.23 & 1.02 & $62^{+23}_{-16}$ & $57^{+27}_{-17}$ & [1,9] \\
J1921+1929 & $-3.2(8)$ & $-11.0(1.0)$ & $39.649263750(20)$ & 3.24 & 2.43 & $100^{+30}_{-26}$ & $86^{+32}_{-23}$ & [32] \\
J1923+2515 & $-6.97(4)$ & $-14.17(8)$ & -- & 1.63 & 1.20 & $92^{+28}_{-23}$ & $77^{+31}_{-20}$ & [1] \\
J1933--6211 & $-5.54(7)$ & $10.70(20)$ & $12.8194067183(8)$ & 0.52 & 0.65 & $66^{+20}_{-15}$ & $76^{+23}_{-17}$ & [33] \\
J1939+2134 & $0.070(4)$ & $-0.401(5)$ & -- & 3.56 & 2.90 & $89^{+31}_{-27}$ & $76^{+33}_{-24}$ & [2,1] \\
J1944+0907 & $14.07(4)$ & $-22.73(9)$ & -- & 1.79 & 1.22 & $200^{+47}_{-44}$ & $148^{+39}_{-36}$ & [1] \\
J1946+3417 & $-7.01(23)$ & $4.51(21)$ & $27.01994783(5)$ & 5.14 & 6.94 & $233^{+64}_{-58}$ & $272^{+70}_{-66}$ & [34] \\
J1949+3106 & $-2.894(31)$ & $-5.09(4)$ & $1.94953755(20)$ & 6.52 & 7.47 & $46^{+37}_{-20}$ & $56^{+41}_{-26}$ & [35] \\
J1950+2414 & $-2.12(18)$ & $-3.64(19)$ & $22.19137127(6)$ & 5.57 & 7.27 & $56^{+31}_{-18}$ & $64^{+32}_{-18}$ & [35] \\
J1955+2908 & $-1.12(11)$ & $-4.21(19)$ & $117.34909722(6)$ & 4.64 & 6.30 & $54^{+32}_{-19}$ & $66^{+32}_{-18}$ & [1,9] \\
J2010--1323 & $2.40(30)$ & $-5.60(30)$ & -- & 1.03 & 1.16 & $48^{+24}_{-15}$ & $55^{+26}_{-17}$ & [16,1] \\
J2010+3051 & $-10.3(9)$ & $-3.0(1.0)$ & $23.358895750(20)$ & 5.53 & 6.45 & $180^{+48}_{-41}$ & $194^{+44}_{-43}$ & [32] \\
J2017+0603 & $2.21(8)$ & $0.15(19)$ & $2.19848113613(12)$ & 1.57 & 1.40 & $61^{+27}_{-18}$ & $58^{+32}_{-18}$ & [1,9] \\
J2019+2425 & $-9.41(12)$ & $-20.60(15)$ & $76.511634790(20)$ & 1.49 & 1.16 & $124^{+35}_{-31}$ & $105^{+32}_{-27}$ & [36] \\
J2033+1734 & $-5.9(4)$ & $-9.8(5)$ & $56.30779531(7)$ & 2.00 & 1.74 & $79^{+26}_{-20}$ & $74^{+31}_{-19}$ & [1,9] \\
J2039--5617 & $4.20(30)$ & $-14.90(30)$ & $0.2279798050(30)$ & 0.94 & 1.71 & $85^{+26}_{-24}$ & $152^{+47}_{-41}$ & [37] \\
J2042+0246 & $15.1(1.2)$ & $-14.1(2.5)$ & $77.20058060(30)$ & 0.83 & 0.64 & $88^{+28}_{-24}$ & $72^{+28}_{-22}$ & [19] \\
J2043+1711 & $-5.722(11)$ & $-10.831(19)$ & $1.482290786394(15)$ & 1.76 & 1.48 & $79^{+26}_{-19}$ & $73^{+31}_{-20}$ & [1,9] \\
J2053+4650 & $-2.8(8)$ & $-5.4(5)$ & $2.45249901140(20)$ & 4.11 & 3.81 & $55^{+34}_{-22}$ & $55^{+39}_{-23}$ & [38] \\
J2124--3358 & $-14.14(4)$ & $-50.08(9)$ & -- & 0.27 & 0.36 & $86^{+29}_{-24}$ & $114^{+37}_{-32}$ & [8,1] \\
J2129--5721 & $9.25(4)$ & $-9.58(4)$ & $6.6254930923(13)$ & 1.36 & 6.17 & $98^{+29}_{-25}$ & $343^{+72}_{-107}$ & [8] \\
J2145--0750 & $-9.49(5)$ & $-9.11(8)$ & $6.83890261536(5)$ & 0.57 & 0.69 & $53^{+26}_{-13}$ & $62^{+28}_{-16}$ & [16,8] \\
J2229+2643 & $-2.0(5)$ & $-5.7(7)$ & $93.01589270(15)$ & 1.43 & 1.80 & $54^{+31}_{-15}$ & $64^{+32}_{-18}$ & [1,9] \\
J2234+0611 & $25.300(20)$ & $9.71(5)$ & $32.001401630(8)$ & 0.68 & 0.85 & $94^{+27}_{-25}$ & $116^{+34}_{-30}$ & [39] \\
J2302+4442 & $-0.05(14)$ & $-5.91(18)$ & $125.93529697(13)$ & 1.18 & 0.86 & $45^{+34}_{-17}$ & $44^{+42}_{-22}$ & [1,40] \\
J2317+1439 & $-1.30(30)$ & $3.6(5)$ & $2.459331465168(18)$ & 0.83 & 2.16 & $59^{+25}_{-14}$ & $103^{+24}_{-21}$ & [16,9] \\
J2322+2057 & $-18.4(4)$ & $-15.4(5)$ & -- & 0.80 & 1.01 & $90^{+25}_{-21}$ & $109^{+30}_{-25}$ & [2,1] \\
J2322--2650 & $-2.40(20)$ & $-8.3(4)$ & $0.322963997(6)$ & 0.32 & 0.76 & $40^{+27}_{-12}$ & $64^{+23}_{-16}$ & [41] \\
            \hline
        \caption{Parameters used in the creation of the PKV distributions for each MSP system. The MSPs do not have measured systemic radial velocities so there is no corresponding column. PKVs are the potential kick velocity, see Section \ref{sec:results}. The PKV uncertainties correspond to the 15.9 and 84.1 percentiles of the PKV distribution.\\
        \textbf{References:} [1] \citet{Manchester05}; [2] \citet{Desvignes16}; [3] \citet{Kerr12}; [4] \citet{Martinez19}; [5] \citet{Du14}; [6] \citet{Jennings18}; [7] \citet{Ransom14}; [8] \citet{Reardon16}; [9] \citet{Arzoumanian18}; [10] \citet{Guillemot16}; [11] \citet{Ransom11}; [12] \citet{Deller09}; [13] \citet{Kramer06}; [14] \citet{Cromartie20}; [15] \citet{Ng14}; [16] \citet{Deller19}; [17] \citet{Bhattacharyya21}; [18] \citet{Konacki03}; [19] \citet{Sanpa-Arsa16}; [20] \citet{Swiggum17}; [21] \citet{Spiewak20}; [22] \citet{Camilo15}; [23] \citet{Lewandowski04}; [24] \citet{Lynch18}; [25] \citet{Zhu15}; [26] \citet{Freire12}; [27] \citet{Ferdman10}; [28] \citet{Ng20}; [29] \citet{Freire11}; [30] \citet{Gonzalez11}; [31] \citet{Liu20}; [32] \citet{Parent19}; [33] \citet{Graikou17}; [34] \citet{Barr17}; [35] \citet{Zhu19}; [36] \citet{Nice01}; [37] \citet{Clark21}; [38] \citet{Berezina17}; [39] \citet{Stovall19}; [40] \citet{Fonseca16}; [41] \citet{Spiewak18}.}
        \label{tab:msp_table}
\end{longtable}
\end{center}
\clearpage
\twocolumn

\section{Results}
\label{sec:results}
Here we present PKV probability distributions for 145 NS binary systems, as discussed in section \ref{sec:data}. We report the PKV distribution for distance estimates from both NE2001 and YMW16 for the systems with DM distance as the best distance estimate available. Therefore, we present a total of 258 PKV probability distributions.

The parametric fits to the total sample and subsample distributions were determined using the method outlined in Section \ref{sec:model_fit}. The choice of DM model does not significantly change the parametric fits (they agree within errors) or the resulting interpretation. Therefore, here and moving forward, we focus our discussion based on the NE2001 fits. As detailed in Section \ref{sec:results_all} and Table \ref{tab:fits}, the Beta distribution had the lowest AICc and highest LOO weight for the full sample and subsamples, and is therefore the preferred model in all cases. The parameters of the best-fit beta distribution for the full sample and each subsample are reported in Table \ref{tab:fits_samples}.

\begin{table}
    \begin{center}
        \begin{tabular}{lcccc}
            \hline
            Sample & $\alpha$ & $\beta$ & $s$ (\si{\km \per \second}) \\
            \hline
            All systems & $3.05^{+0.32}_{-0.30}$ & $14.6^{+2.2}_{-2.1}$ & $563^{+72}_{-68}$ \\
            Redbacks & $8.4^{+1.3}_{-1.3}$ & $18.5^{+3.3}_{-3.2}$ & $433^{+69}_{-62}$ \\
            NS LMXBs & $1.87^{+0.29}_{-0.28}$ & $10.8^{+1.8}_{-1.8}$ & $1110^{+200}_{-180}$ \\
            Black Widows & $2.44^{+0.38}_{-0.37}$ & $12.7^{+2.1}_{-2.1}$ & $860^{+150}_{-140}$ \\
            MSPs & $2.81^{+0.30}_{-0.29}$ & $12.8^{+2.2}_{-2.3}$ & $414^{+77}_{-71}$ \\
            \hline
        \end{tabular}
        \caption{Parametric fits to the PKV distributions for all systems and the different subclasses of systems. The Beta distribution fits are presented for each.}
        \label{tab:fits_samples}
    \end{center}
\end{table}

\subsection{Redback Pulsars}
\begin{figure}
	\includegraphics[width=\columnwidth]{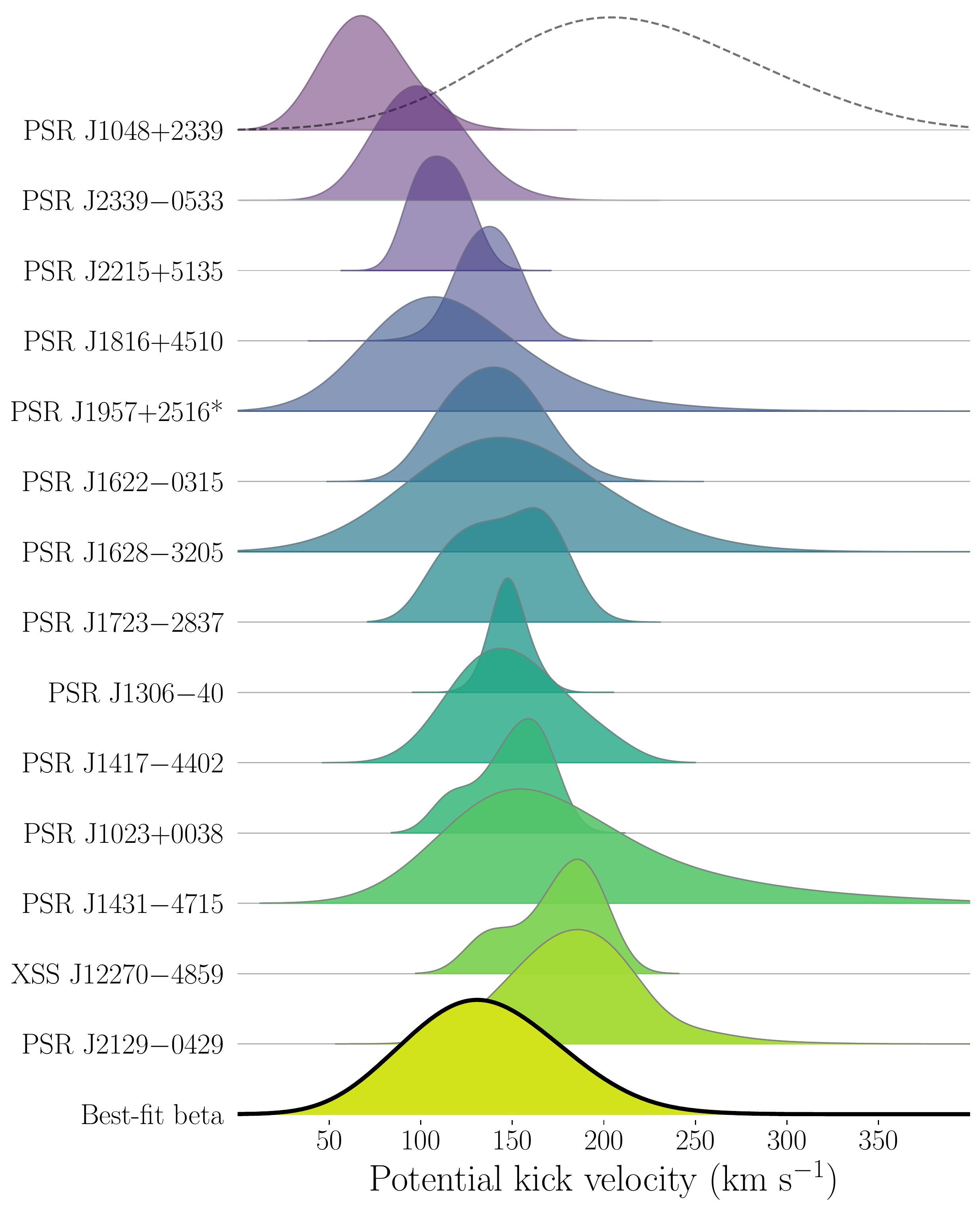}
    \caption{Ridgeline plots showing scaled PDFs of the PKV distribution for each redback system. An asterisk appended to the source's name indicates that the radial velocity for this system was estimated using the method outlined in Section \ref{sec:RV_prior}. For PSR J1048+2339 the shaded curve corresponds to the PKV distribution determined using the NE2001 DM distance estimate. The dotted curve represents the same thing for the YMW16 DM distance estimate. The `Best-fit beta' curve is the best-fit beta distribution to the full sample of redbacks.}
    \label{fig:rb_ridge}
\end{figure}

A probabilistic representation of the redback PKV distributions is presented in Figure \ref{fig:rb_ridge}. Qualitatively, eight of the 14 systems have PKV distributions centred around 150 \si{\km \per \second}, with two centred slightly above and four below. The best-fit Beta distribution for the redbacks has $\alpha = 8.4^{+1.3}_{-1.3}$, $\beta = 18.5^{+3.3}_{-3.2}$, and $s = 433^{+69}_{-62}$ \si{\km \per \second}. This distribution has a slight positive skew, with a mean and mode of $135$ and $129$ \si{\km \per \second}, respectively. The PDF corresponding to the best-fit parameters can be found in Figure \ref{fig:beta_pdf_cdf}. See Table \ref{tab:rb_table} for a summary of the parameters and kicks for each redback system. Figure \ref{fig:cdf_90_rbs} compares the redback PKV MC realisations to the posterior of the beta model fitting.

\subsection{Neutron Star Low-Mass X-ray Binaries}
\begin{figure}
	\includegraphics[width=\columnwidth]{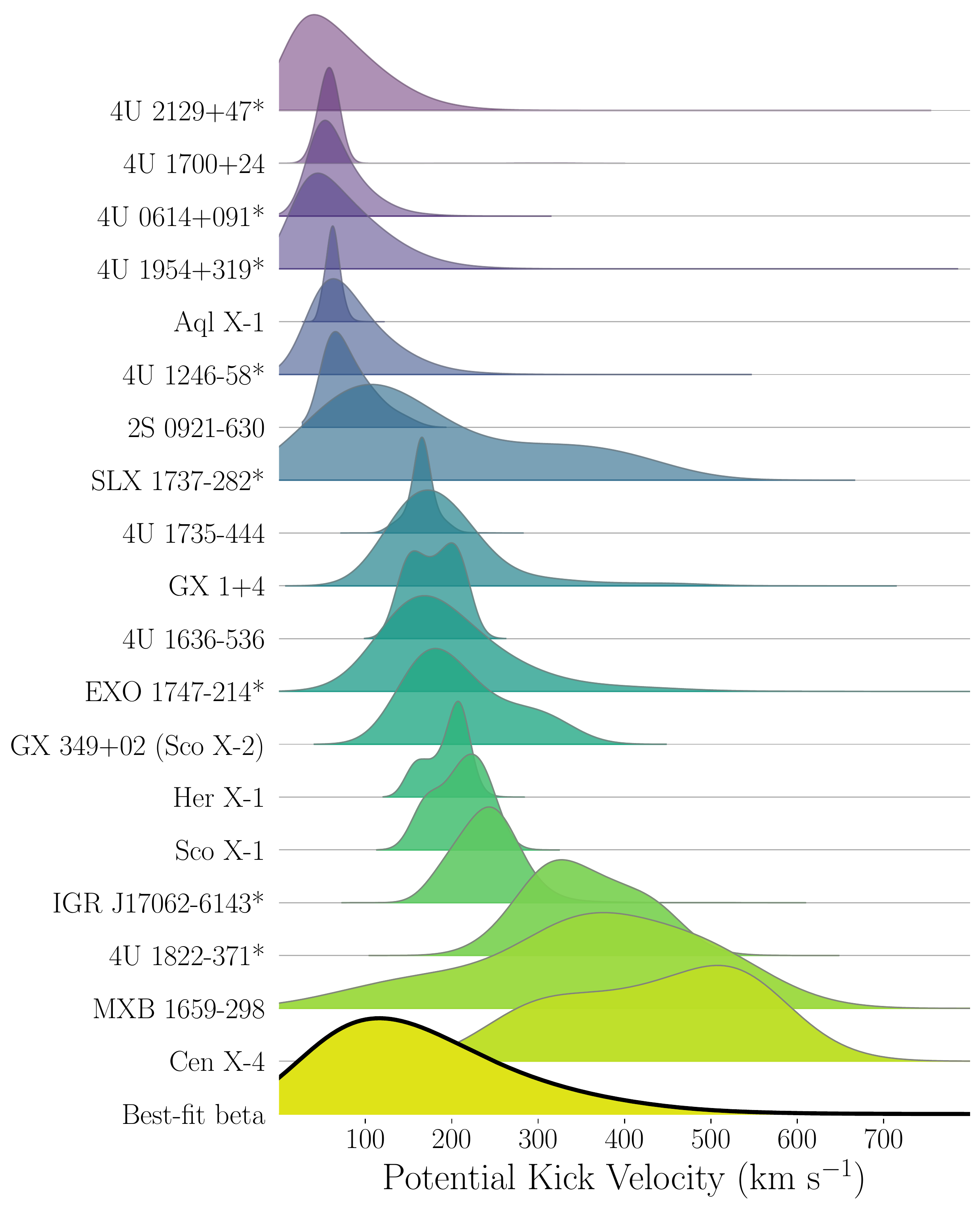}
    \caption{Ridgeline plots showing scaled PDFs of the PKV distribution for each NS LMXB system. An asterisk appended to the source's name indicates that the radial velocity for this system was estimated using the method outlined in Section \ref{sec:RV_prior}. Note that for some systems the curve appears to go below zero; this is an artefact of the smoothing process as all PKV MC realisations are $>0$ \SI{}{\km\per\second}. The `Best-fit beta' curve is the best-fit beta distribution to the full sample of NS LMXBs.}
    \label{fig:nsxb_ridge}
\end{figure}

The NS LMXB PKV distributions are more varied than the redback systems. Figure \ref{fig:nsxb_ridge} presents the PKV distribution for each system. The PKV distributions for eight of the 19 systems peak around 100 \si{\km \per \second}. Eight of the remaining 11 systems peak just above 200 \si{\km \per \second}, with the remaining three systems peaking at approximately 300, 400, and 500 \si{\km \per \second}. Most PKV distributions are relatively narrow, however SLX 1737--282, MXB 1659--298, and Cen X-4 span greater than 400 \si{\km \per \second}. See Table \ref{tab:nsxb_table} for a summary of system parameters and PKV estimates. The best-fit Beta distribution for the NS LMXBs has $\alpha = 1.87^{+0.29}_{-0.28}$, $\beta = 10.8^{+1.8}_{-1.8}$, and $s = 1110^{+200}_{-180}$ \si{\km \per \second}. The corresponding distribution has a positive skew, with a mean and mode of $164$ and $90$ \si{\km \per \second}, respectively. The PDF corresponding to the best-fit parameters can be found in Figure \ref{fig:beta_pdf_cdf}. Figure \ref{fig:cdf_90_nsxbs} compares the NS LMXB PKV MC realisations to the posterior of the beta model fitting.

\subsection{Black Widow Pulsars}
The shapes of the individual black widow PKV distributions are relatively homogeneous. The lowest PKV is at $65^{+61}_{-33}$ \si{\km \per \second}, which then increases smoothly to $302^{+118}_{-92}$ \si{\km \per \second} when using the NE2001 DM distance model. YMW16 distances give slightly higher velocities; the lowest PKV is at $83^{+79}_{-47}$ \si{\km \per \second}, and the largest is $383^{+154}_{-124}$ \si{\km \per \second}. The PKVs calculated from each DM model are typically different by $ <10$ \si{\km \per \second}, which is well within error. See Table \ref{tab:bw_table} for a summary of system parameters and PKV estimates. The best-fit Beta distribution for the black widows has $\alpha = 2.44^{+0.38}_{-0.37}$, $\beta = 12.7^{+2.1}_{-2.1}$, and $s = 860^{+150}_{-140}$ \si{\km \per \second}. The corresponding distribution has a positive skew, with a mean and mode of $138$ and $94$ \si{\km \per \second}, respectively. The PDF corresponding to the best-fit parameters can be found in Figure \ref{fig:beta_pdf_cdf}. Figure \ref{fig:cdf_90_bws} compares the black widow PKV MC realisations to the posterior of the beta model fitting.

\begin{figure}
	\includegraphics[width=\columnwidth]{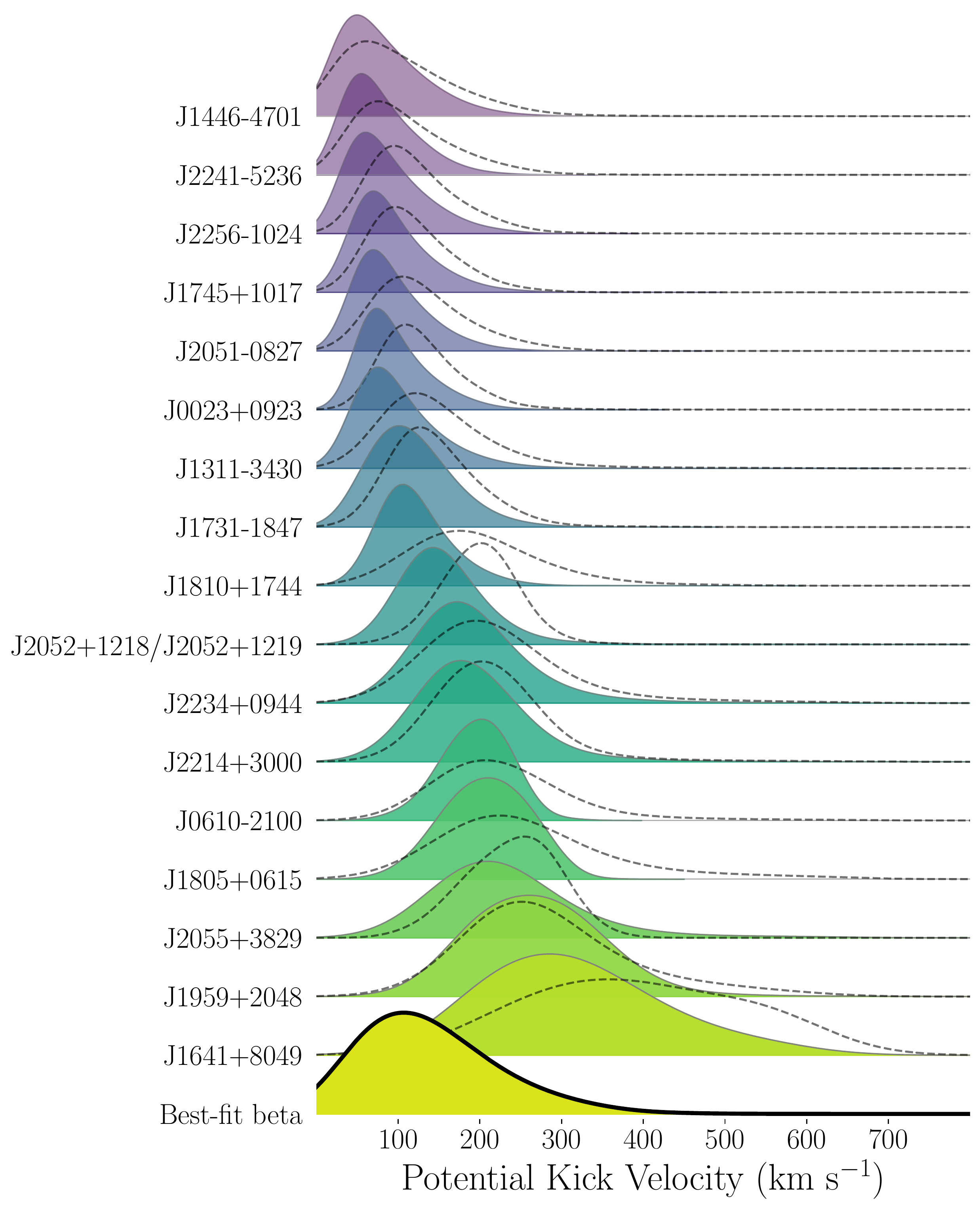}
    \caption{Ridgeline plots showing scaled PDFs of the PKV distribution for each black widow system calculated using the NE2001 (solid line and filled) and YMW16 (dashed line) DM distance model. The `Best-fit beta' curve is the best-fit beta distribution to the full sample of black widows with NE2001 DM distances.}
    \label{fig:bw_ridge}
\end{figure}

\subsection{Millisecond Pulsars}
Similar to the black widows, the 95 MSPs appear relatively homogeneous, with the peaks of the PKV distribution varying smoothly across the sample. See Figures \ref{fig:msp_ridge_1}, \ref{fig:msp_ridge_2}, and \ref{fig:msp_ridge_3} in Appendix \ref{appendix:MSP} for the MSP ridgeline plots. The minimum and maximum PKVs using the NE2001 model are $29^{+26}_{-13}$ \si{\km \per \second} and $233^{+64}_{-58}$ \si{\km \per \second}, respectively. For the YMW16 model the minimum and maximum PKVs are $32^{+29}_{-17}$ \si{\km \per \second} and $355^{+103}_{-118}$ \si{\km \per \second}. Note that while the NE2001 and YMW16 PKVs do not agree for a few MSPs (see Table \ref{tab:msp_table}), the sample-level best-fit parameters for the NE2001 and YMW16 PKVs agree within errors. See Table \ref{tab:msp_table} for a summary of system parameters and PKVs. The best-fit Beta distribution for the MSPs has $\alpha = 2.81^{+0.30}_{-0.29}$, $\beta = 12.8^{+2.2}_{-2.3}$, and $s = 414^{+77}_{-71}$ \si{\km \per \second}. The corresponding distribution has a positive skew, with a mean and mode of $74$ and $55$ \si{\km \per \second}, respectively. The PDF corresponding to the best-fit parameters can be found in Figure \ref{fig:beta_pdf_cdf}. Figure \ref{fig:cdf_90_msp} compares the MSP PKV MC realisations to the posterior of the beta model fitting.

\subsection{All Systems}
\label{sec:results_all}
\begin{figure*}
    \begin{subfigure}{.5\textwidth}
        \includegraphics[width=\columnwidth]{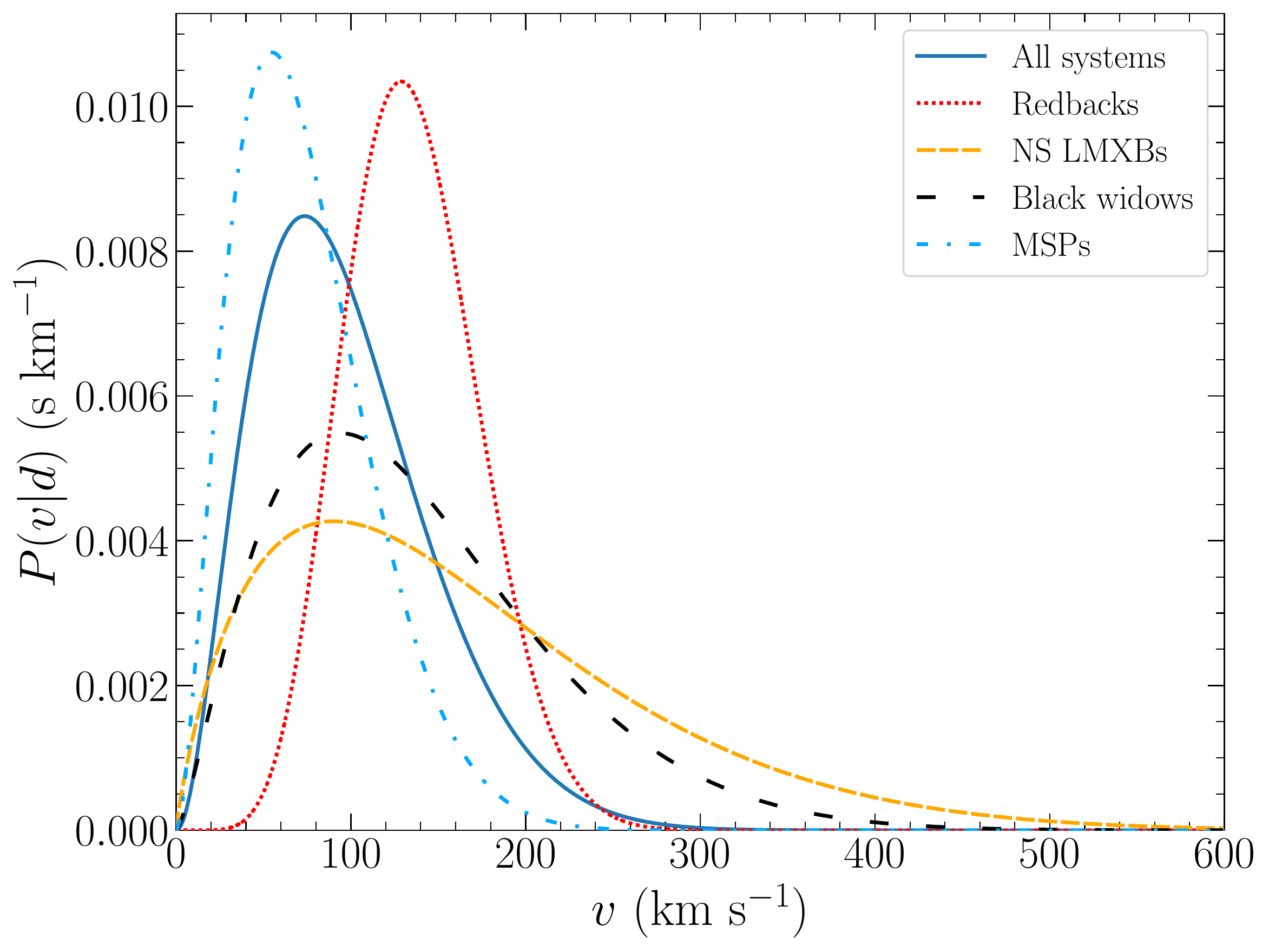}
    \end{subfigure}%
    \begin{subfigure}{.5\textwidth}
        \includegraphics[width=\columnwidth]{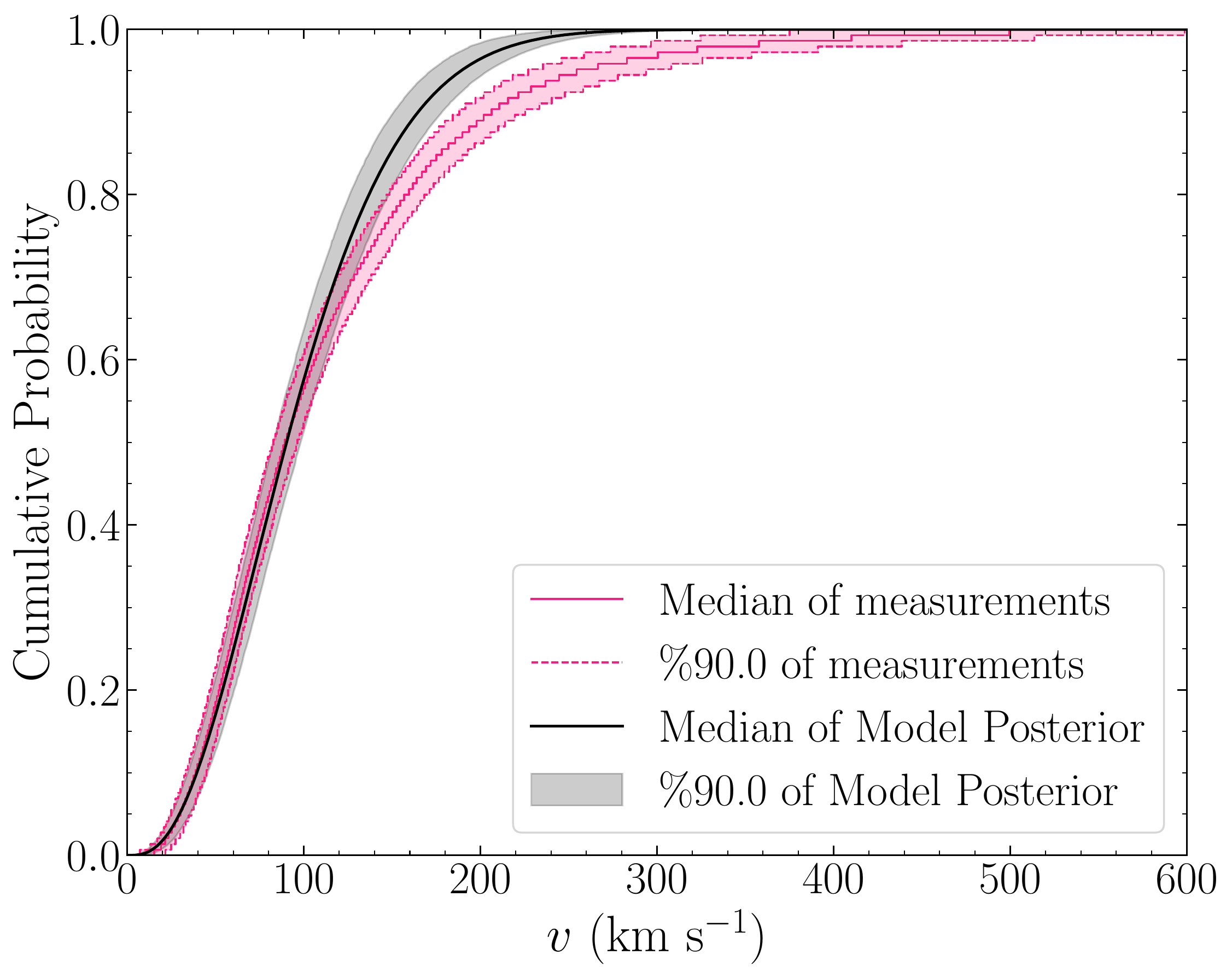}
    \end{subfigure}%
	
    \caption{Left panel: PDFs corresponding to the best-fit Beta distributions for individual sub-samples as well as the whole sample. The transparent black curves correspond to Beta distributions whose parameters came from randomly sampling the posterior, and therefore reflect the statistical uncertainty of the fit. Right panel: eCDF created using the PKV distributions for all 145 systems. One PKV (MC realisation) was randomly sampled from the PKV distribution of each system. An indiviudal eCDF was then constructed using the 145 PKVs. This was repeated 1000 times. The dashed line and filled region describes a 90\% quantile of the individual eCDFs. The central solid line indicates the median of the 1000 eCDFs. Plotted over as a solid black line is the CDF corresponding to the best-fit Beta distribution fit to all systems, with the shaded region encompassing a 90\% quantile of the model posterior.}
    \label{fig:beta_pdf_cdf}
\end{figure*}

We combined all 145 systems and fit the four models -- unimodal truncated Gaussian, bimodal truncated Gaussian, Maxwellian, and Beta -- discussed in Section \ref{sec:model_fit}. The best-fit parameters for each model can be found in Table \ref{tab:fits}. Located in the same table are the statistics for model comparison, AICc and LOO, as discussed in Section \ref{sec:model_comp}. As can be seen, the Beta distribution is the preferred model, with only the Maxwellian distribution having non-negligible probability. Furthermore, qualitatively, when comparing the different models to empirical cumulative distribution functions (eCDFs) created using the system PKV distributions, the Beta distribution follows the data much better than the other three models. The right panel of Figure \ref{fig:beta_pdf_cdf} presents a comparison of the posterior of the beta model fitting against a 90\% eCDF created using the full sample of systems. The same plot for the Maxwellian (Figure \ref{fig:maxw}), unimodal truncated Gaussian (Figure \ref{fig:unigaus}), and bimodal truncated Gaussian (Figure \ref{fig:bigaus}) distributions can be found in Appendix \ref{appendix:model_comp}. As can be seen, there is good agreement between the Beta model distribution and data. Therefore, we do not consider the truncated Gaussian or Maxwellian models further. However, it is worth noting that the Beta model does not have the high-velocity tail present in the data, as can be seen in Figure \ref{fig:beta_pdf_cdf}. This level of confidence is echoed for each sub-sample when we repeat the same fitting. The best-fit Beta distribution for the full sample of systems has $\alpha = 3.05^{+0.32}_{-0.30}$, $\beta = 14.6^{+2.2}_{-2.1}$, and $s = 563^{+72}_{-68}$ \si{\km \per \second}. The corresponding distribution has a positive skew, with a mode of $73.8^{+5.3}_{-5.4}$ \si{\km \per \second} and a mean of $97.3^{+4.9}_{-4.7}$ \si{\km \per \second}. The PDF corresponding to the best-fit parameters can be found in Figure \ref{fig:beta_pdf_cdf}, as can a comparison between the PKV MC realisations of the entire sample and the posterior of the beta model fitting.

\begin{center}
    \begin{table*}
        \begin{tabular}{lcccccc}
            \hline
             Model&  \multicolumn{3}{c}{Parameters}  & AICc (Prob) & LOO weight \\
            \hline
            \textbf{Beta} & $\pmb{\alpha = 3.05^{+0.32}_{-0.30}}$ & $\pmb{\beta = 14.6^{+2.2}_{-2.1}}$ & $\pmb{s = 563^{+72}_{-68}}$ & $\pmb{1561.9}$ $\textbf{(1.0)}$ & $\textbf{0.9565}$ \\
           Maxwellian & $\sigma_m = 61.6^{+2.8}_{-2.6}$ & & & $1564.8$ $(0.2306)$ & $0.0435$ \\
           Unimodal truncated Gaussian & $\mu = 92.8^{+5.5}_{-6.3}$ & $\sigma_g = 50.3^{+6.7}_{-5.5}$ & & $1578.2$ $(0.0003)$ & $2.0317\times10^{-14}$ \\
           Bimodal truncated Gaussian & $\mu_1 = 63.9^{+3.4}_{-2.5}$, $\mu_2 = 149.9^{+13.8}_{-13.2}$ & $\sigma_{g1} = 4.9^{+5.1}_{-2.8}$, $\sigma_{g2} = 46.7^{+9.0}_{-8.8}$ & $w_1 = 0.63^{+0.07}_{-0.08}$ & -- & -- \\
            \hline
        \end{tabular}
        \caption{Parametric fits to the PKV distributions for all 145 systems, ordered from best at the top to worst at the bottom. For each distribution the parameters of the parametric fit are presented, along with the AICc value and respective LOO weight. Means, standard deviations, dispersion, and \textit{s} are in \si{\km \per \second}. A smaller AICc value indicates a more appropriate fit. Bracketed next to the AICc value is the relative probability compared to the best model. The LOO weights sum to 1, and a higher value indicates a preference for that model. The bimodal truncated Gaussian has no reported AICc or LOO weight as it was identified to not be a robust model for the data and thus unreliable; it is strongly affected by a single data point. }
        \label{tab:fits}
    \end{table*}
\end{center}

\subsection{Statistical Comparison of Samples}
\label{sec:stat_comp}
The output of the \citet{Atri19} methodology is a list of MC realisations that form the PKV distribution. For each system in each collection, we randomly sampled one realisation from the full distribution. We then performed two-sample Anderson-Darling (AD; \citealt{Scholz1987}) and Kolmogorov-Smirnov (KS; \citealt{hodges1958}) tests to compare each collection to each other, under the hypothesis that they come from the same underlying distribution. We then repeated this process of randomly sampling one realisation and performing AD and KS tests to form a distribution of significance value (p-values). 

\begin{figure*}
    \centering
    \begin{subfigure}{.5\textwidth}
    	\includegraphics[width=\columnwidth]{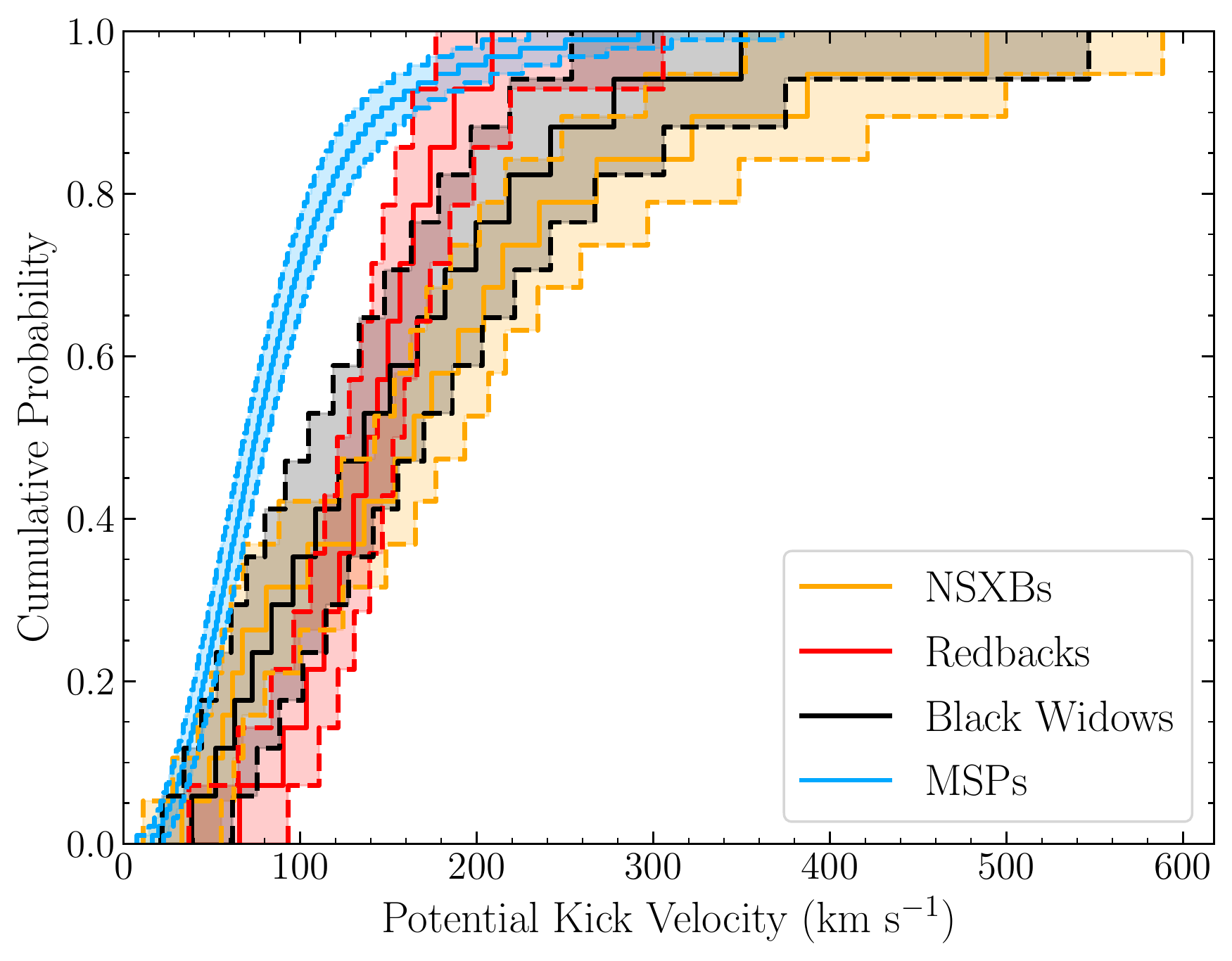}
        \caption{NE2001}
        \label{fig:cdf_90_all_NE2001}
    \end{subfigure}%
    \begin{subfigure}{.5\textwidth}
    	\includegraphics[width=\columnwidth]{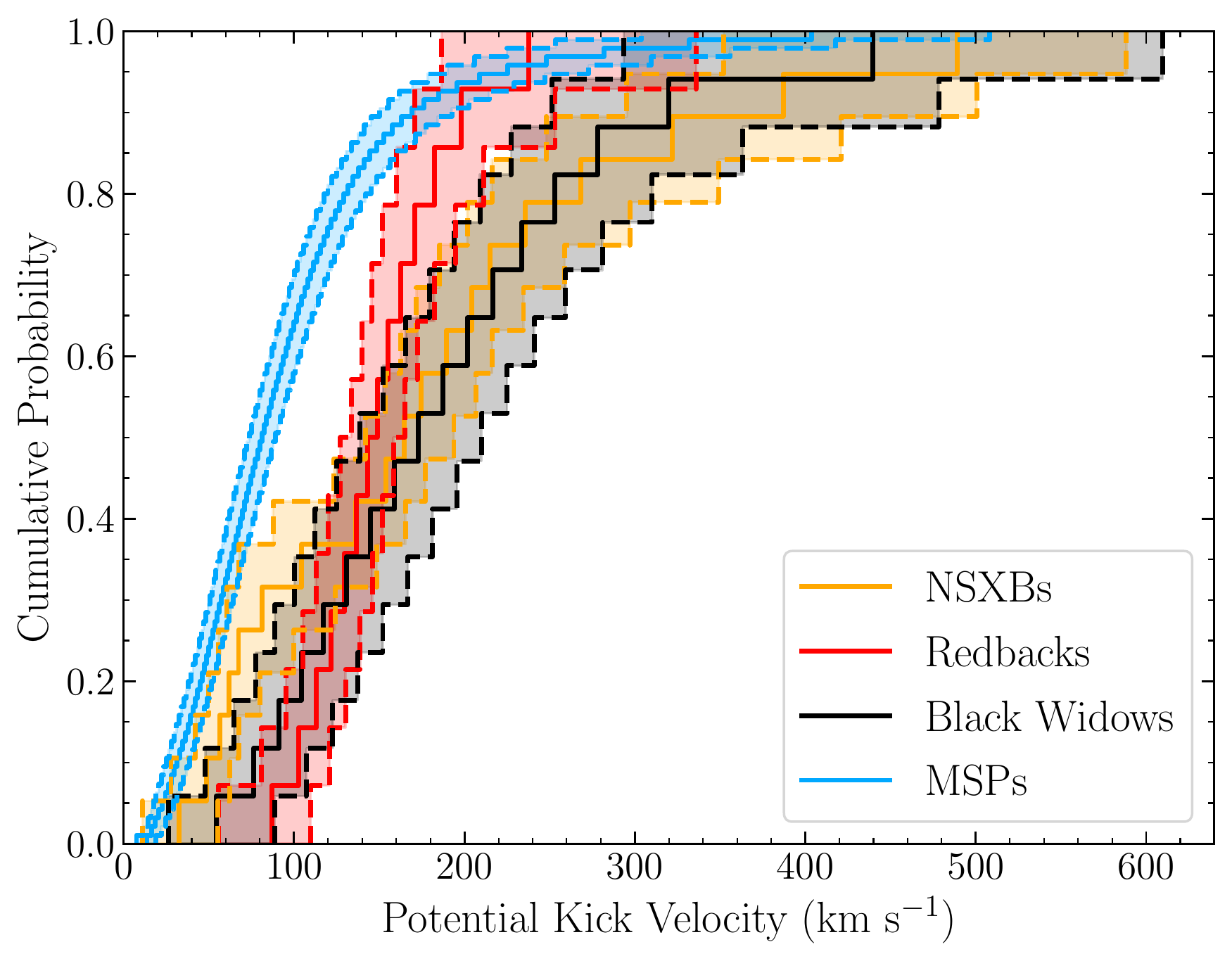}
        \caption{YMW16}
        \label{fig:cdf_90_all_YMW16}
    \end{subfigure}
    \caption{eCDFs comparing the NS LMXBs, redbacks, black widows, and MSPs. The eCDFs were constrcuted in the same manner as described in Figure \ref{fig:beta_pdf_cdf}. The left and right panels indicate PKVs estimated using NE2001 and YMW16 DM distances, respectively. The dashed lines and shaded region indicate the 90\% quantile of eCDFs, with the central solid line indicating the median of the 1000 eCDFs.}
\end{figure*}

We find no statistical difference between the redbacks, NS LMXBs and the black widows. However, the underlying distribution of the MSP PKV estimates is shown to be statistically different compared to redbacks, NS LMXBs and the black widows. This can be visualised in Figures \ref{fig:cdf_90_all_NE2001} and \ref{fig:cdf_90_all_YMW16}, which show empirical cumulative distribution functions (eCDFs) for each collection, with the dotted lines showing the 90\% interval as a result of the random sampling.

\section{Discussion}
\label{sec:discussion}

It is important to note that for the NS LMXBs, rebacks, black widows, and MSPs we are estimating the binary kick. The inertia of the companion means the binary kick we see is slower than the NS's true natal kick. However, it is not simple to correct for this. The companion's mass at the time of the supernova is unknown, and over the time since, the companion has been accreted from, and potentially ablated, which decreases its mass to what is seen today. The impact this has on the kick we estimate depends on whether the mass transfer is conservative or non-conservative. If the mass transfer is conservative, all the matter that has been lost from the companion will be transferred to the NS. Alternately, if the mass transfer is non-conservative, like in redbacks and black widows, the accretor will have gained less total mass than the companion lost. The companions in the systems studied in this work are currently $\lesssim 1 $ M$_\odot$. As LMXBs are, in general, old ($\gtrsim 1$ Gyr), we do not expect the low-mass companions to have originally been more massive than 1--2 M$_\odot$.

\subsection{PKV Distributions of Subsamples}
The sample of 145 systems assembled in this work forms the largest kinematic catalogue of NSs in binaries to date. However, whilst we have a large sample, we are limited in the comparisons we can make between subsamples. This is primarily due to large uncertainties on the estimated PKVs, and, with the exception of the MSPs, small sample sizes. In this section, we discuss ways the sample could be improved such that comparing the PKVs of the redbacks, black widows, and NS LMXBs might become possible. Comparisons between the four subclasses are interesting because while these systems experienced phases of mass transfer at some point in their evolution, probably from low-mass companions, they currently appear very differently. There are likely similarities in their evolutionary histories, however what causes their evolution to diverge is not well understood.

\subsubsection{Redbacks}
The redbacks are, overall, well characterised for the purposes of this analysis. The kick of PSR J1048+2339 is the most poorly constrained, indicated by the presence of a secondary, dotted curve. The best existing distance estimate for this system is from DM, and NE2001 and YMW16 give conflicting distances, which greatly affects the PKV. As always, better constrained distances, and preferably parallax distances, would help refine our results. The systems that would benefit most from better distances are PSR J1048+2339, PSR J1622--0315, PSR J1628--3205, and PSR J1957+2516, as they all rely on DM distance estimates. Measuring a systemic radial velocity for PSR J1957+2516 would remove the need for the radial velocity prior and thus help refine the kick estimate.

\subsubsection{Neutron Star Low-Mass X-ray Binaries}

When considering the width of the PKV distributions of NS LMXBs there are three binaries which clearly have much broader PKV distributions: Cen X-4, MXB 1659--298, and SLX 1737--282. However, this can be explained simply. Cen X-4 has a high Galactic latitude ($ b \approx 24\degree $), and whilst its distance is reasonably constrained ($D = 1.3 \pm 0.3$ kpc), the upper and lower bounds would place the system in different parts of the Galactic potential, which can drastically affect the PKVs. \citet{gonzalezhernandez2005} investigated the kinematics of this system and, based on metallicity and chemical abundance ratios in the companion star, found it unlikely the system formed in a globular cluster and subsequently ejected. They concluded its high space velocity arose from a large kick. The distances of both MXB 1659--298 and SLX 1737--282 place them in the vicinity of the Galactic centre, plausibly on either side, which results in the wide range of PKVs we see.

There are clear opportunities for the improvement of this sample. First, measuring systemic radial velocities for the LMXBs without existing measurements would greatly improve the sample. Secondly, more accurate distances, particularly for Cen X-4, MXB 1659--298, and SLX 1737--282 would greatly improve kick estimates. It would also be beneficial to expand the sample; there are a lot of known NS LMXBs, but there are only the 19 systems studied here that have sufficient kinematic measurements and distance estimates to estimate their kick.

\subsubsection{Black Widows and Millisecond Pulsars}
The PKV curves look similar for all black widows and MSPs due to their proximity and lack of well characterised distances and radial velocities. Their distances are all DM distances, and the systemic radial velocities for all systems come from the radial velocity prior discussed above in Section \ref{sec:RV_prior}. The uncertainties in these parameters dominate, with the measured proper motions slightly altering the individual curves. Radio parallaxes of the pulsars in these systems would help improve distance estimates and provide invaluable model-independent distances. Where possible, measuring systemic radial velocities of the systems would also be of benefit. However, for the black widows and some of the MSPs, the companion is so small this becomes quite challenging. For the MSPs with no detectable companion, measuring the radial velocity would be impossible.

Of particular interest is the double pulsar PSR J0737--3039 \citep{Burgay03,Lyne04}, the single known DNS system in the MSP sample studied in this work. PSR J0737--3039 is one of the few known Galactic DNS systems that will merge within a Hubble time due to emitting gravitational waves, with a merger timescale of 85 Myr \citep{Burgay03}. The important difference between DNS systems and the other systems studied in this work is that the binary has remain bound despite undergoing two supernovae, and thus experiencing two kicks. The PKV of this system is $36^{+38}_{-24}$ \si{\km \per \second} and $41^{+42}_{-24}$ \si{\km \per \second} for the NE2001 and YMW16 DM models, respectively. These PKVs are among the lowest in the full 145 system sample, and consistent with the lowest within error. In line with other works on kicks (e.g., \citealt{Tauris17,Atri19,Mandel20,Willcox21}), we use $\approx 50$ \si{\km \per \second} as the cutoff between low and high kicks, with kicks $\leq 50$ \si{\km \per \second} assumed to be consistent with low, or no, kick. Under this definition, PSR J0737--3039 is consistent with having received low to zero natal kicks. Small to no kicks have previously been suggested for this DNS system \citep[and references therein]{Tauris17}, which our results support.

\subsection{Selection Effects}
It is important to note that the samples presented in this work are not complete and suffer from selection effects. The most dominant selection effect that all subsamples share is that they are not sensitive to large kicks, as a large kick would have unbound the binary. The NS LMXB, redback, black widow, and MSP samples each have different observational selection effects, with the various MSP classes sharing several. The subsamples also potentially have different selection effects due to different evolutionary pathways.

\citet{Lorimer08} provides a good discussion on the three main observational selection effects relevant to MSPs. First is the inverse square law; surveys are most sensitive to the closest and brightest pulsars. There is then pulse smearing due to dispersion and multipath scattering by free electrons in the interstellar medium. This means surveys are less sensitive to more distant pulsars and pulsars in the Galactic plane where there is a large number of free electrons along the line of sight. The third, relevant to pulsars in binaries, is due to orbital acceleration, which causes surveys to lose sensitivity to pulsars in tight binaries when the integration time becomes large. Searching \textit{Fermi}-LAT GeV $\gamma$-ray sources for MSPs has also proved a fruitful method of finding new MSPs (e.g., \citealt{Ray12}), particularly for finding new spider pulsars, with 75\% of the confirmed and strong candidate redback systems being found this way. Identification through \textit{Fermi}-LAT GeV $\gamma$-ray sources is biased away from the Galactic plane. Whilst redback and black widow candidate identification suffers from this $\gamma$-ray selection effect, subsequently finding the radio MSP is not as affected by the latter two pulsar selection effects discussed above; these pulsar searches are targeted to a small region of interest.

NS LMXBs have different selection effects to the MSPs. Discovery of NS LMXBs is primarily through X-ray all-sky surveys. This implies the known population of these systems is biased towards those binaries with short outburst recurrence times or systems that are persistent, such that they are bright enough to be detected. As proper motion measurements of NS LMXBs come from the optical companion, the sample is selected where this is feasible. This selects against systems in the Galactic plane and bulge where extinction is high, systems with light companions, systems in tight orbits (P < few hrs), and systems at large distances.

These selection effects can bias the PKV distributions in different ways. For example, a bias against the Galactic plane might overestimate the number of systems that received large kicks as these systems are likely to travel further out of the plane. This same bias could create a bias against NSs with massive companions due to the extra inertia. A larger natal kick is then required to have the same binary kick as a NS with a lighter companion. The bias against systems at large distances could underestimate the number of systems that received large kicks if it is assumed the system formed in the Galactic plane; systems with higher velocities will travel further, in the same time, than those that received low kicks. There are many selection effects which may introduce many different biases, and it is not obvious how these manifest in the systems and estimated PKVs.

\subsection{Sub-Sample PKV Comparisons}
\label{sec:dist_comp}
As reported in Section \ref{sec:stat_comp}, there appears to be a statistically significant difference between the kicks of MSPs and the kicks of redbacks, NS LMXBs, and black widows. It is not immediately clear if this is a result of selection effects, or if their difference is truly physical. While the redbacks, NS LMXBs, black widows, and MSPs are biased against the plane, they're biased for different reasons. Systems that receive smaller kicks are more likely to currently be closer to the Galactic Plane than systems that receive a large kick. This is because, on average, a small kick will result in the system reaching a lower distance perpendicular to the plane than a high kick. Indeed, we see that MSPs have lower kicks than the other sub-samples. Similarly, if we compute the current distance from the plane, $\abs{\mathrm{z}}$, for each system, the MSPs are found closest to the plane. We made a cut to the MSP sample such that it has a similar distribution of $\abs{\mathrm{z}}$ distances to the other samples and repeated the same statistical comparison as is discussed in Section \ref{sec:stat_comp}. The results of that section are echoed here, as kicks of the MSPs remained statistically distinct from the kicks of the redbacks, NS LMXBs, and black widows.

We then investigated if the use of DM distances for MSPs could be the leading cause of the distinction between the MSPs and the redbacks, NS LMXBs, and black widows, as DM distances are known to be unreliable in some cases (e.g. \citealt{Deller19,Price21}). To test this we retrieved all the MSPs from the ATNF pulsar catalogue \footnote{\url{https://www.atnf.csiro.au/research/pulsar/psrcat/}} \citep{Manchester05} with at least $3\sigma$ parallax measurements, producing a sample of 35 MSPs (hereafter parallax-MSPs). We then estimated the PKVs of each system, and performed the same statistical comparisons as before. The PKVs of the MSPs with DM distances and the PKVs of the parallax-MSPs are statistically indistinguishable. The PKVs of the parallax-MSPs are statistically distinct from the redbacks and NS LMXBs, as they were for the MSPs with DM distances. There were two outliers in the sample of 35 MSPs. If they are included, the PKVs of the black widows are not statistically distinct from the PKVs of the parallax-MSPs. However, if these two are removed then the black widows and parallax-MSPs are distinct (removing the two systems does not affect the distinction from the redbacks and NS LMXBs or the indistinguishability of the parallax-MSPs and MSPs with DM distances). As the PKVs of the parallax-MSPs are consistent with the PKVs using DM distances, we use DM distances for all MSPs for consistency.

The above tests suggest that the difference is physical in nature. Alternately, this difference may hint at unappreciated selection effects.

It is interesting to question if there is any reason why the kicks of redbacks, NS LMXBs, and black widows should be similar. The kicks of redbacks look to occur over a much narrower range than either NS LMXBs or black widows (Figures \ref{fig:cdf_90_all_NE2001} and \ref{fig:cdf_90_all_YMW16}). However, the small sample sizes for each of these sub-samples limits statistical comparison. Whether they appear to be the same because of sample size, selection effects, or because they are actually the same is a difficult question. Larger samples of redbacks, NS LMXBs, and black widows would help to properly address this question.

\subsubsection{Kicks of Isolated and Binary MSPs}
\begin{figure}
    	\includegraphics[width=\columnwidth]{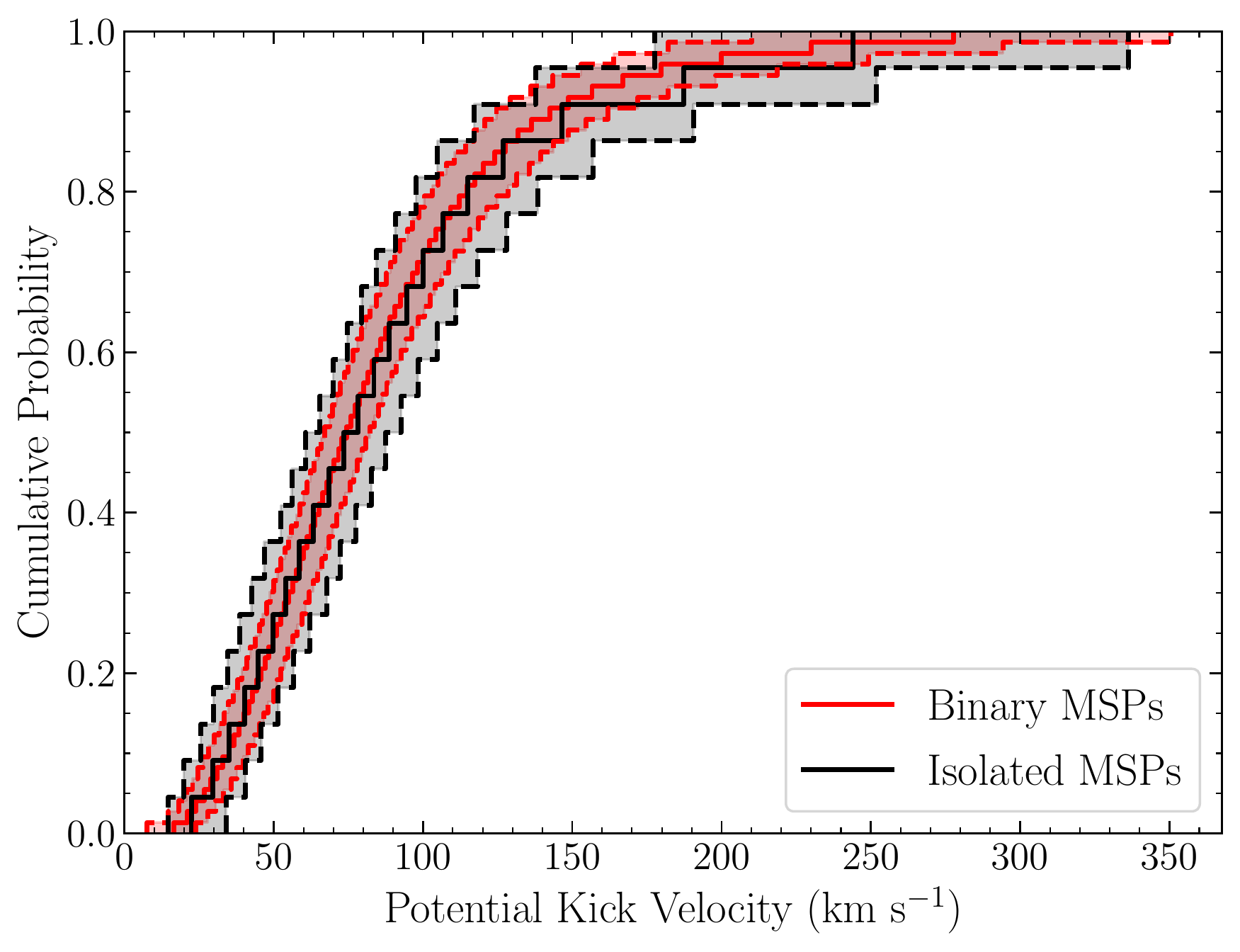}
        \caption{eCDFs comparing the PKVs of the 73 MSPs with known binary companions and the 22 MSPs without known binary companions. The eCDFs were constrcuted in the same manner as described in Figure \ref{fig:beta_pdf_cdf}. The dashed lines and shaded region indicate the 90\% quantile of eCDFs.}
        \label{fig:cdf_90_msps_bin_iso}
\end{figure}

\citet{Toscano99} found the distribution of 2D transverse velocities of MSPs with a known binary companion to be distinct from the velocities of MSPs without a known binary companion. However, studies since have not found any difference (e.g. \citealt{Gonzalez11}, \citealt{Desvignes16}). The sample of 95 MSPs in this work contains 73 MSPs with known binary companions and 22 with no known binary companion. We replicated the analysis described in Section \ref{sec:stat_comp} using these two samples of MSPs. We find no evidence for the kicks coming from different distributions. Indeed, as can be seen in Figure \ref{fig:cdf_90_msps_bin_iso}, the eCDFs look almost identical.

\subsection{Comparison With Other Neutron Star Velocity Studies}
\label{sec:NS_velocity_studies}

\subsubsection{Natal Kicks of Isolated Neutron Stars}
\label{sec:NS_kicks_comp}
The natal kicks of young, isolated pulsars have been studied extensively in the literature, as summarised in Section \ref{sec:intro}. Here we will briefly discuss the most relevant works and compare with our results for NSs in binaries. The $\sigma = 265$ \si{\km \per \second} Maxwellian (mode at 370 \si{\km \per \second})  kick velocity distribution of \citet{Hobbs05} is the most widely used distribution for NS kicks (see Section \ref{sec:intro}). Their sample contained 46 pulsars with characteristic ages $<3$ Myr, with most proper motions coming from pulsar timing and most distances coming from DM. They combined these 2D transverse velocities with a novel deconvolution technique to estimate their 3D birth velocities, to which they fit a Maxwellian. The most recent young, isolated pulsar velocity distribution comes from \citet{Igoshev20}, updating the work of \citet{Verbunt17} to take advantage of the complete PSR$\pi$ survey \citep{Deller12}. \citet{Igoshev20} exclusively used pulsars with VLBI parallaxes and proper motions. They performed the same analysis on the full sample of 69 pulsars, as well as on a smaller sample of 21 objects with characteristic ages $<3$ Myr. In both cases they find a bimodal Maxwellian to be a superior fit to the 2D velocities than a unimodal Maxwellian, with the young sample distribution having best-fit parameters of $\sigma_1 = 56^{+25}_{-15}$ \si{\km \per \second} (mode at 80 \si{\km \per \second}), $\sigma_2 = 336 \pm 45$ \si{\km \per \second} (mode at 480 \si{\km \per \second}), and $w_1 = 0.2^{+0.11}_{-0.10}$. Whilst \citet{Willcox21} do not perform model fitting to velocities, they compile a sample of 81 pulsars with VLBI parallaxes and proper motions (also taking advantage of the complete PSR$\pi$ survey, \citealt{Deller12}) with their study focusing on the implications for weak natal kicks. \citet{Kapil22} used the same single pulsars as used by \citet{Willcox21} to calibrate the momentum-preserving natal kick model proposed by \citet{Mandel20}. All works excluded pulsars with GC associations, MSPs, pulsars known to be in binaries, or pulsars with evidence of recycling.

Due to the selection bias against large kicks present in the sample of NS binaries in this work (i.e., binary disruption), we would expect the most significant difference to be that the best-fit model in this work predicts significantly lower velocities. Furthermore, the population of isolated NSs is likely dominated by NSs whose progenitor was a part of a binary system, but also includes those that formed from isolated stars. Either sufficient mass loss or a large enough kick is required to disrupt the binary, as discussed in Section \ref{sec:intro}. This introduces a bias towards high kicks in the population of isolated NSs. \citet{Hobbs05}, \citet{Igoshev20}, and \citet{Willcox21} find that $<0.2$\%, $5$\%, and $\lessapprox 5$\% of single-pulsar velocities are $\leq50$ \si{\km \per \second}, respectively. We find that $18.6^{+3.4}_{-3.4}$\% of binaries containing NSs have velocities $\leq50$ \si{\km \per \second} based on the PKV distributions of all 145 systems. 

The distributions of \citet{Hobbs05} and \citet{Igoshev20} extend to greater than 1000 \si{\km \per \second}, whereas the best-fit model presented in this work only extends to $\approx 300$ \si{\km \per \second}. The mode of the first Maxwellian component of the bimodal distribution from \citet{Igoshev20} is at 80 \si{\km \per \second}, which is not too different to the mode of the best-fit model in this work that is at $68^{+2}_{-2}$ \si{\km \per \second}. Because of the selection effect against large kicks due to binary disruption and as the model is fit to binary kicks, it is unsurprising that the best-fit model in this work yields lower velocities.

\subsubsection{Current Velocities of Millisecond Pulsars}
\label{sec:current_vel_MSPs}

We now look to previous studies that measured MSP transverse velocities and compare with our results. The sample sizes, selection criteria, and velocities of the works summarised in Section \ref{sec:intro} are described in Table \ref{tab:MSP_velocities}. Proper motions for all systems came from pulsar timing, and the vast majority of distances are DM distances. \citet{Lommen06} and \citet{Gonzalez11} estimated the MSPs' peculiar velocities by correcting for their LSR, whereas \citet{Toscano99}, \citet{Hobbs05}, \citet{Desvignes16}, and \citet{Lynch18} did not make this correction. It is worth noting that whilst studies of MSP velocities have existed for many years, they have not yet been incorporated into the majority of observationally-motivated kick distributions (see Section \ref{sec:intro}).

\begin{table*}
    \begin{center}
        \begin{tabular}{lcccc}
            \hline
              & Criteria & N & $V_{T\rm{,NE2001}}$ (\si{\km \per \second}) & $V_{T\rm{,YMW16}}$ (\si{\km \per \second}) \\
            \hline
            \citet{Toscano99}   &  & 23 & $85 \pm 13$ &  \\
            \citet{Hobbs05}     & $P < 0.1$ s, $\Dot{P} < 10^{-17}$ s s$^{-1}$ & 35 & $87 \pm 13$ &  \\
            \citet{Lommen06}    & $P < 0.01$ s & 29 & $91 \pm 28$ &  \\
            \citet{Gonzalez11}  & $P < 0.01$ s & 37 & $108 \pm 15$ &  \\
            \citet{Desvignes16} & $P < 0.02$ s & 76 & $92 \pm 10$ &  \\
            \citet{Lynch18}     &  & 5  & $152 \pm 48$ & $234 \pm 143$ \\
            \hline
        \end{tabular}
        \caption{Summary of works discussed in Sections \ref{sec:intro} and \ref{sec:current_vel_MSPs} that provide transverse velocities for MSPs. The criteria column refers to selection criteria listed in the paper that was used when assembling the sample. $N$ is the number of MSPs in the sample. $V_{T}$ (\si{\km \per \second}) is the mean and standard deviation of the transverse velocities of the MSPs in the sample. The subscript NE2001 and YMW16 indicates if the velocities were calculated using distances from the NE2001 or YMW16 DM model.}
        \label{tab:MSP_velocities}
    \end{center}
\end{table*}

There are several important things to note before comparing the MSP velocities in Table \ref{tab:MSP_velocities} to the results in this work. First, these are 2D transverse velocities and thus put a lower limit on the MSPs' true 3D space velocities. In this work, we have estimated the binary kicks of NS binaries in 3D. For an isotropically distributed 3D velocity vector, projecting to 2D and integrating over all projection angles gives the ratio of mean 3D to 2D speed of $4/\pi$. Therefore, multiplying the mean 2D speeds in Table \ref{tab:MSP_velocities} by $4/\pi$ gives an estimate of the mean 3D speed. Second, peculiar velocity is not a conserved quantity as the source orbits within the Galaxy, as the system experiences acceleration in the Galactic potential. Therefore, the velocities measured by these works should not be assumed to be indicative of the birth velocity. As these works reported the mean velocities of their samples, we shall compare with the mean of the best-fit model of the full sample and MSP subsample in this work.

Making the correction based on the assumption of isotropy, all of the mean velocities listed in Table \ref{tab:MSP_velocities} agree with the mean of the best-fit Beta distribution for the full sample ($97.3^{+4.9}_{-4.7}$\SI{}{\km\per\second}) within error, excluding the NE2001 and YMW16 mean velocities from \citet{Lynch18}. If we then compare the mean MSP velocities to the mean of the best fit distribution for only the MSPs ($74.4^{+5.2}_{-4.8}$ \SI{}{\km\per\second}), none of the sample mean velocities agree within error. However, this is not unexpected as we would not expect peculiar velocity to be a conserved quantity.

\subsubsection{Kicks of High-Mass X-ray Binaries}
\citet{Igoshev21} investigated the 2D velocity distribution of Be XRBs. These XRBs are HMXBs with Be companions with masses typically around 8 M$_\odot$. \citet{Igoshev21} compile a sample of 45 such Be XRBs using proper motions and parallaxes from eDR3. When computing the transverse velocities of these objects \citet{Igoshev21} subtracted the contribution of the LSR. They find the transverse velocities of these systems can be described as the sum of two Maxwellians with $\sigma_1 = 11$ \si{\km \per \second} (mode at 16 \si{\km \per \second}), $\sigma_2 = 44$ \si{\km \per \second} (mode at 62 \si{\km \per \second}), and $w_1 = 0.8$.

The Be XRB systems studied by \citet{Igoshev21} are most likely younger than $100$ Myr which, depending on the kick they received, is more than enough time for them to leave the region in which they formed. However, due to the mass of the companion, and the binary remaining bound, they are unlikely to have received a strong binary kick. This is reflected in the small velocities predicted by the bimodal Maxwellian fit in \citet{Igoshev21}. As a result, these velocities are likely indicative of the system's binary kick. The velocities predicted by this distribution are clearly lower than those coming from our best-fit Beta distribution.

With the exception of Her X-1 and 4U 1954+319, all NSs studied in this work have companion masses less than $1$ M$_\odot$. It is likely that the inertia of the massive Be companions limit how much the natal kick influences the motion of the binary, and thus the observed binary kick. Furthermore, the NSs in many Be XRBs are likely born in very low-kick ECSNe (e.g., \citealt{Vinciguerra20}). 4U 1954+319, with a companion mass of $9^{+6}_{-2}$ M$_\odot$ \citep{Hinkle20}, is consistent with having a low binary kick ($55^{+55}_{-30}$ \si{\km \per \second}; note the radial velocity prior was used for this source) as would be expected for something with such a massive companion. We find that Her X-1 received a large binary kick of $201^{+10}_{-38}$ \si{\km \per \second}, however, this is not unreasonable given its companion is estimated to have a mass of $\approx2$ M$_\odot$ \citep{Rawls11}. It is likely that the difference in binary kicks can be attributed to multiple factors, including the difference in companion masses, different evolutionary histories, and particularly ECSNe after the NS progenitor has been stripped.

\citet{Igoshev21} then combine these Be XRBs and the young, isolated pulsars from \citet{Igoshev20} to perform joint fitting. In their best model, a bimodal Maxwellian, the second Maxwellian component is held constant using $\sigma_2$ from \citet{Igoshev20}. They find $\sigma_1 = 45^{+25}_{-15}$ \si{\km \per \second} (mode at 64 \si{\km \per \second}). The mode of $\sigma_1$ is consistent with the mode of the best-fit model of all 145 systems in this work, within error. 

Recently, \citet{Fortin22} estimated the natal kicks of 35 NS HMXBs. Combining data from EDR3 and the literature, they compiled a sample of 35 HMXBs with measured orbital periods, positions, proper motions, and parallaxes. Of the 35, they found systemic radial velocities for 17 of them. For the remaining 18, they assumed a uniform prior. Using this information, \citet{Fortin22} estimated the system's peculiar velocity (their Figure 2), which due to the young age of HMXBs, should be representative of the binary kick. As expected, and similar to the findings of \citet{Igoshev21} for Be XRBs, these binary kicks are much lower than what was found in this work for LMXBs. They extended their work to estimate the natal kick received by the NS by modelling the pre- and post-supernova system and employing an MCMC methodology to estimate the unknown parameters, which includes the kick velocity. \citet{Fortin22} found a Maxwellian distribution to insufficiently model the data. Instead, they fit a Gamma distribution with mean and skew of $116^{+16}_{-16}$ \si{\km \per \second} and $1.7^{+0.2}_{-0.2}$, respectively. The mode of this gamma distribution is at $32$ \si{\km \per \second}, with a tail extending out to above $450$ \si{\km \per \second} (99 percentile is $\approx 456$ \si{\km \per \second}). The mode of their natal kick distribution is approximately half the mode of the binary kick distribution of all 145 systems reported in this work, and also extends to slightly higher velocities. While outside the scope of this paper, determining and comparing the natal kick distributions for the LMXBs in this work to the natal kick distribution for the HMXBs in \citet{Fortin22} would be enlightening.

\subsection{Interpretation}
\subsubsection{Millisecond Pulsar Kick Differences}

\begin{figure}
	\includegraphics[width=\columnwidth]{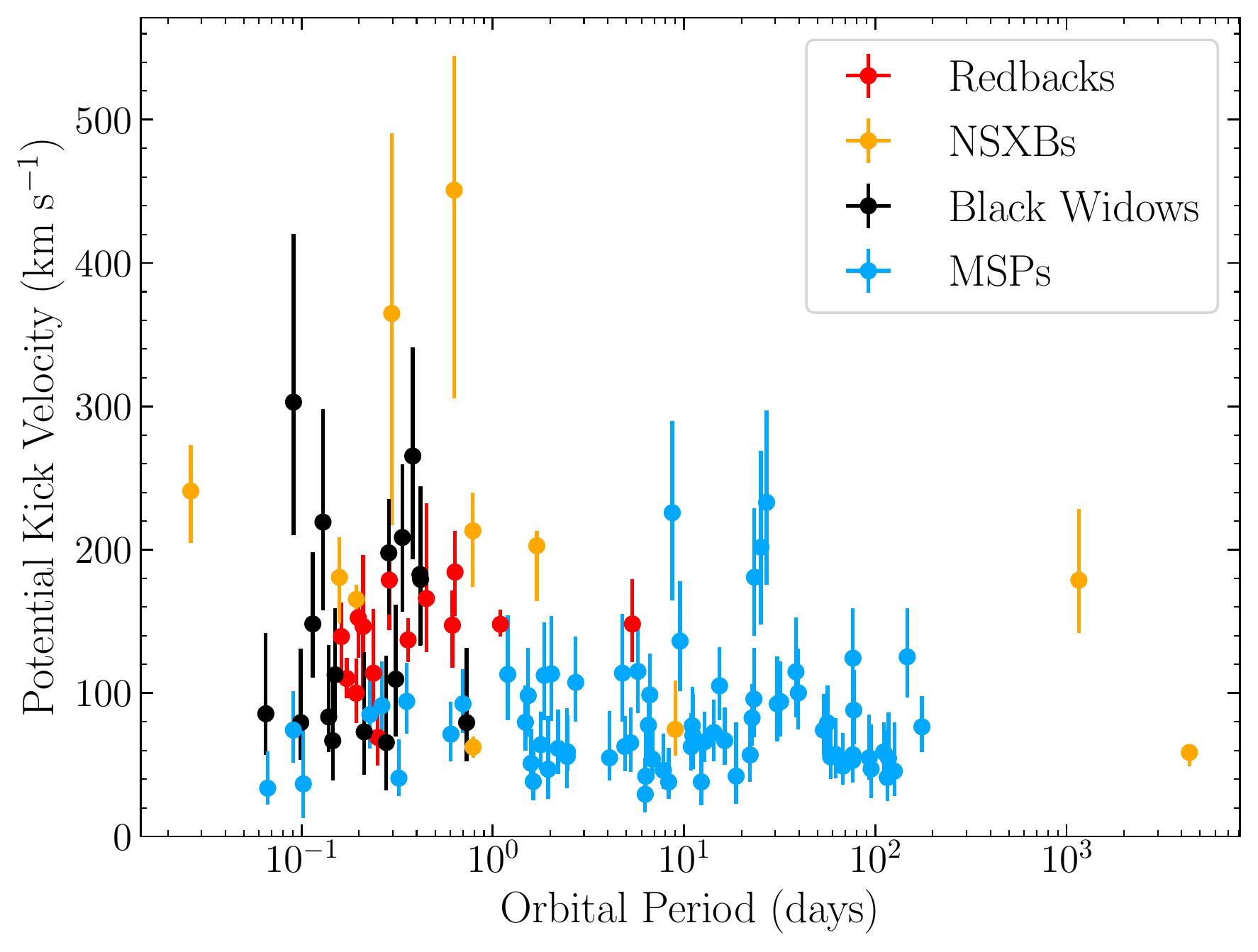}
    \caption{Estimated PKV against binary orbital period for all systems that have a measured orbital period. Note that some NS LMXBs do not have constrained periods (Table \ref{tab:nsxb_table}), and that 22 MSPs do not have known binary companions.}
    \label{fig:pkv_orbital_period}
\end{figure}

We have not identified a clear explanation for the difference in kick velocities of the MSPs when compared to the redbacks, NS LMXBs, and black widows. We explored one main avenue not related to selection effects. The Blauuw kick is strongly correlated with the pre-supernova orbital velocity \citep{Blaauw61}, which in turn relates to the orbital period. These binaries typically have well measured orbital periods, so the orbital period could be used as a proxy for orbital velocity, which should correlate with kick. We investigated this possibility and found no apparent correlation (see Figure \ref{fig:pkv_orbital_period}). However, the uncertainties on the kicks are not negligible in comparison to the difference between the systems with the largest and smallest kicks, which could obscure any correlation. Currently there is not a sufficient lever arm to make a definitive statement. However, with better data in the future it should become possible to interrogate this relationship further.

It is important to note that using these proxies assumes the orbit of the binary does not widen significantly post-supernova. However, accretion onto the neutron star post supernova could widen the binary by an order of magnitude or more. If $q_i$ and $q_f$ are, respectively, the initial and final mass ratio of the donor to the accretor and the mass transfer is completely non-conservative, with the neutron star isotropically re-emitting all accreted material, then the binary widens by a factor of $e^{2(q_f - q_i)} (q_i/q_f)^2 (q_i + 1)/(q_f + 1)$  (e.g., \citealt{Postnov14}).  For example, an initially $1$ M$_\odot$ companion donating mass onto a $1.4$ M$_\odot$ NS until the donor's mass decreases to $0.2$ M$_\odot$ will widen the binary by a factor of ~12. This makes it challenging to infer the pre-supernova orbital velocity from current observations without more detailed binary evolution modelling.

\subsubsection{Supernova Mechanisms and Kick Magnitudes}
\label{sec:supernova_mechanisms}

Different supernova mechanisms can intrinsically produce different kick magnitudes. Iron core-collapse supernovae (Fe CCSNe) can explain kick magnitudes up to  $1000$ \si{\km \per \second}, and are likely responsible for the large magnitude kicks received by some NSs \citep{Wongwathanarat13,Muller2020}. However, \citet{Janka17} and references therein, showed, based on simulations, that there are other processes which will always produce very low kicks ($\leq 50$ \si{\km \per \second}). These are typically associated with electron-capture supernovae (ECSNe; \citealt{Nomoto84,Nomoto87}) and ultra-stripped supernovae (USSNe; \citealt{Tauris13,Tauris15}). ECSNe occur when an ONeMg core collapses due to electron-capture. Many modern population synthesis codes incorporate a low-kick Maxwellian attributed to ECSNe into their assumed kick distributions, with the dispersion of the Maxwellian coming from theoretical modelling (see Section \ref{sec:intro}). The supernova explosion in these systems is believed to be fast, ejecting the envelope before instabilities can grow and accelerate the NS, resulting in kicks with low velocities.

2D and 3D simulations performed by \citet{Gessner18} found ECSNe provided kicks on the order of a few \si{\km \per \second}, noting that it seems incredibly unlikely that any current ECSNe models can explain kicks larger than 100 \si{\km \per \second}. In USSNe mass transfer strips a star of its envelope leaving a naked core behind. \citet{Tauris15} find it likely that USSNe produce small kicks as only $\sim0.1$ M$_\odot$ is ejected in the supernova, compared to several M$_\odot$ in standard supernova explosions. Furthermore, even weak outgoing shocks can expel the star's envelope before large anisotropies can build up \citep{Tauris15}. Note that small kicks have been suggested to come from Fe CCSNe when the iron core is small (\citealt{Podsiadlowski04}, and references therein).

Considering the kick mechanisms and the kick magnitudes Fe CCSNe, ECSNe, and USSNe can produce, it would not be unreasonable to expect the intrinsic NS kick distribution to be bimodal. Indeed, this has been suggested theoretically (e.g., \citealt{Katz75,Katz83,Podsiadlowski04,Schwab10,Beniamini16}) and observationally based on pulsar velocities (e.g., \citealt{Arzoumanian02,Brisken03,Verbunt17}). However, we find no evidence for bimodality in the sample of binaries in this work. As can be seen in Figure \ref{fig:beta_pdf_cdf}, the eCDFs vary smoothly from $\approx0-300$ \si{\km \per \second} with no evidence or suggestion of multiple components. Similarly, we find a unimodal distribution sufficient to model the data. Note that while we do not find evidence for bimodality, we cannot rule it out for the true underlying NS kick distribution.

We find that a significant fraction of the systems in this work are consistent with having low binary kicks, such as those that can come from ECSNe and USSNe. Based on the PKV distributions of all 145 systems we find the fraction of binaries that receive binary kicks $\leq 50$ \SI{}{\km\per\second} is $18.6^{+3.4}_{-3.4}$\%. Modelling by \citet{Podsiadlowski04} found that for isolated stars and stars in wide binaries the fraction of stars that undergo ECSNe is very low due to dredge up of the helium core. This is supported observationally by \citet{Willcox21} who studied isolated pulsars and found ECSNe to be rare. However, \citet{Podsiadlowski04} found that for close binaries that experience mass transfer, specifically the stripping of the star's hydrogen-rich envelope, ECSNe are comparatively much more common. With the exception of three NS LMXBs (4U 1700+24, GX 1+4, and possibly 4U 1954+319; see Table \ref{tab:nsxb_table}) that accrete via the companion's wind, the remaining systems studied in this work all accrete via RLOF. Without further consideration, and if we make the rather extreme assumption that all kicks $\leq 50$ \si{\km \per \second} come from ECSNe, our results support the aforementioned findings of \citet{Podsiadlowski04}. The fraction of systems that receive binary kicks $\leq 50$ \SI{}{\km\per\second} in this work ($18.6^{+3.4}_{-3.4}$\%) is significantly larger than that coming from isolated pulsars ($\lessapprox 5$\%; \citealt{Willcox21}). However, as discussed in Section \ref{sec:intro}, it is likely a large fraction of isolated pulsars were originally in binaries that were disrupted upon formation. Moreover, mass transfer likely occurred in a subset of these binaries, in which case the isolated pulsar population cannot be assumed to exclusively have supernova arising from effectively single star evolution. Furthermore, low kicks are much less likely to disrupt a binary than large kicks, further contaminating the comparison.

\citet{Wanajo11} constrained the fraction of all CCSNe that are ECSNe over Galactic history to be $\approx4$\%. There are several reasons why the fraction of low kicks predicted by \citet{Wanajo11} is lower than what is found in this work. The systems we study are most likely old ($\gtrsim 10^9$ yr), and \citet{Wanajo11} found that the fraction could have been higher at earlier Galactic epochs, consistent with when the NSs in these systems formed. There is also the possibility that the low kicks do not come exclusively from ECSNe, and that USSNe and small kicks from small iron CCSNe also contribute. However, we think the most likely reason for the difference is that the fraction of NSs that receive ECSNe is going to be higher for a sample of binary systems. ECNSe produce lower kicks than typical iron CCSNe, thus naturally producing the higher fraction of low kicks found in this work. Investigating bimodality and the contributions of different supernovae mechanisms more conclusively will become possible in the future with better constrained PKVs and larger samples.

\subsection{Comparison with Black Hole Natal Kicks}
It is interesting to compare the results for the NS LMXBs in this work with the BH XRBs of \citet{Atri19}. Their best-fit to the BH PKVs is a unimodal Gaussian with $\mu = 107\pm16$ \si{\km \per \second}. It is important to note that \citet{Atri19} did not test a Beta distribution when performing their fitting. We suggest testing a Beta distribution and comparing it to the best-fit Gaussian would be wise, as we find that their Gaussian model does not fit the data as well as it should. This can be seen clearly in the PKVs of 1A 0620--00, Cyg X-1, and V404 Cygni. These three systems all have kicks $<60$ \si{\km \per \second}. The predicted fraction of kicks $<60$ \si{\km \per \second} is $<0.2$\% using the best-fit Gaussian. However, we know from the PKV estimates that three out of 16 sources have PKVs $<60$ \si{\km \per \second}, in strong contrast to the model prediction.

\begin{figure}
	\includegraphics[width=\columnwidth]{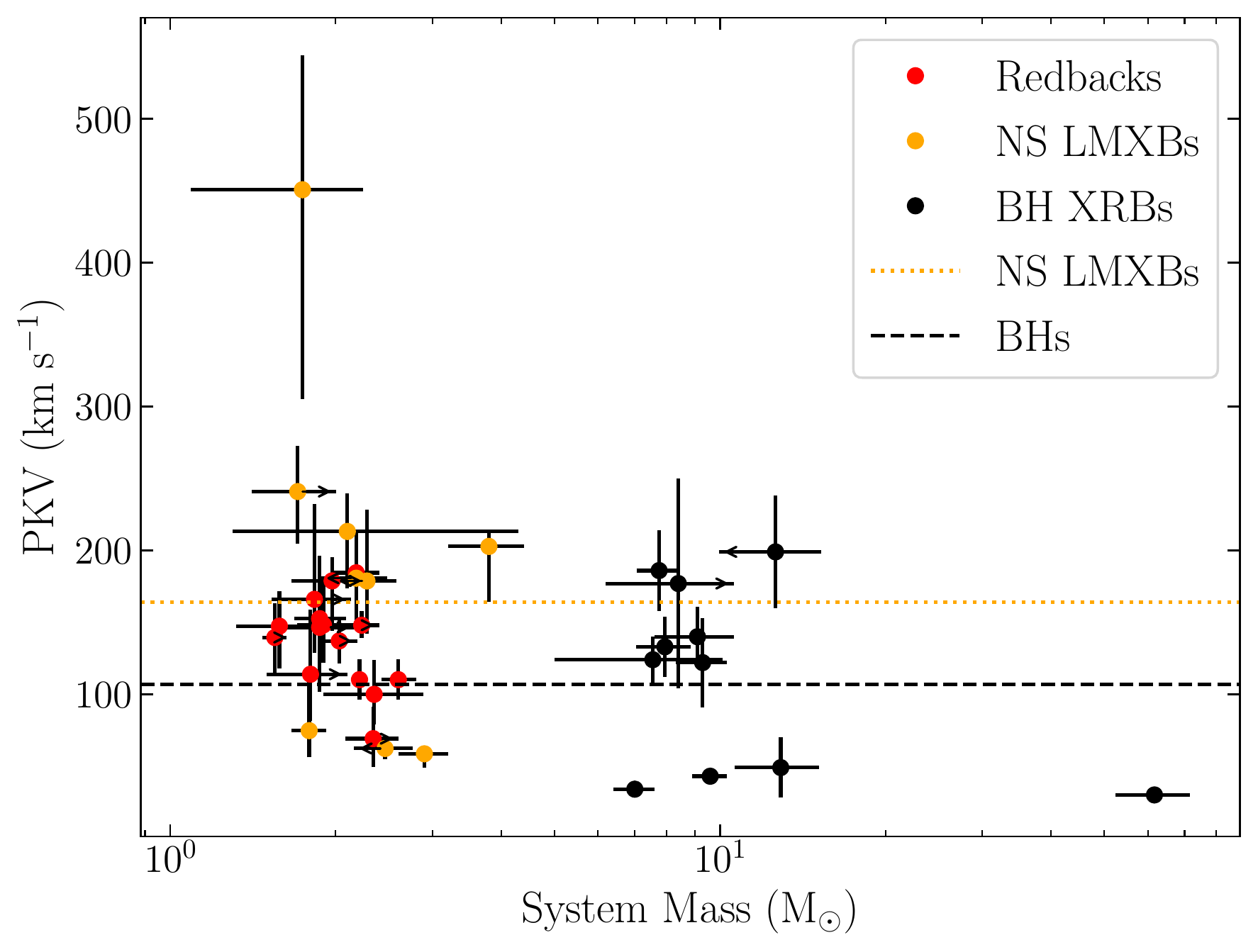}
    \caption{Estimated PKV against total system mass (accretor + donor). Where there was no mass estimate for a NS, a mass of $1.7\pm0.3$ M$_\odot$ was used. The horizontal lines indicate the mean of the best-fit distributions for the NS LMXBs and BHs. See Table \ref{tab:mass_table} in Appendix \ref{appendix:mass_table} for the compact object and stellar companion masses used.}
    \label{fig:pkv_mass_comp}
\end{figure}

Nevertheless, it is still useful to compare the fit distributions. Although the mode of the best-fit model for the NS LMXBs presented in this work ($89^{+23}_{-23}$ \si{\km \per \second}) is agrees with the mode of the BH XRB Gaussian within error, the best-fit model in this work has a tail that goes to significantly higher velocities. The tail of the Beta distribution stretches to $\approx 600$ \si{\km \per \second}, whereas their Gaussian model is negligible above $\approx 150$ \si{\km \per \second}. Looking instead at the PKV distributions of each BH XRB, there is little evidence for any of the BH XRBs receiving binary kicks larger than $\approx 400$ \si{\km \per \second}. This is not entirely unexpected. BHs are much more massive than NSs, and as such, if the same momentum is imparted to both NSs and BHs, the NSs will have the higher velocity. \citet{Repetto12} ruled out BH kicks where the BH kick velocity is simply reduced by the factor $M_{BH}/M_{NS}$ compared to NS kicks to high statistical significance. The BHs in \citet{Atri19} are in the range of 2--15 times more massive than the canonical NS mass. Comparing the means of the two distributions, the NS LMXBs have a mean of $162.8^{+26.5}_{-22.1}$ \si{\km \per \second}, only $\approx50$\% larger than the BH XRBs. This can be visualised in Figure \ref{fig:pkv_mass_comp}, where PKV is plotted against total system mass for the redbacks, NS LMXBs, and BH XRBs. Note that with the exception of Cygnus X-1, the systems studied by \citet{Atri19} are all LMXBs, and therefore their companions are similar in mass to those of the NS LMXBs studied in this work. This supports the suggestion that BH kicks are not purely momentum conserving. 

\subsection{Implications for Modelling the Formation of Double Neutron Star Systems}
NS kicks are a critical component of modelling the formation of DNSs which are not yet sufficiently understood \citep{Janka12,Vigna-Gomez18}. As discussed in Section \ref{sec:NS_kicks_comp}, the kicks received by isolated pulsars are, on average, higher than those received by NSs that remain in binaries post supernova. Furthermore, as discussed in the same Section, the \citet{Hobbs05} kick distribution (the most commonly used distribution in population synthesis) appears to underestimate the fraction of isolated pulsars that receive low kicks by more than an order of magnitude. Small kicks are likely to preserve the initially wide massive binary from disruption during the first supernova, and therefore, a distribution that predicts lower kicks is likely to increase the DNS merger rate. The under-representation of low kicks in kick distributions fit to, or modelled on, samples of only isolated pulsars will therefore have ramifications when used to model the formation of DNS systems (e.g., \citealt[and references therein]{Tauris17}). Whilst there are likely issues with some of the other physical assumptions in population synthesis (e.g., common-envelope evolution), an appropriate NS kick distribution is an important part of reconciling the observed DNS merger rate observed using gravitational waves with the significantly lower rate estimated using population synthesis (e.g., \citealt{Kapil22}). The creation, and adoption, of an up-to-date, observationally-motivated NS kick distribution for use with population synthesis by combining the kick distributions of young, isolated pulsars, binary NSs, and other constraints should be a priority.

\subsection{Implications for Neutron Star Kick Distributions and Population Synthesis Models}
This work highlights the importance of considering more than just young isolated pulsar velocities when fitting for the kick distribution of NSs. However, it is not straightforward to combine the binary kicks of NSs with isolated pulsar velocities to construct a new NS kick distribution. Disentangling information about the natal kicks of NSs in binaries has three main challenges. First, are assumptions related to evolution. For example, did these systems always have a low-mass companion, or were they intermediate-mass XRBs where the companion has since lost most of its mass (e.g., \citealt{Podsiadlowski04})? Furthermore, in order to disentangle the Blaauw kick from the natal kick, the system's pre-supernova orbital velocity is needed. While \"Opik's law could be applied for the orbital velocity distribution \citep{Opik24}, NS LMXBs and all classes of MSPs very likely experienced mass transfer, and potentially a common envelope phase, prior to the NS's birth. In this case, assumptions about evolutionary history must still be made. Second, in the case where the natal kick velocity is much less than the pre-supernova orbital velocity, the binary kick may well be due to the Blaauw kick rather than the natal kick. In that case, the binary kick distribution reported here is an upper limit on the natal kick distribution of NSs in binaries. Third, the fraction of NSs that receive low kicks (and end up remaining in binaries) or high kicks (and end up as isolated pulsars) is not known, which makes combining the NS natal kick distribution of NSs in binaries and isolated pulsars very challenging. Further complicating the problem is that the observable timescales of mildly recycled pulsars ($\sim 10^8$ yr) is significantly less than the observable timescales of NS LMXBs and MSPs ($\gtrsim 10^9$ yr).

However, the PKVs presented in this work are useful as an observational constraint without extracting the underlying natal kick distribution. Combining multiple data sets to constrain kick distributions is exemplified in \citet{Richards22}. To update the momentum conserving kick first presented in \citet{Bray16}, they consider the gravitational wave merger rates, observations of Galactic DNSs, isolated pulsar velocities, and the kick velocities of USSNe. Combining the different constraints of each significantly narrows the permissible parameter space for the free parameters in their kick model. We cannot trivially compare the PKVs presented in this work to the kick distribution presented in \citet{Richards22} as we do not know the ejecta mass or, in most cases, the remnant mass. However, the binary kicks of the systems in this work form another data set against which to calibrate future kick distributions.

While devising a new NS kick distribution is outside the scope of this work, we would like to emphasise the importance of a new NS kick distribution. The majority of population synthesis codes incorporate the $\sigma = 265$ \si{\km \per \second} Maxwellian distribution from \citet{Hobbs05} as the main component of their assumed kick distribution. Whilst there has been an increase in kick prescriptions moving away from simple fits to isolated pulsar velocities (e.g., \citealt{Richards22}, \citealt{Kapil22}), these same velocities are often the dominant observational constraint. The results presented in this work are an important data set that should be studied holistically with other data that can provide natal kick constraints (e.g. the works discussed in Section \ref{sec:NS_velocity_studies} and \citealt[and references therein]{Richards22}) to create a modern, observationally-consistent NS kick distribution.

\section{Conclusions}
We have presented a catalogue of NSs in binaries, containing 145 systems with distances (coming from radio and optical parallaxes, optical light curve distances, PRE X-ray bursts, and DM models), measured proper motions and where possible, systemic radial velocities. Using this sample we estimated the binary kick each system may have received at birth, using a fully 3D treatment. This full sample can be split into subsamples of NS LMXBs (19), redbacks (14), black widows (17), and MSPs (95).

We compared the kicks of the sub-samples, finding that there is a statistically significant difference between the kicks of MSPs, and the kicks of NS LMXBs, redbacks, and black widows. The difference cannot be simply explained by known observational selection effects, suggesting this difference may be physical in nature. Whilst there is no statistically significant difference between the kicks of NS LMXBs, redbacks, and black widows, this is potentially due to the small sample sizes and may not mean that the intrinsic kick distributions are truly the same. 

We modelled the kick distributions of both the full sample, and each sub-sample, testing a unimodal and bimodal truncated Gaussian, a unimodal and bimodal Maxwellian, and a Beta distribution. In all cases the Beta distribution was determined to be the superior model. The full sample best-fit Beta distribution (Equation \ref{eq:beta}) has parameters $\alpha = 3.05^{+0.32}_{-0.30}$, $\beta = 14.6^{+2.2}_{-2.1}$, and $s = 563^{+72}_{-68}$ \si{\km \per \second}. This distribution has a mode of $73.8^{+5.3}_{-5.4}$ \si{\km \per \second} and a mean of $97.3^{+4.9}_{-4.7}$ \si{\km \per \second}.

We find that the binary kicks of NS binaries suggest that NSs in binaries receive significantly lower natal kicks than isolated pulsars. This is expected due to the selection effect against large natal kicks in binaries, as large kicks will disrupt the binary. Ignoring the significant difference between natal and binary kick, distributions fit to isolated NSs will predict $\approx 4-100$ times fewer NSs receiving low kicks (defined in this work to be $\leq50$ \si{\km \per \second}), depending on the specific model used, than when using the best-fit model presented in this work.

Existing studies of MSP velocities report mean transverse 2D speeds. Note that the current velocities of old systems are not representative of the kick they received due to acceleration in the Galactic potential. With the exception of \citet{Lynch18}, the mean velocities of all other examined works are consistent with the best fit model fit of both the full sample and the MSP susbsample.

Comparing the binary kicks of NS LMXBs in this work to BH LMXBs in \citet{Atri19} shows that while the BHs in the sample are 2--15 times more massive than NSs, the mean of the NS LMXB binary kicks are only $\approx50$\% larger. This supports the theory that the kicks of BHs are not purely rescaled from NS kicks by the remnant mass.

We find the standard NS kick distribution assumed in the literature, the $\sigma=265$ \si{\km \per \second} Maxwellian from \citet{Hobbs05}, severely underestimates the fraction of NSs that receive low kicks. This underestimation is clear when comparing to modern kick distributions fit to young, isolated pulsars, clearer still when compared with the binary kicks from this work. We emphasise the importance and need for a new NS natal kick distribution for binary modelling, specifically DNS formation and DNS merger rates, as kicks can both disrupt binaries and significantly alter future binary evolution.

\section*{Acknowledgements}

The authors thank Frank~Verbunt, Danny~C.~Price, N.~D.~Ramesh Bhat, Adam~T.~Deller and Samuel~J.~McSweeney for helpful discussions. The authors also thank the reviewer, Andrei~P.~Ioghsev, for his constructive comments that helped improve this work. T.N.O'D~was supported by a Forrest Research Foundation Scholarship, and an Australian Government Research Training Program (RTP) Stipend and RTP Fee-Offset Scholarship. I.M.~acknowledges support from the Australian Research Council Centre of Excellence for Gravitational  Wave  Discovery  (OzGrav), through project number CE17010004. I.M.~is a recipient of the Australian Research Council Future Fellowship FT190100574. Part of this work was performed at the Aspen Center for Physics, which is supported by National Science Foundation grant PHY-1607611.  The participation of I.M.~at the Aspen Center for Physics was partially supported by the Simons Foundation. P.A.~was supported by Vici research program 'ARGO' with project number 639.043.815, financed by the Dutch Research Council (NWO). J.S.~acknowledges support from NSF grant AST-2205550 and the Packard Foundation. This work was supported by the Australian government through the Australian Research Council’s Discovery Projects funding scheme (DP200102471). We acknowledge extensive use of the SIMBAD database \citep{simbad}, NASA’s Astrophysics Data System, and arXiv. This work has made use of data from the European Space Agency (ESA) mission
{\it Gaia} (\url{https://www.cosmos.esa.int/gaia}), processed by the {\it Gaia}
Data Processing and Analysis Consortium (DPAC,
\url{https://www.cosmos.esa.int/web/gaia/dpac/consortium}). Funding for the DPAC
has been provided by national institutions, in particular the institutions
participating in the {\it Gaia} Multilateral Agreement.

The analysis and visualisation presented in this paper have been performed using the following packages: 
\textsc{aladin} \citep{aladin1}
\textsc{Galpy} \citep{galpy}, 
\textsc{iPython} \citep{ipython}, 
\textsc{Matplotlib} \citep{matplotlib}, 
\textsc{Numpy} \citep{numpy}, 
\textsc{Pandas} \citep{pandas}, 
\textsc{PyMC3} \citep{pymc3}, 
\textsc{SAO DS9} \citep{ds9}, 
and \textsc{Scipy} \citep{scipy}.

%%%%%%%%%%%%%%%%%%%%%%%%%%%%%%%%%%%%%%%%%%%%%%%%%%
\section*{Data Availability}
All data used in this work are publicly available in the ATNF Pulsar catalogue, Gaia EDR3/DR3 archive, and in the literature \citep{Strader19}. The algorithm used for estimating kicks is accessible in \citet{Atri19}. All the results presented in this work are tabulated throughout the manuscript and are also available in machine-readable format with the online version of the work.

%%%%%%%%%%%%%%%%%%%% REFERENCES %%%%%%%%%%%%%%%%%%

% The best way to enter references is to use BibTeX:

\bibliographystyle{mnras}
\bibliography{ref} % if your bibtex file is called example.bib

%%%%%%%%%%%%%%%%%%%%%%%%%%%%%%%%%%%%%%%%%%%%%%%%%%

%%%%%%%%%%%%%%%%% APPENDICES %%%%%%%%%%%%%%%%%%%%%

\appendix

\section{Radial Velocity Prior Derivation}
\label{appendix:RV_prior}
Here, we derive the observed radial component of an object in the plane using the same notation as \citet{Verbunt17}.

This can be achieved by rotating the $\textbf{U}$, $\textbf{V}$ coordinate system such that $\textbf{U}^\prime$ is parallel to line joining \textit{S} to \textit{P}, i.e.,
\begin{equation}
\begin{aligned}
    \textbf{U}^\prime &= \textbf{U}\cos l + \textbf{V}\sin l,\\ 
    \textbf{V}^\prime &= -\textbf{U}\sin l + \textbf{V}\cos l.
\end{aligned}
\end{equation}
Thus the observed radial velocity is simply the $\textbf{U}^\prime$ component. Therefore,
\begin{equation}
    v_r = [\cos l,\sin l,0] \cdot (\textbf{v}_p - \textbf{v}_\odot).
\end{equation}
We are interested in the component of the radial velocity that comes from the Sun's peculiar velocity and from Galactic rotation, without the contribution from the object's peculiar velocity. Thus the estimate of systemic radial velocity used in our prior is
\begin{equation}
    v_{r,G} = -U \cos l -(V + v_R(R_0))\sin l + v_R(R)\sin \theta_l.
\end{equation}
As the source may be closer to the Galactic centre where $v_R(R_0) \neq v_R(R)$, we estimate $v_R(R)$ assuming the MWPotential2014 model for the Galactic potential as implemented in \textsc{Galpy} \citep{galpy}.

\section{Millisecond Pulsar Ridgeline Plots}
\label{appendix:MSP}

\begin{figure}
	\includegraphics[width=\columnwidth]{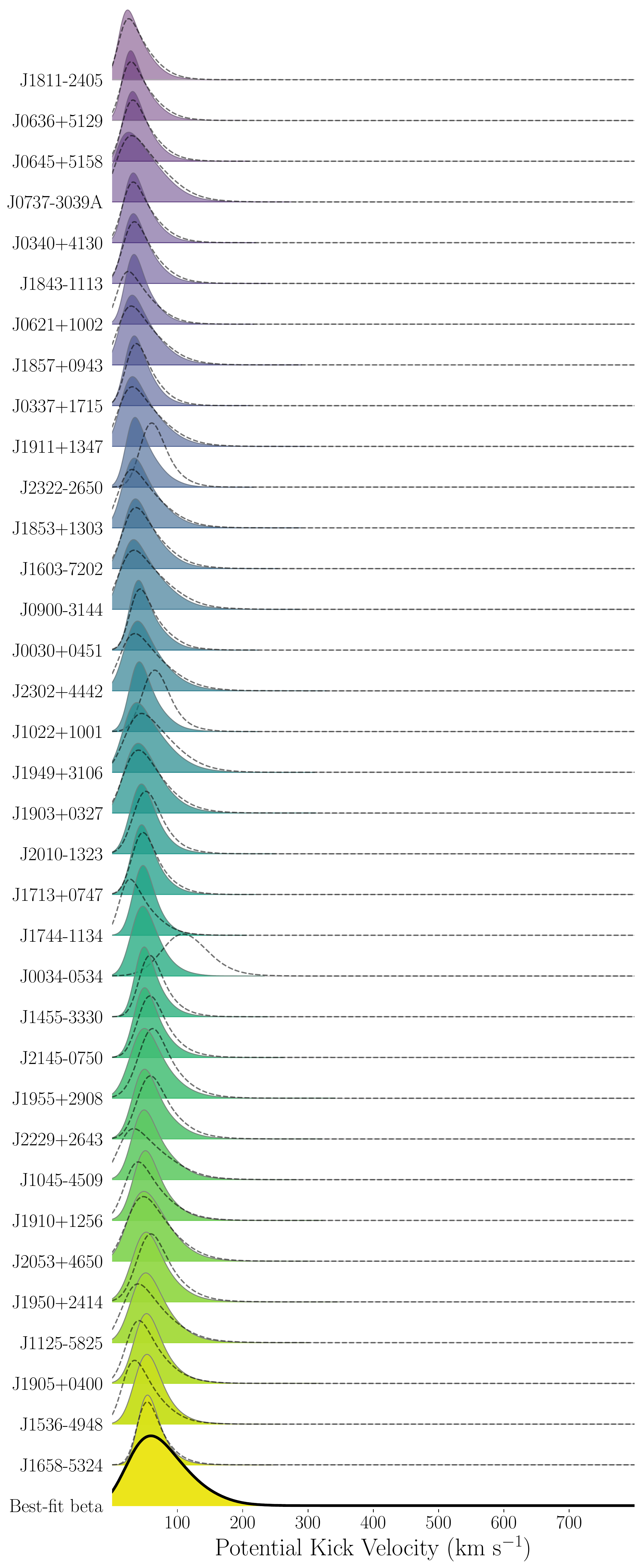}
    \caption{Ridgeline plots showing scaled PDFs of the PKV distribution for the 35 MSP systems with the lowest PKV estimate, calculated using the NE2001 (solid line and filled) and YMW16 (dashed line) DM distance model. Here, and in Figures \ref{fig:msp_ridge_2} and \ref{fig:msp_ridge_3}, the order is based on median PKV based on the NE2001 DM model. The `Best-fit beta' curve is the best-fit beta distribution to the full sample of MSPs with NE2001 DM distances.}
    \label{fig:msp_ridge_1}
\end{figure}

\begin{figure}
	\includegraphics[width=\columnwidth]{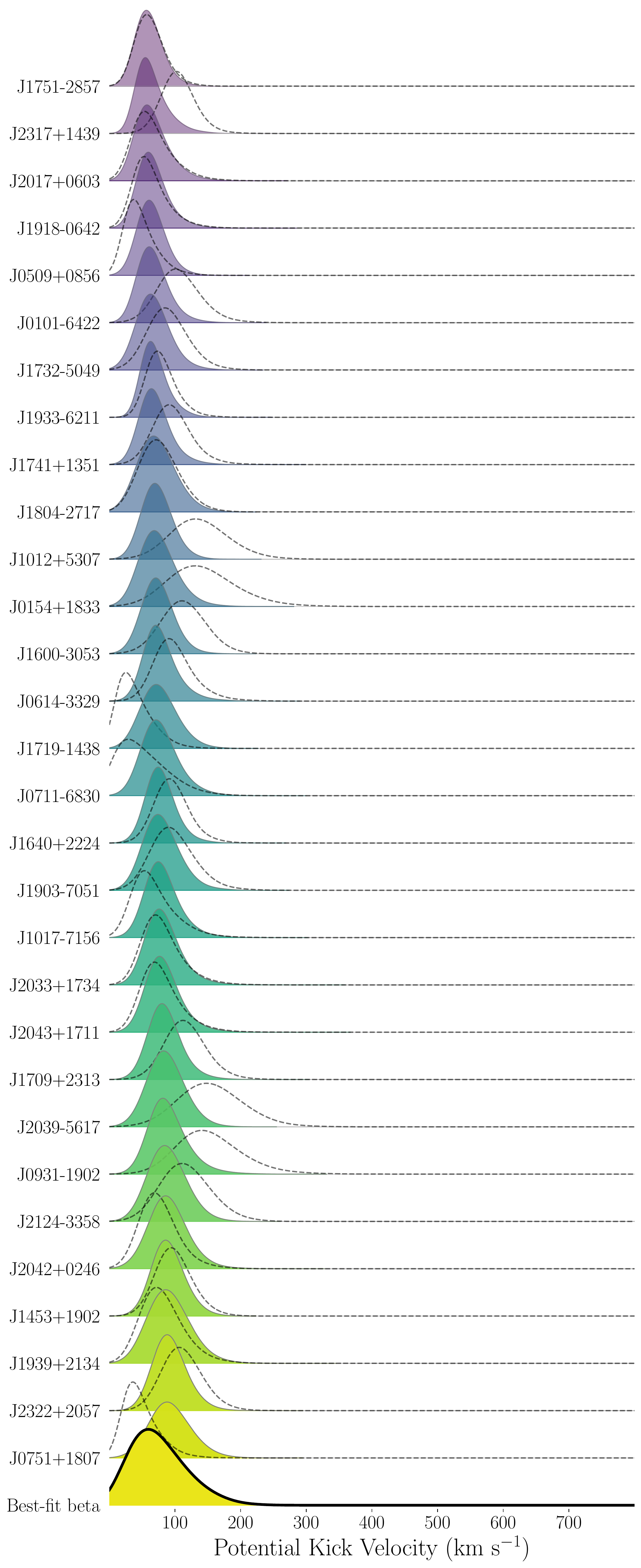}
    \caption{Continuation of Figure \ref{fig:msp_ridge_1} showing scaled PDFs of the PKV distribution for the next 30 MSP systems, calculated using the NE2001 (solid line and filled) and YMW16 (dashed line) DM distance model. The `Best-fit beta' curve is the best-fit beta distribution to the full sample of MSPs with NE2001 DM distances.}
    \label{fig:msp_ridge_2}
\end{figure}

\begin{figure}
	\includegraphics[width=\columnwidth]{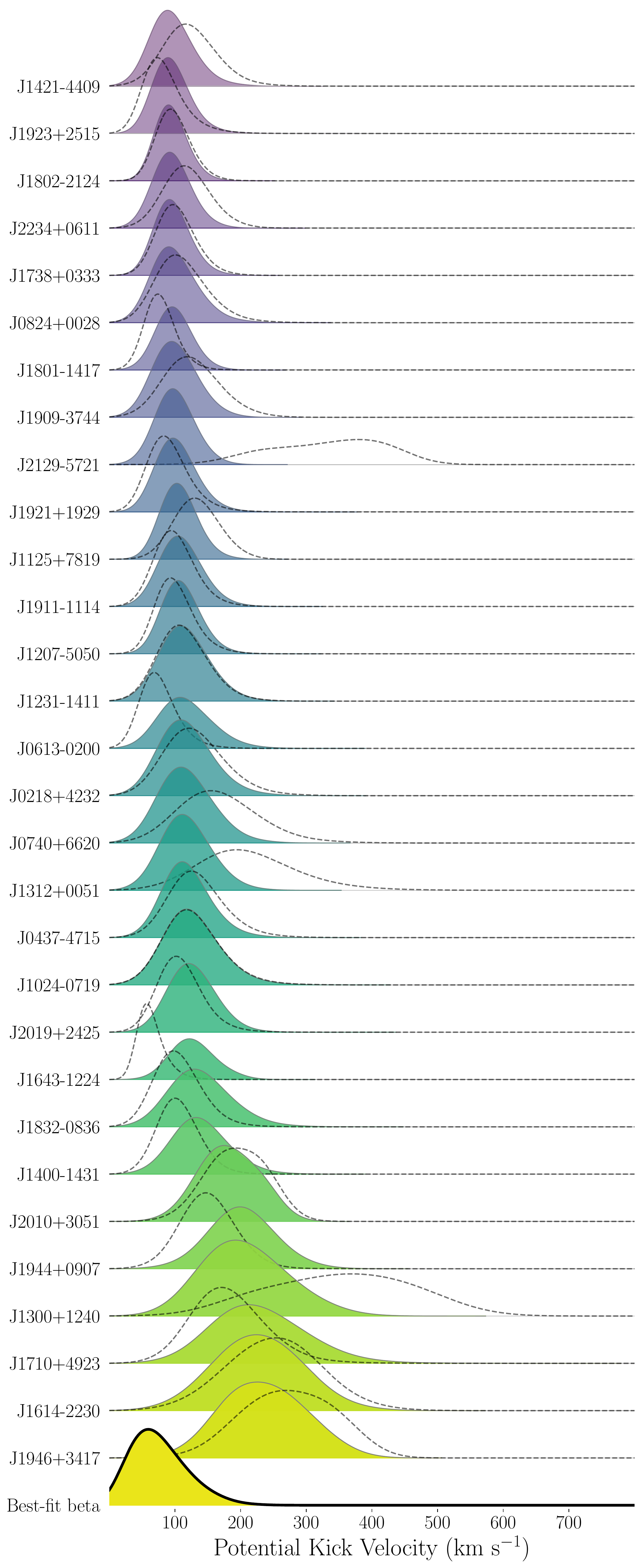}
    \caption{Ridgeline plots showing scaled PDFs of the PKV distribution for the 30 MSP systems with the highest PKV estimate, calculated using the NE2001 (solid line and filled) and YMW16 (dashed line) DM distance model. The `Best-fit beta' curve is the best-fit beta distribution to the full sample of MSPs with NE2001 DM distances.}
    \label{fig:msp_ridge_3}
\end{figure}

\section{Subsample Model Fit Comparisons}
\label{appendix:subsample_model_comp}

\begin{figure}
    	\includegraphics[width=\columnwidth]{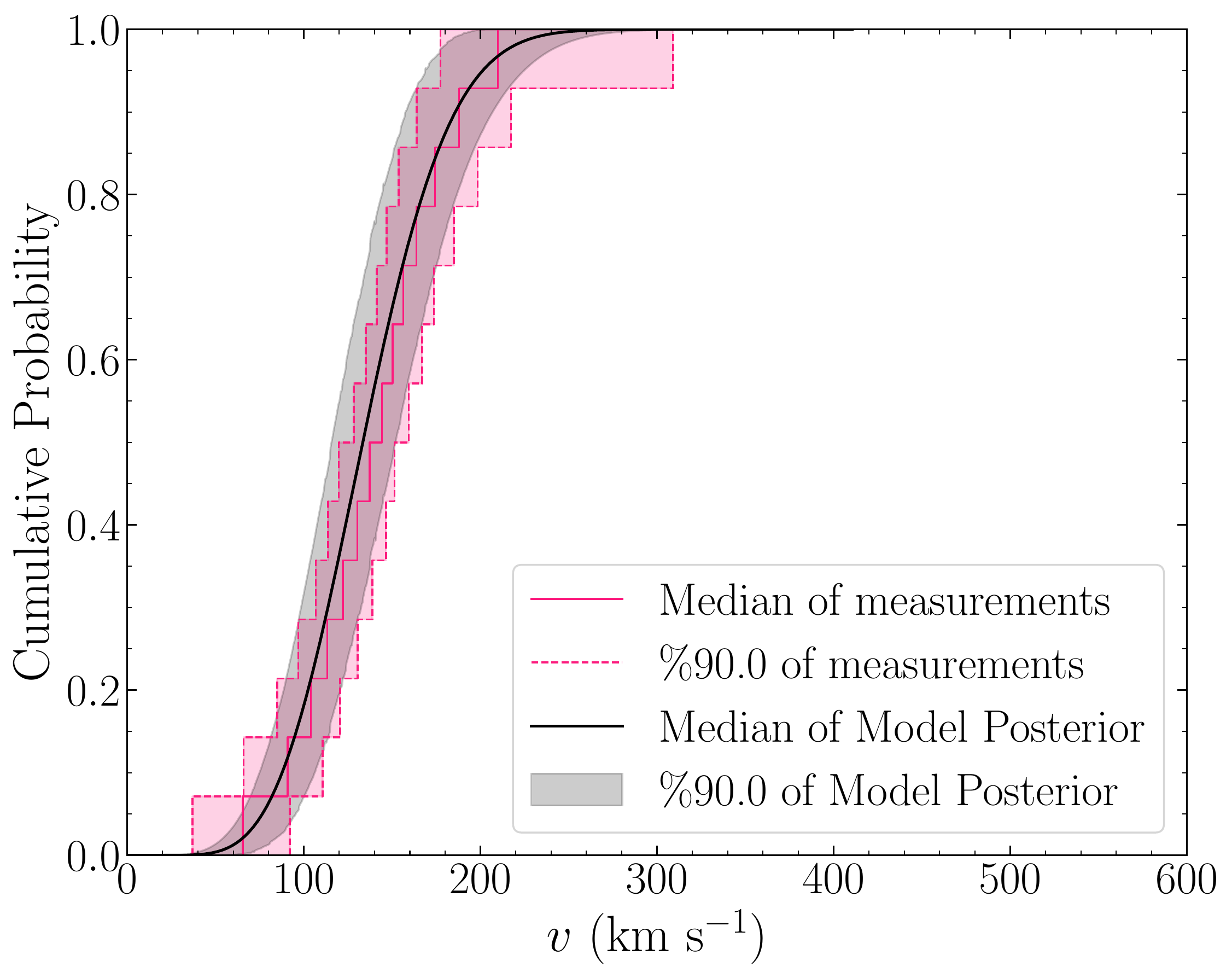}
        \caption{Comparing eCDFs constructed from the redback PKV MC realisations to the posterior of the beta model fitting. This Figure was constructed in the same way as the right panel of Figure \ref{fig:beta_pdf_cdf}.}
        \label{fig:cdf_90_rbs}
\end{figure}

\begin{figure}
    	\includegraphics[width=\columnwidth]{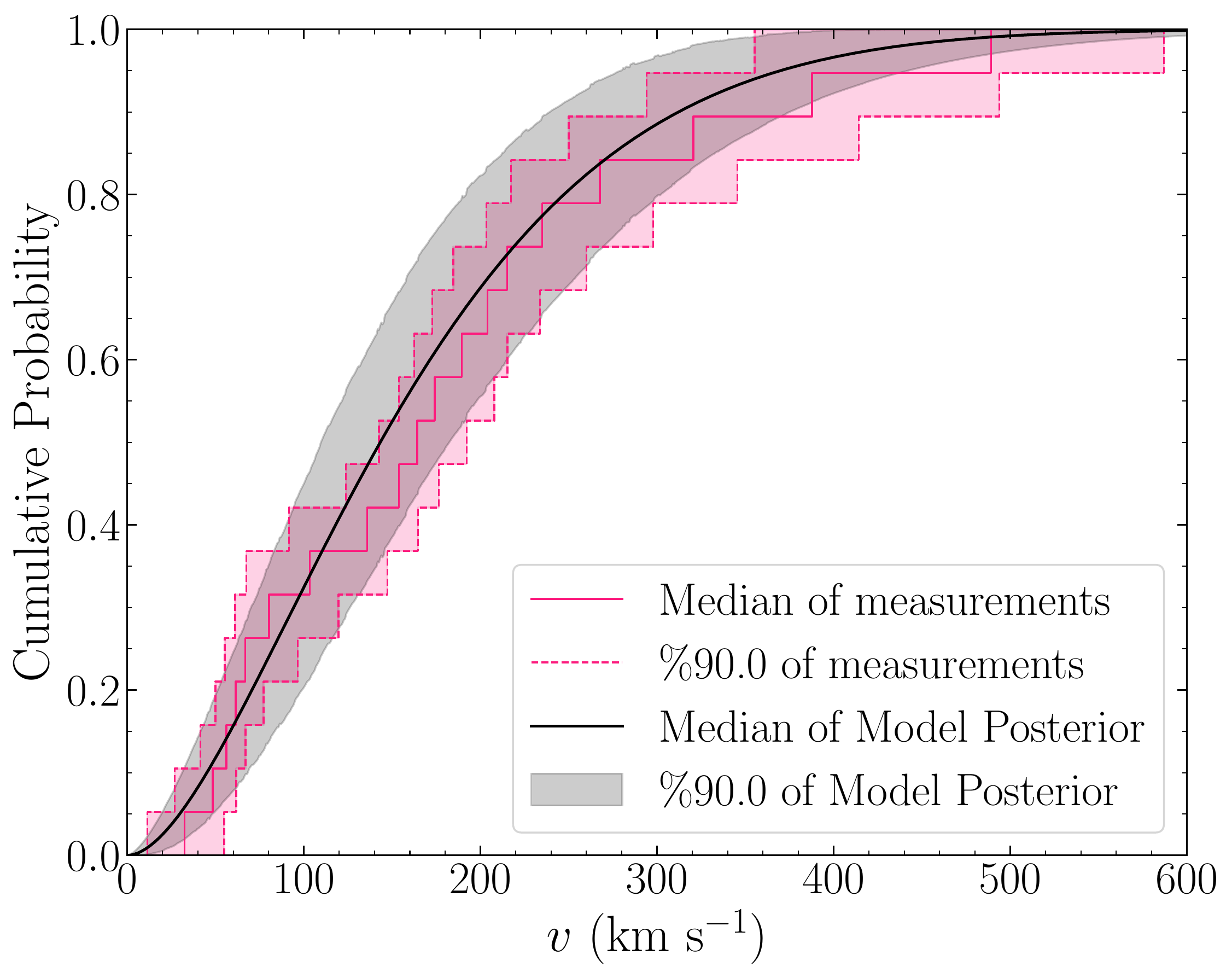}
        \caption{Comparing eCDFs constructed from the NS LMXB PKV MC realisations to the posterior of the beta model fitting. This Figure was constructed in the same way as the right panel of Figure \ref{fig:beta_pdf_cdf}.}
        \label{fig:cdf_90_nsxbs}
\end{figure}

\begin{figure}
    	\includegraphics[width=\columnwidth]{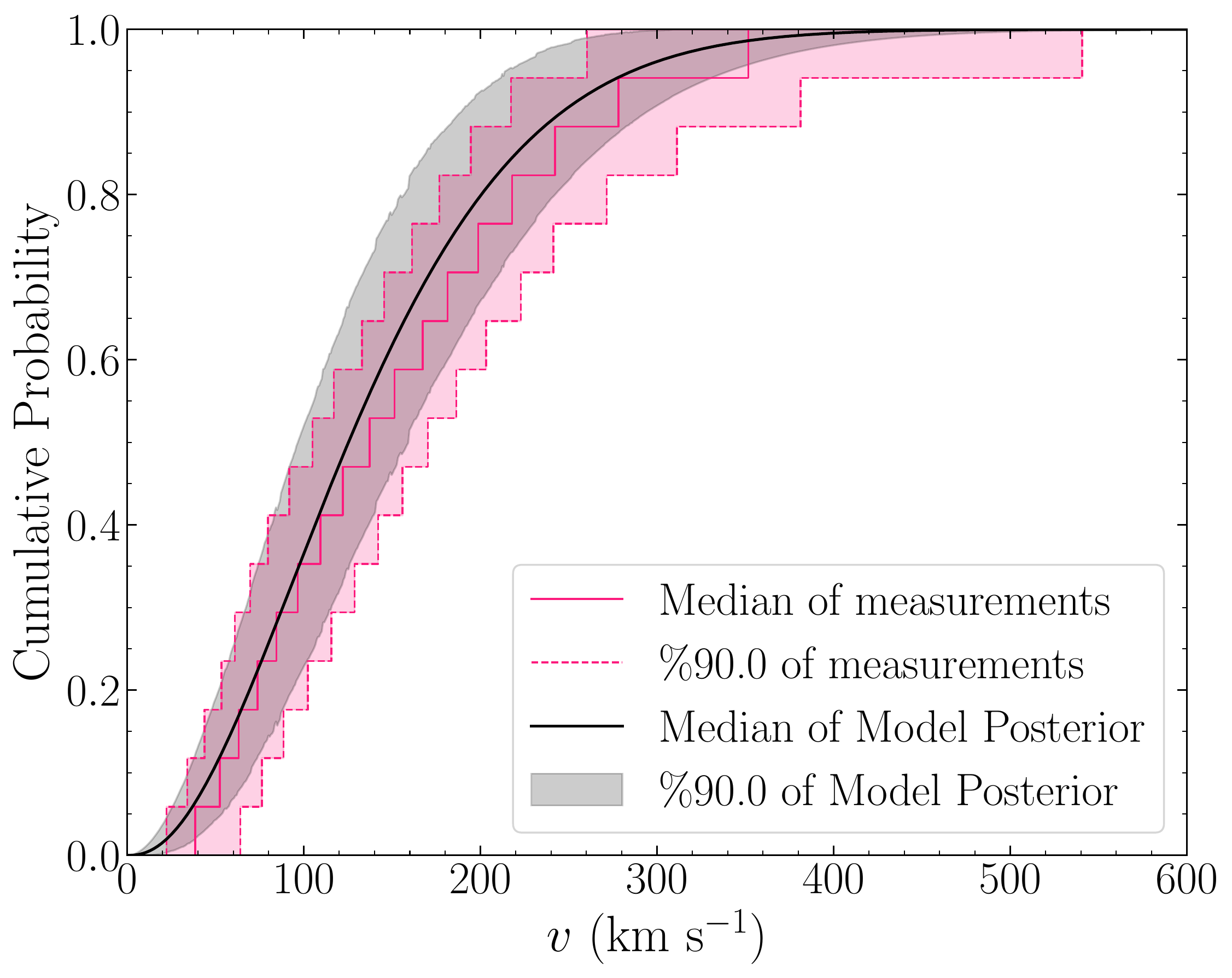}
        \caption{Comparing eCDFs constructed from the black widow PKV MC realisations to the posterior of the beta model fitting. This Figure was constructed in the same way as the right panel of Figure \ref{fig:beta_pdf_cdf}.}
        \label{fig:cdf_90_bws}
\end{figure}

\begin{figure}
    	\includegraphics[width=\columnwidth]{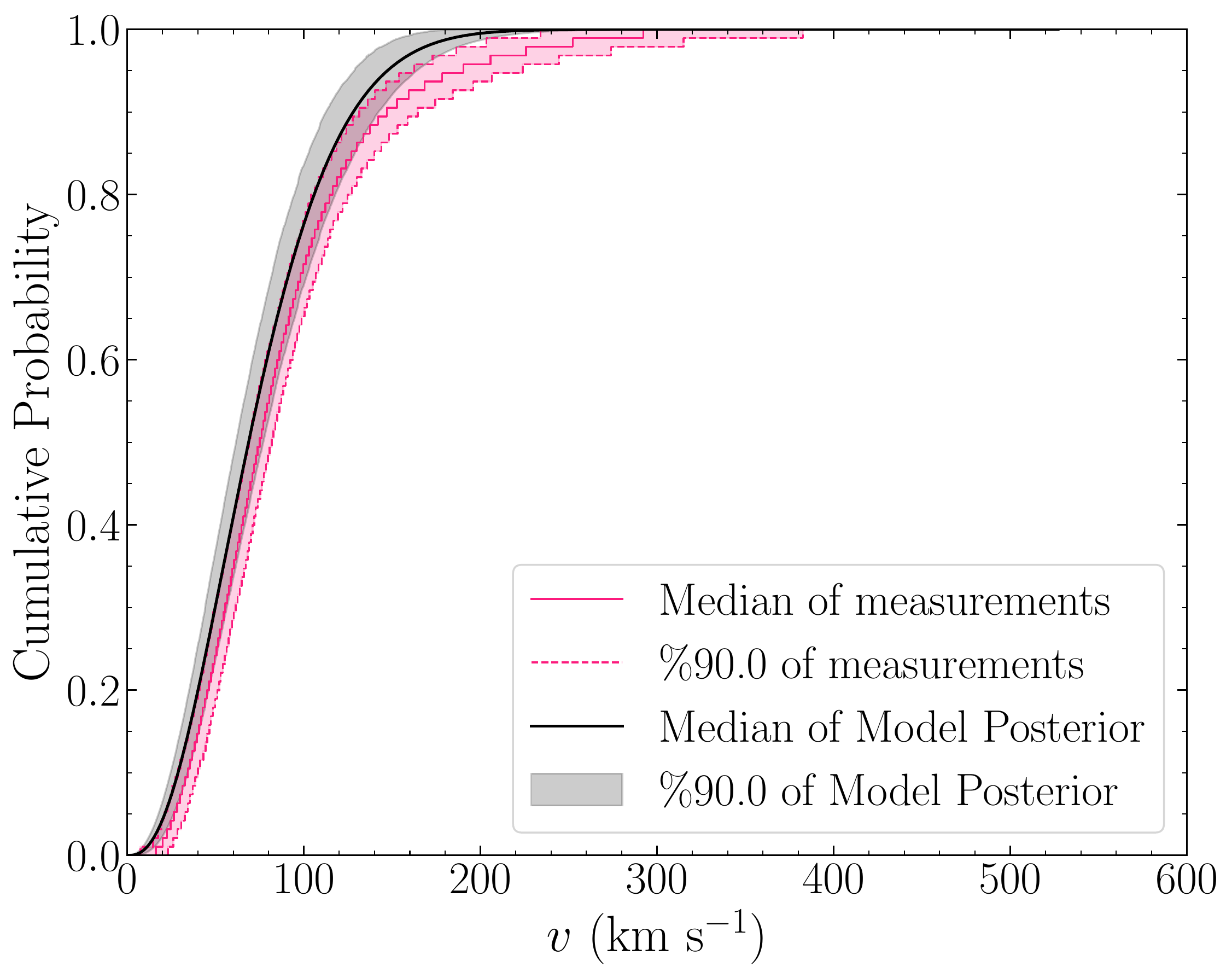}
        \caption{Comparing eCDFs constructed from the MSPs PKV MC realisations to the posterior of the beta model fitting. This Figure was constructed in the same way as the right panel of Figure \ref{fig:beta_pdf_cdf}.}
        \label{fig:cdf_90_msp}
\end{figure}

\section{Model Comparisons for the Whole Sample}
\label{appendix:model_comp}
\begin{figure}
    \includegraphics[width=\columnwidth]{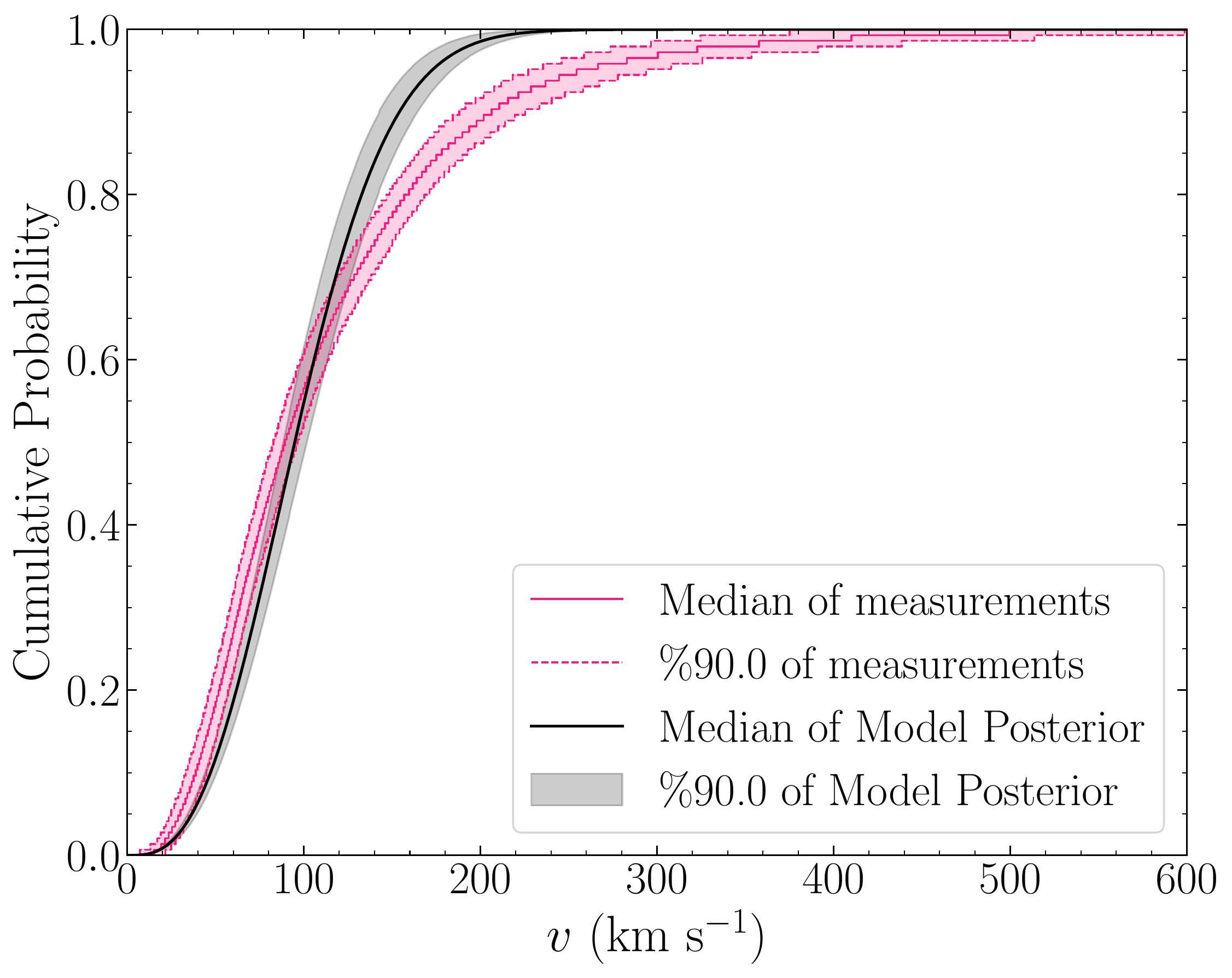}
        \caption{Comparing the posterior from Maxwellian model fitting to eCDFs constructed from the PKV MC realisations of the full sample. This Figure was constructed in the same way as the right panel of Figure \ref{fig:beta_pdf_cdf}, except for the Maxwellian modelling.}
        \label{fig:maxw}
\end{figure}

\begin{figure}
    \includegraphics[width=\columnwidth]{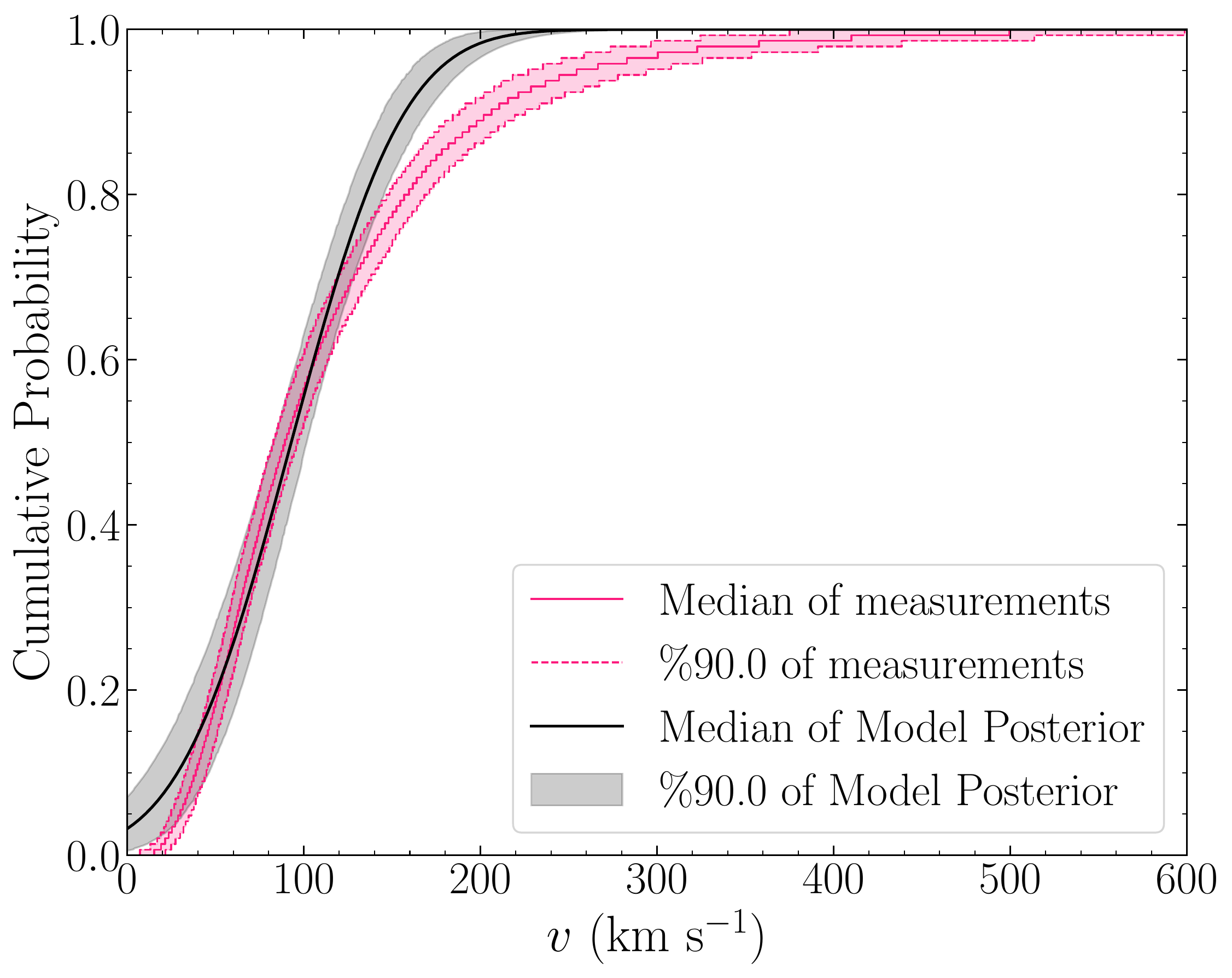}
        \caption{Comparing the posterior from unimodal truncated Gaussian model fitting to eCDFs constructed from the PKV MC realisations of the full sample. This Figure was constructed in the same way as the right panel of Figure \ref{fig:beta_pdf_cdf}, except for the unimodal truncated Gaussian modelling.}
        \label{fig:unigaus}
\end{figure}

\begin{figure}
    \includegraphics[width=\columnwidth]{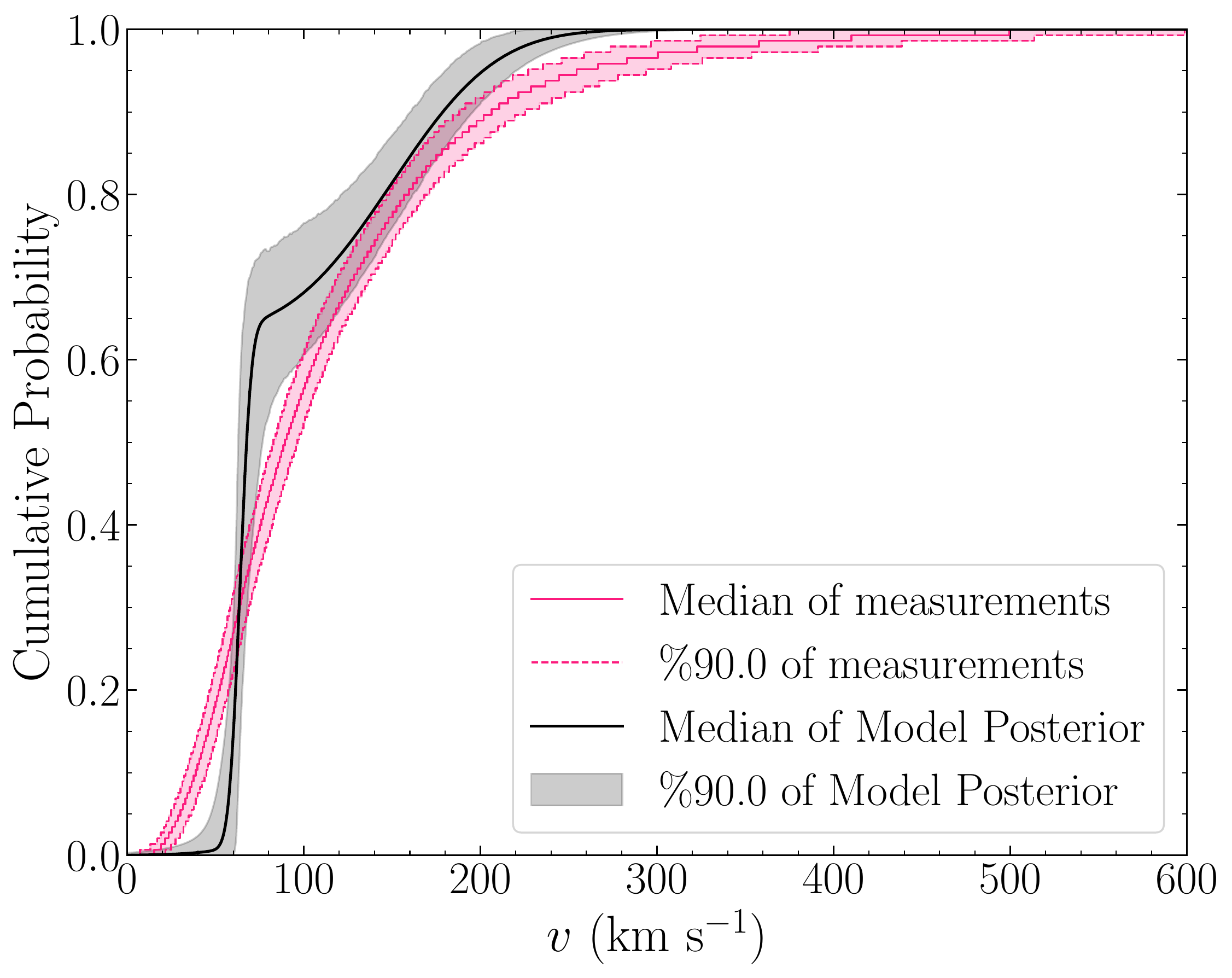}
        \caption{Comparing the posterior from bimodal truncated Gaussian model fitting to eCDFs constructed from the PKV MC realisations of the full sample. This Figure was constructed in the same way as the right panel of Figure \ref{fig:beta_pdf_cdf}, except for the bimodal truncated Gaussian modelling.}
        \label{fig:bigaus}
\end{figure}

\section{Binary Masses}
\label{appendix:mass_table}
\begin{center}
    \begin{table*}
        \begin{tabular}{lcccc}
            \hline
            Source & $M_{\mathrm{CO}}$ ($M_\odot$) & $M_{\mathrm{star}}$ ($M_\odot$) & $q$ & References\\
            \hline
            2S 0921-630 & $1.44 \pm0.1$         & $0.35\pm0.3$ & --         & [1]\\
            Cen X-4     & $1.51^{+0.4}_{-0.55}$ & $0.23\pm0.1$ & --         & [2]\\
            Sco X-1     & $1.4^{+1.4}_{-0.5}$   & $0.7^{+0.8}_{-0.3}$ & --  & [3]\\
            4U 1636-536 & --                    & -- & $0.76\pm0.47$               & [4]\\
            Her X-1     & $1.5\pm0.3$           & $2.3\pm0.3$ &     --     & [5]\\
            IGR J17062-6143  & --                    & $0.006\pm0.001$ & --      & [6]\\
            4U 1700+24  & --                    & -- & $0.00214\pm0.00047$               & [7]\\
            GX 1+4      & --                    & -- & $0.371\pm0.026$               & [8]\\
            Aql X-1 & --                    & -- & $0.43\pm0.44$               & [9]\\
            GRS 1915+105 & $12.4^{+2.0}_{-1.8}$ & -- & $0.042 \pm 0.024$ & [10,11] \\
            Cyg X-1 & $21.1\pm2.2$ & $40.6^{+7.7}_{-7.1}$ & -- & [12] \\
            \hline
        \end{tabular}
        \caption{Masses of the compact object ($M_{\mathrm{CO}}$) and companion star ($M_{\mathrm{star}}$) for the systems included in Figure \ref{fig:pkv_mass_comp}. The mass ratio ($q$) is only reported if it was used to estimate the total system mass in conjunction with either $M_{\mathrm{CO}}$ or $M_{\mathrm{star}}$. Masses for all the redback systems were adapted from \citet[and references therein]{Strader19}. Unless otherwise specified, all BH XRB masses were adapted from \citet[and references therein]{Atri19}. NS LMXB sources with only a reported $q$ had $M_{\mathrm{CO}}$ assumed to be $1.7\pm0.3$, and then $M_{\mathrm{star}}$ calculated using $q$ and  $M_{\mathrm{CO}}$.\\
        \textbf{References:} [1] \citet{Ashcraft12}; [2] \citet{Hammerstein18}; [3] \citet{Wang18}; [4] \citet{Casares06}; [5] \citet{Reynolds97}; [6] \citet{matasanchez17}; [7] \citet{Hinkle19}; [8] \citet{Hinkle06}; [9] \citet{Strohmayer18}; [10] \citet{Reid14}; [11] \citet{Steeghs13}; [12] \citet{Miller-Jones21}.}
        \label{tab:mass_table}
    \end{table*}
\end{center}

%%%%%%%%%%%%%%%%%%%%%%%%%%%%%%%%%%%%%%%%%%%%%%%%%%

% do not change these lines
\bsp	% typesetting comment
\label{lastpage}
\end{document}